\documentclass[preprintnumbers, showpacs, floatfix,onecolumn,preprintnumbers, letterpaper, superscriptaddress,nofootinbib]{revtex4}
\usepackage{amsfonts}
\usepackage{amsmath}
\usepackage{amssymb,epsf}
\usepackage{latexsym}
\usepackage{graphicx,epsfig}
\usepackage{amssymb}
\usepackage{float}
\usepackage{subfigure}
\usepackage{epstopdf}
\usepackage[colorlinks=true,citecolor=blue,linkcolor=blue,urlcolor=black]{hyperref}
\usepackage{dcolumn}
\usepackage{psfrag}
\usepackage{wrapfig}
\usepackage{makeidx}
\usepackage{epsf}
\begin{document}
\title{Conductivity of the holographic $p$-wave superconductors with higher order corrections}
\author{Mahya Mohammadi}
\email{mahya689mohammadi@gmail.com}
\affiliation{Physics Department and Biruni Observatory, Shiraz University, Shiraz 71454,
Iran}

\author{Ahmad Sheykhi}
\email{asheykhi@shirazu.ac.ir}
\affiliation{Physics Department and Biruni Observatory,
 Shiraz University, Shiraz 71454,
Iran} \affiliation{Research Institute for Astronomy and
Astrophysics of Maragha (RIAAM), P. O. Box: 55134-441, Maragha,
Iran}

\begin{abstract}
We investigate the holographic $p$-wave superconductors in the
presence of the higher order corrections on the gravity as well as
on the gauge field side. On the gravity side, we add the
Gauss-Bonnet curvature correction terms, while on the gauge field
side we take the nonlinear Lagrangian in the form $\mathcal{F}+b
\mathcal{F}^{2}$, where $\mathcal{F}$ is the Maxwell Lagrangian
and $b$ indicates the strength of nonlinearity. We employ the
shooting method for the numerical calculations in order to obtain
the ratio of the critical temperature $T_{c}$ over
$\rho^{1/(d-2)}$. We observe that by increasing the values of the
mass and the nonlinear parameters the critical temperature
decreases and thus the condensation becomes harder to form. In
addition, the stronger Gauss-Bonnet parameter $\alpha$ hinders the
superconducting phase in Gauss-Bonnet gravity. Furthermore, we
calculate the electrical conductivity based on the holographic
setup. The real and imaginary parts are related to each other by
Kramers-Kronig relation which indicates a delta function and pole
in low frequency regime, respectively. However, at enough large
frequencies the trend of real part can be interpreted by
$Re[\sigma]=\omega^{(d-4)}$. Moreover, in holographic model the
ratio $\omega_{g}/T_{c}$ is always much larger than the BCS value
$3.5$ due to the strong coupling of holographic superconductors.
In both gravity kinds, decreasing the temperature or increasing
the effect of nonlinearity shifts the gap frequency toward larger
values. Besides, the gap frequency is occurred at larger values by
enlarging the Gauss-Bonnet parameter. In general, the behavior of
conductivity depends on the choice of the mass, the nonlinear and
the Gauss-Bonnet parameters.
\end{abstract}

\maketitle

\section{introduction}\label{section0}
In recent years, the gauge/gravity duality which connects a weak
gravitational system in $(d+1)$-dimension to the strong coupling
conformal field theory in $d$-dimension grabs a lot of attentions
because it provides a powerful theoretical methods to study strong
interacting systems such as high temperature superconductors
\cite{Maldacena,H08,G98,W98,HR08,R10}. One of the famous
consequence of AdS/CFT correspondence is the appearance of a
revolutionary theory which is called holographic superconductor
\cite{H08}. The idea of holographic superconductor was proposed in
$2008$ \cite{H08} by applying AdS$4$/CFT$3$ correspondence on the
probe limit in Einstein gravity. Based on this theory in order to
describe a superconductor on the boundary, we need a transition
from hairy black hole to a no hair black hole in the bulk for
temperatures below and upper the critical value, respectively
\cite{H11}. The appearance of hair corresponds to the spontaneous
$U(1)$ symmetry breaking \cite{H11}. This theory opened up a new
horizon in condensed matter physics to study the high temperature
superconductors as well as unconventional types \cite{H08,H09}.
Furthermore, holographic superconductors have widely investigated
in the presence of nonlinear electrodynamics which involve more
information than the usual Maxwell state. There are different
kinds of nonlinear electrodynamics in the literatures such as
Born-Infeld \cite{25}, Exponential \cite{hendi}, Logarithmic
\cite{log} and Power-Maxwell\cite{SSh16} electrodynamics. Many
investigations have been devoted to disclose analytically as well
as numerically different aspects of holographic superconductors in
the presence of various kinds of
 (see e.g.
\cite{Hg09,Gu09,HHH08,JCH10,SSh16,SH16,cai15,SHsh(17),
Ge10,Ge12,Kuang13,Pan11,CAI11, SHSH(16),shSh(16),Doa, Afsoon,
cai10,yao13,n4,n5,n6,Gan1,mahya}).

Besides the $s$-wave superconductors which can be described very
well by the BCS theory, there are unconventional superconductors
such as $p$-wave and $d$-wave superconductors\cite{BCS57,superp}.
Unconventional superconductors can be classified as high
temperature and strong coupling superconductors. It is expected
that the holographic hypothesis may shed some light on this field
by introducing holographic $p$-wave as well as $d$-wave
superconductors. In the past decade, several researches are done
to investigate the properties of unconventional superconductors
based on the holographic hypothesis (see e.g.
\cite{Caip,cai13p,Donos,Gubser,chaturverdip15,Roberts8,zeng11,cai11p,pando12,momeni12p,gangopadhyay12,mahyap,francessco1,francessco2}).
The holographic $p$-wave superconductors can be interpreted as odd
parity or triplet superconductors which are constructed by
coupling of electrons with parallel spins by exchange of the
electronic excitations with angular momentum $\ell =
1$\cite{superp}. The holographic $p$-wave superconductors have
been explored from different points of view such as condensation
of a complex or real charge vector field in the bulk which
corresponds to the vector order parameter in the boundary or the
spin-$1$ order parameter can be originated from the condensation
of a $2$-form field in the gravity side
\cite{Caip,cai13p,Donos,Gubser,chaturverdip15}. Introducing a
$SU(2)$ Yang-Mills gauge field in the bulk in which one of the
gauge degrees of freedom corresponds to the vector order parameter
at the boundary, is another method to study this topic.
Superconducting phase at the boundary for this type of holographic
superconductors corresponds to appearance of vector hair outside
the horizon by decreasing the temperature below the critical
value\cite{mahyap}.

In this work, we are going to investigate the holographic $p$-wave
superconductors in all higher dimensions by taking into account
the higher order corrections both on the gravity as well as on the
gauge field sides. While, most previous works on the holographic
$p$-wave superconductors have been investigated in the presence of
the linear Maxwell field, it's interesting to examine the effects
of the nonlinear electrodynamics on the properties of holographic
$p$-wave superconductors. For the correction to the gauge field
side, we consider the general nonlinear electrodynamics with
higher order correction term, namely $\mathcal{L}=\mathcal{F}+b
\mathcal{F}^{2}$. We shall do the numerical calculations for
different values of the mass $m$ and the nonlinear parameter $b$
in each dimension to disclose the effects of these terms on the
critical temperature. In all cases, we find the relation between
critical temperature $T_{c}$ and $\rho^{1/(d-2)}$ where $\rho$ is
regarded as charge density and plot the behavior of condensation
as a function of temperature. We shall also explore the electrical
conductivity by applying an appropriate perturbation on the gauge
field in the background. We obtain the electrical conductivity
formula and plot the behavior of the real and imaginary parts of
conductivity as a function of frequency for different values of
the mass and nonlinearity parameter in $d=4,5$ and $6$. Not only
the trend of figures differs by dimension but also our choice of
mass and nonlinearity has a straight effect on the behavior of
conductivity. Although the obvious differences, all of them follow
some universal behaviors. For example, the Kramers-Kronig relation
relates the real and imaginary parts of conductivity. In low
frequency regime, we observe a delta function behavior for real
part while the imaginary part has a pole. The infinite DC
conductivity is a feature of superconducting phase. Furthermore,
we find the universality $\omega_{g} \simeq 8 T_{c}$ is totally
dominated but deviates in higher dimensions which is logical.
Moreover, the gap frequency depends on the mass and nonlinearity
parameters. Afterwards, by following the same procedure as before
we do our study in Gauss-Bonnet gravity with higher order
corrections in gravity and gauge fields in $d$-dimensional
spacetime. Holographic $p$-wave superconductor in Gauss-Bonnet
gravity previously studied in \cite{caipp,gaussp1}. We analyze the
vector condensation and find that the critical temperature reduces
not only by rising the mass and nonlinear parameter but also by
enlarging the Gauss-Bonnet parameter $\alpha$. Finally, we
consider the electrical conductivity for holographic $p$-wave
superconductor in Gauss-Bonnet gravity with higher order
corrections. The global trends are also seen in this case. In this
gravity, the gap frequency depends on the mass, nonlinearity and
the Gauss-Bonnet parameters and the ratio of $\omega_{g}/T_{c}$
deviates from the universal value $8$ by increasing the effect of
$\alpha$, too.

This work is outlined as follows. In section \ref{section1} we
introduce the holographic $p$-wave model in Einstein gravity
through condensation a vector field. In section \ref{section2}, we
describe the procedure to calculate electrical conductivity based
on the AdS/CFT correspondence. Section \ref{section3} is devoted
to conductor/superconductor phase transition in holographic setup
in Gauss-Bonnet gravity. We calculate the electrical conductivity
in the Gauss-Bonnet gravity with higher order corrections in
section \ref{section4}. Finally, we summarize our results in
section \ref{section5}.

\section{Holographic $p$-wave superconductor in Einstein gravity}
\subsection{The holographic model and condensation of the vector field}\label{section1}
We adopt the following form for the action to describe a
holographic $p$-wave superconductor with a vector field
$\rho_{\mu}$ with mass $m$ and charge $q$
\begin{eqnarray}
&&S =\int d^{d}x\sqrt{-g} \left[\mathcal{L}_{G}+\mathcal{L}_{m}\right], \notag \\
&& \mathcal{L}_{G}= R-2 \Lambda , \ \ \ \
\mathcal{L}_{m}=
\mathcal{L}_{\mathcal{NL}}-\frac{1}{2}\rho_{\mu\nu}^{\dagger}
\rho^{\mu\nu}-m^{2} \rho_{\mu}^{\dagger} \rho^{\mu} + i q \gamma
\rho_{\mu} \rho_{\nu}^{\dagger} F^{\mu\nu} ,\label{act}
\end{eqnarray}%
where $g$ and $R$ are metric determinant and Ricci scalar,
respectively, $l$ is the radius of the AdS spacetime, which is
related to the negative cosmological constant via
\begin{equation}
\Lambda=-\frac{(d-1)(d-2)}{2 l^2}.
\end{equation}
Hereafter, for simplicity we set $l=1$. The Lagrangian density of
nonlinear electrodynamics $\mathcal{L}_{\mathcal{NL}}$ in the
Lagrangian of the
matter field, $\mathcal{L}_{m}$, is given by %
\begin{equation}\label{eqnon}
\mathcal{L}_{\mathcal{NL}}=\mathcal{F}+b \mathcal{F}^{2}, \ \   \
\ \mathcal{F}=-\frac{1}{4} F_{\mu\nu}F^{\mu\nu}.
\end{equation}%
When the nonlinearity parameter tends to zero, $b\rightarrow0$, it
reduces to the standard Maxwell Lagrangian, namely
$\mathcal{L}_{\mathcal{NL}}\rightarrow-1/4 F_{\mu\nu}F^{\mu\nu}$
where $F_{\mu\nu}=\nabla_{\mu} A_{\nu}-\nabla_{\nu} A_{\mu}$ and
$F_{\mu\nu}=\nabla_{\mu} A_{\nu}-\nabla_{\nu} A_{\mu}$. The term
$b\mathcal{F}^{2}$ is the first order leading nonlinear correction
term to the Maxwell field. There are several motivation for
choosing the nonlinear Lagrangian in the form of (\ref{eqnon}).
First, the series expansion of the three well-known Lagrangian of
nonlinear electrodynamics such as Born-Infeld, Logarithmic and
Exponential nonlinear electrodynamics have the form of
(\ref{eqnon}) \cite{Hendi1}. Second, calculating one-loop
approximation of QED, it was shown \cite{Ritz} that the effective
Lagrangian is given by (\ref{eqnon}). Besides, if one neglect all
other gauge fields, one may arrive at the effective quadratic
order of $U(1)$ as $\mathcal{F}^{2}$ \cite{Liu,Kats}. Furthermore,
considering the next order correction terms in the heterotic
string effective action one can obtain the $\mathcal{F}^{2}$ term
as a corrections to the bosonic sector of supergravity, which has
the same order as the Gauss-Bonnet term \cite{Liu,Kats,An}.

With the help of covariant derivative $D_{\mu}=\nabla_{\mu}- i q
A_{\mu}$, we can define $\rho_{\mu\nu}=D_{\mu} \rho_{\nu}-D_{\nu}
\rho_{\mu}$. The last term in the matter Lagrangian can be ignored
in our work because it characterizes the strength of interaction
between $\rho_{\mu}$ and $A_{\mu}$ with $\gamma$ as the magnetic
moment in the case with an applied magnetic field.

Varying action (\ref{act}) with respect to the gauge field
$A_{\mu}$ and the vector field $\rho_{\mu}$, we obtain the
equations of motion as

\begin{equation}\label{eqmax}
\nabla ^{\nu }\left[(1+2 b \mathcal{F})F_{\nu \mu }\right] =i q \left(\rho ^{\nu }
\rho ^{\dagger }{}_{\nu \mu }- \rho ^{\nu \dagger } \rho _{\nu \mu }\right)+
i q \gamma \nabla^{\nu} \left(\rho _{\nu } \rho ^{\dagger }{}_{\mu }-\rho ^{\dagger }{}_{\nu } \rho _{\mu } \right),
\end{equation}
\begin{equation}\label{eqvector}
D ^{\nu } \rho_{\nu \mu }-m^{2} \rho_{\mu }+ i q \gamma \rho^{\nu
} F_{\nu\mu} =0.
\end{equation}

\begin{figure*}[t]
\centering
\subfigure[~$m^{2}=0$]{\includegraphics[width=0.4\textwidth]{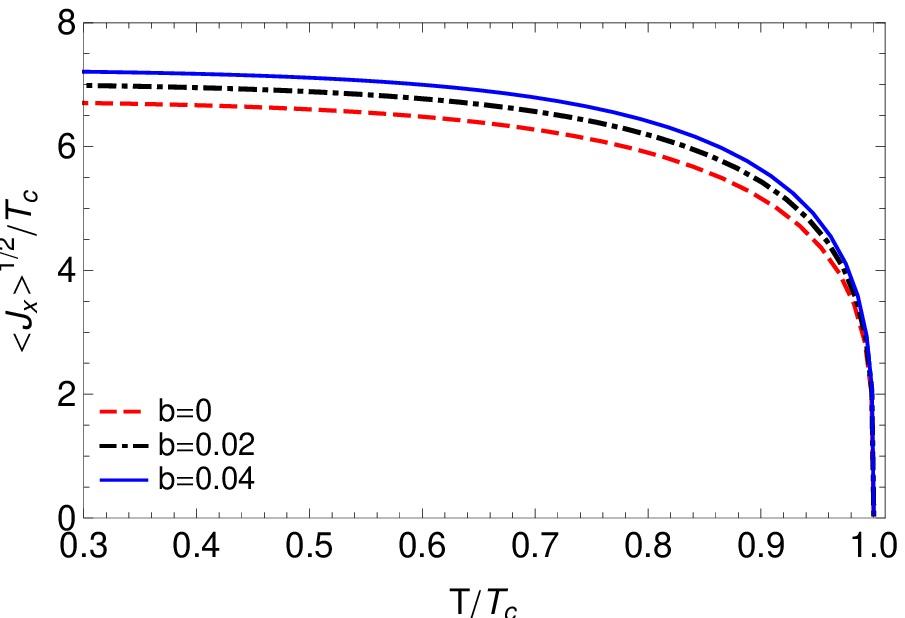}} \qquad %
\subfigure[~$m^{2}=3/4$]{\includegraphics[width=0.4\textwidth]{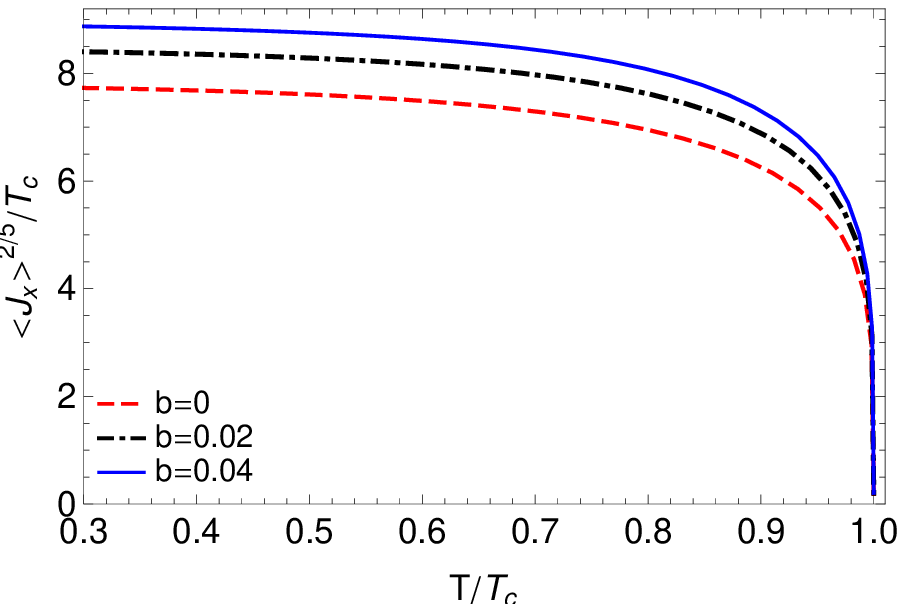}} \qquad %
\caption{The behavior of the condensation parameter as a function
of the temperature for different values of the mass and
nonlinearity parameters in $d=4$.} \label{fig1}
\end{figure*}

\begin{figure*}[t]
\centering
\subfigure[~$m^{2}=-3/4$]{\includegraphics[width=0.4\textwidth]{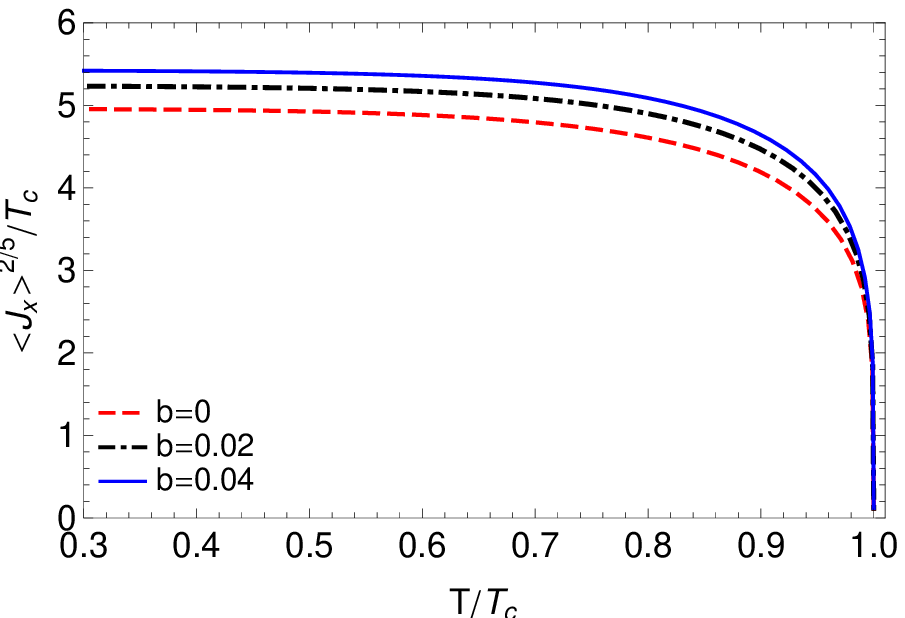}} \qquad %
\subfigure[~$m^{2}=5/4$]{\includegraphics[width=0.4\textwidth]{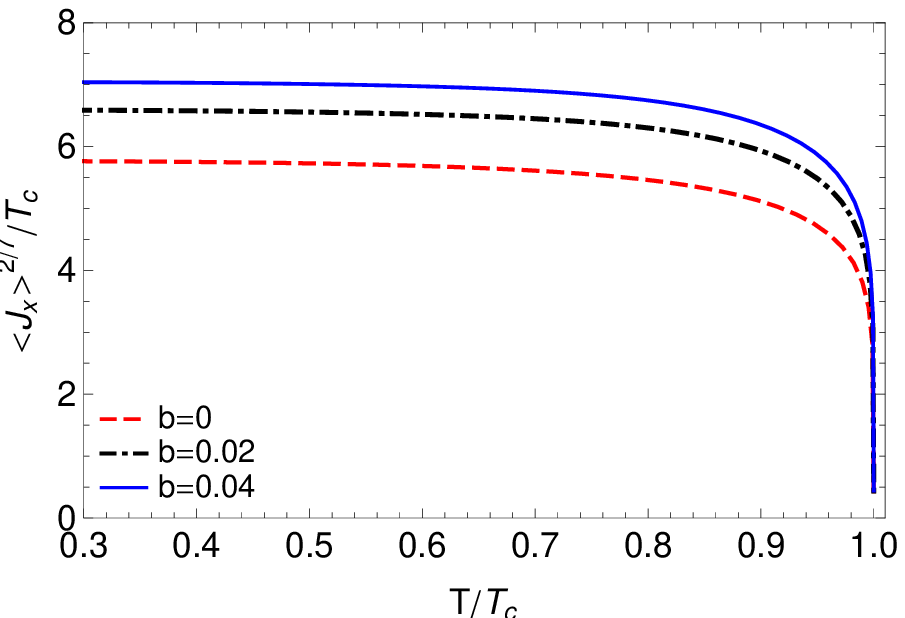}} \qquad %
\caption{The behavior of the condensation parameter as a function
of the temperature for different values of mass and nonlinearity
parameters in $d=5$.} \label{fig7}
\end{figure*}

\begin{figure*}[t]
\centering
\subfigure[~$m^{2}=-2$]{\includegraphics[width=0.4\textwidth]{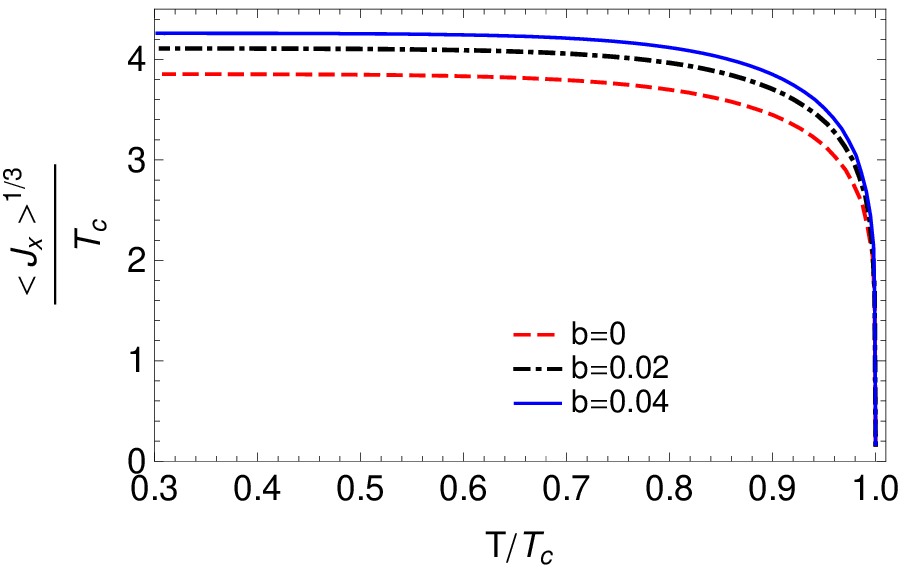}} \qquad %
\subfigure[~$m^{2}=-5/4$]{\includegraphics[width=0.4\textwidth]{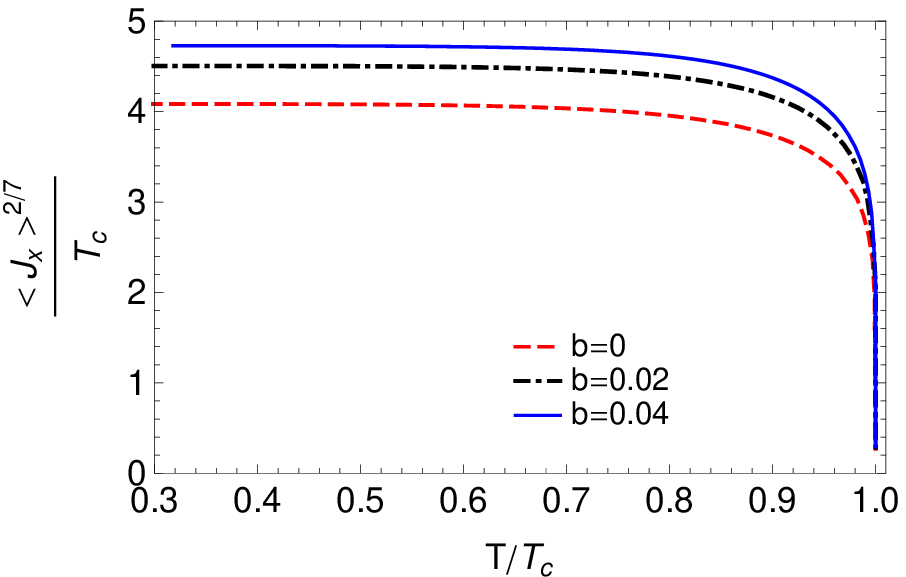}} \qquad %
\caption{The behavior of the condensation parameter as a function
of the temperature for different values of mass and nonlinearity
parameters in $d=6$.} \label{fig12}
\end{figure*}

In order to describe a $d$-dimensional AdS Schwarzschild black
hole with flat horizon, we consider the metric as
\begin{eqnarray} \label{metric}
&&{ds}^{2}=-f(r){dt}^{2}+\frac{{dr}^{2}}{f(r)}+r^{2} \sum _{i=1}^{d-2}{dx_{i}}^{2}%
,\\
&&f(r)=r^2-\frac{r_{+}^{d-1}}{r^{d-3}},\label{eqf} %
\end{eqnarray}%
where $r_{+}$ defines the horizon location obeying $f(r_{+})=0$.
While the vector and the gauge fields are assumed to have the
following form
\begin{eqnarray}
&& \rho_{\nu} dx^{\nu}=\rho_x(r) dx, \ \ \ \ A_{\nu} dx^{\nu}=\phi
(r) dt. \label{rhoA}
\end{eqnarray}%
 The Hawking
temperature of the black hole is given by \cite{mahya}
\begin{equation}
T=\frac{f^{^{\prime }}(r_{+})}{4\pi }=\frac{(d-1) r_{+}}{4\pi}. \label{temp}
\end{equation}%
Inserting relations (\ref{metric}) and (\ref{rhoA}) in the field
equations (\ref{eqmax}) and  (\ref{eqvector}), we arrive at
\begin{equation}\label{eqphi}
\phi ''(r)+\frac{  (d-2)}{r} \left[\frac{1+b \phi '^{2} (r)}{1+3 b
\phi '^{2} (r)}\right] \phi '(r)-\frac{2 \rho _x^{2} (r)}{r^2 f(r)
\left(1+3 b \phi '^{2}(r)\right)} \phi (r)=0,
\end{equation}
\begin{equation}\label{eqrho}
\rho _x''(r)+\left[\frac{(d-4)}{r}+\frac{f'(r)}{f(r)}\right]\rho _x'(r)
+\left[\frac{\phi^2 (r)}{f^2(r)}-\frac{m^2}{f(r)}\right]\rho _x(r)=0.
\end{equation}
These equations have the asymptotic solutions ($r\rightarrow
\infty$) of the form
\begin{equation}\label{eqasym}
\rho _x(r)=\frac{\rho _{x_+}}{r^{\Delta _+}}+\frac{\rho _{x_-}}{r^{\Delta _-}},
 \  \  \ \phi(r) =\mu -\frac{\rho }{r^{d-3}},
\end{equation}
with
\begin{equation}\label{eqasym2}
\Delta _\pm=\frac{1}{2} \left[(d-3)\pm\sqrt{(d-3)^2+4 m^2}\right],
\end{equation}
where our choosing masses should satisfy the Breitenlohner-Freedman (BF) bound as \cite{wen18}
\begin{equation}
m^{2}\geqslant - \frac{(d-3)^{2}}{4}.
\end{equation}
Based on the AdS/CFT correspondence $\rho _{x_-}$ and $\rho
_{x_+}$ are, respectively, interpreted as the source and the
expectation value $\langle J_{x}\rangle$ which plays the role of
the order parameter in the boundary theory. Moreover, $\mu$ and
$\rho$ are regarded as chemical potential and charge density in
the dual field theory. In order to follow our research, we use
shooting method which solves equations (\ref{eqphi}) and
(\ref{eqrho}) numerically by applying suitable conditions. For
this purpose, we introduce a new variable $z=r_{+}/r$. For
convenience, we will set $r_{+}=1$ in the following calculation.
We do our numerical solutions in $d=4, 5$ and $6$ and find the
relation between critical temperature and $\rho^{1/(d-2)}$. We
summarize our results in tables I, II and III. By analyzing the
effects of mass of the vector field, we investigate the properties
of the holographic superconductor for $2$ values of mass in each
dimension. Meanwhile, we consider the effects of nonlinearity for
this model.  Our results show that by increasing the value of
nonlinearity parameter as well as mass makes the condensation
harder to form. Figures \ref{fig1}-\ref{fig12} show the behavior
of the condensation $\langle J_{x}\rangle^{1/(1+\Delta_{+})}$ for
two values of mass as a function of temperature for different
values of the nonlinearity parameter in each dimension. According
to these graphs, the condensation values goes up for stronger
effect of the nonlinearity parameter and mass which means it is
harder to have a superconductor in the presence of nonlinear
electrodynamics for massive vector fields. It was already argued
that holographic $p$-wave superconductor undergoes the first order
phase transition \cite{chaturverdip15,cai13p} in some cases
instead of usual second order type. However, in our work with
these choices for the mass and nonlinear parameters, we don't
observe this phenomenon. It may occur in the presence of
backreaction parameter or for other values of the mass and
nonlinear parameters.
\begin{table*}[t]
\label{tab1}
\begin{center}
\begin{tabular}{|c|c|c|c|c|c|}
\hline
\multicolumn{2}{|c|}{b=0} &\multicolumn{2}{|c|}{b=0.02} & \multicolumn{2}{|c|}{b=0.04} \\
\hline
$m^{2}=0$ & $m^{2}=3/4$ & $m^{2}=0$ & $m^{2}=3/4$ & $m^{2}=0$ & $m^{2}=3/4$ \\
\hline
0.125 $\rho^{1/2}$ & 0.102 $\rho^{1/2}$ & 0.120 $\rho^{1/2}$ & 0.093 $\rho^{1/2}$
 & 0.115 $\rho^{1/2}$ & 0.087 $\rho^{1/2}$ \\
\hline
\end{tabular}
\caption{Numerical results for critical temperature $T_{c}$ in
$d=4$ for different values of the mass and nonlinear parameters.}
\end{center}
\end{table*}
\begin{table*}[t]
\label{tab2}
\begin{center}
\begin{tabular}{|c|c|c|c|c|c|}
\hline
\multicolumn{2}{|c|}{b=0} &\multicolumn{2}{|c|}{b=0.02} & \multicolumn{2}{|c|}{b=0.04} \\
\hline
$m^{2}=-3/4$ & $m^{2}=5/4$ & $m^{2}=-3/4$ & $m^{2}=5/4$ & $m^{2}=-3/4$ & $m^{2}=5/4$ \\
\hline
0.224 $\rho^{1/3}$ &  0.184 $\rho^{1/3}$ &0.212 $\rho^{1/3}$ &0.159 $\rho^{1/3}$ & 0.204 $\rho^{1/3}$ & 0.147 $\rho^{1/3}$ \\
\hline
\end{tabular}
\caption{Numerical results for critical temperature $T_{c}$ in
$d=5$ for different values of the mass and nonlinear parameters.}
\end{center}
\end{table*}
\begin{table*}[t]
\label{tab3}
\begin{center}
\begin{tabular}{|c|c|c|c|c|c|}
\hline
\multicolumn{2}{|c|}{b=0} &\multicolumn{2}{|c|}{b=0.02} & \multicolumn{2}{|c|}{b=0.04} \\
\hline
$m^{2}=-5/4$ & $m^{2}=-2$ & $m^{2}=-5/4$ & $m^{2}=-2$ & $m^{2}=-5/4$ & $m^{2}=-2$ \\
\hline
0.289 $\rho^{1/4}$ &  0.312 $\rho^{1/4}$ &  0.261 $\rho^{1/4}$ & 0.292 $\rho^{1/4}$ & 0.247 $\rho^{1/4}$ &  0.281 $\rho^{1/4}$ \\
\hline
\end{tabular}
\caption{Numerical results for critical temperature $T_{c}$ in
$d=6$ for different values of mass and nonlinearity parameters.}
\end{center}
\end{table*}

\subsection{Electrical conductivity}\label{section2}
\begin{figure*}[t]
\centering
\subfigure[~$b=0$]{\includegraphics[width=0.4\textwidth]{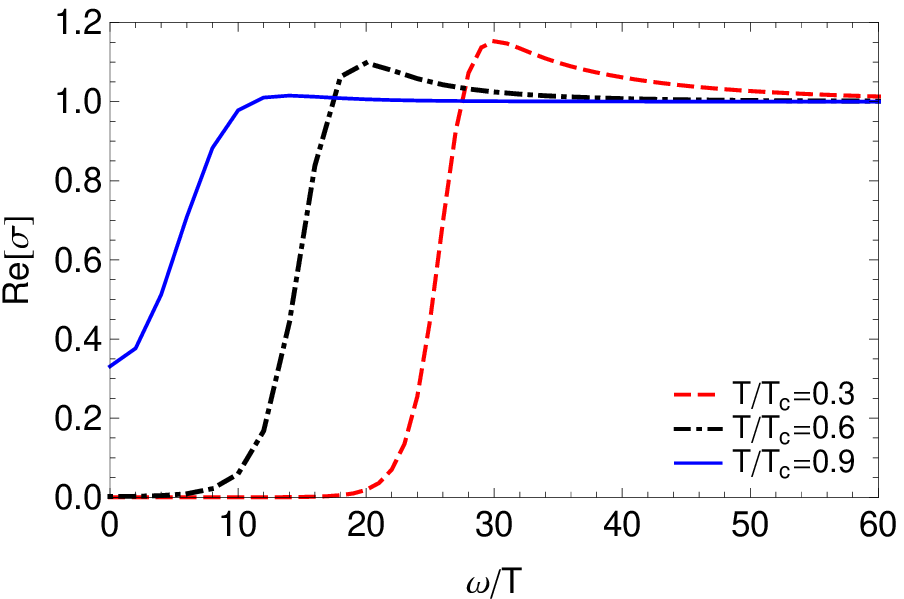}} \qquad %
\subfigure[~$ b=0.04$]{\includegraphics[width=0.4\textwidth]{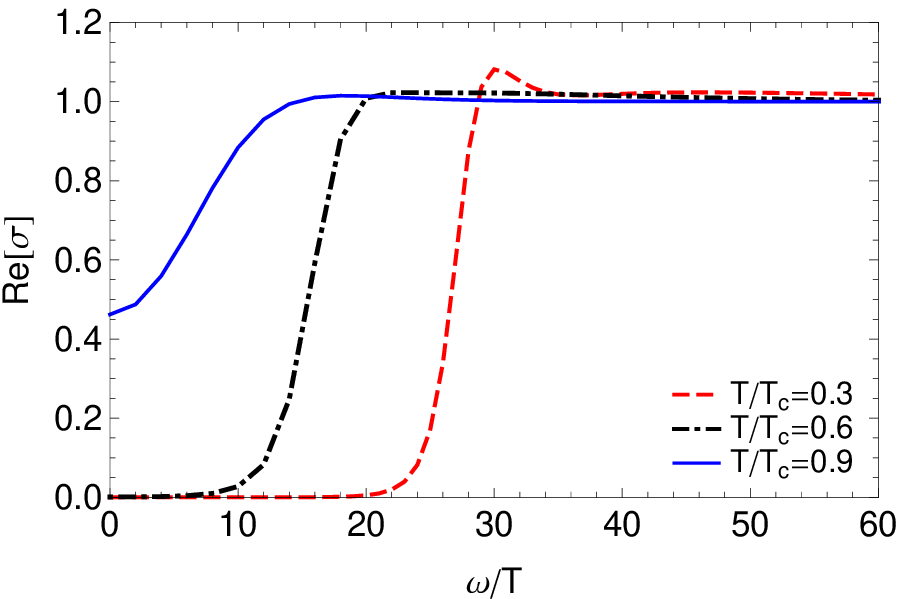}} \qquad %
\subfigure[~$b=0$]{\includegraphics[width=0.4\textwidth]{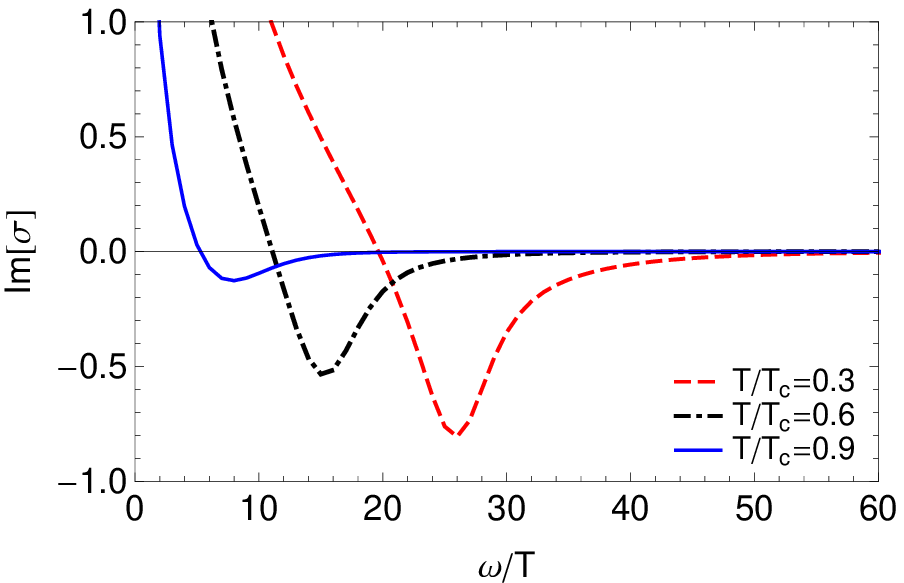}} \qquad %
\subfigure[~$ b=0.04$]{\includegraphics[width=0.4\textwidth]{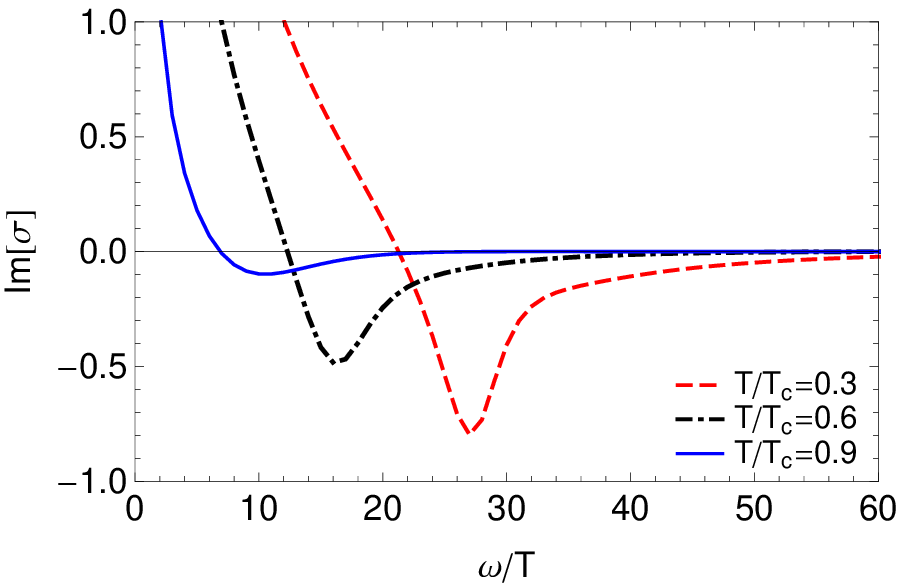}} \qquad %
\caption{The behavior of real and imaginary parts of conductivity
for $m^{2}=0$ in $d=4$.} \label{fig3}
\end{figure*}

\begin{figure*}[t]
\centering
\subfigure[~$b=0$]{\includegraphics[width=0.4\textwidth]{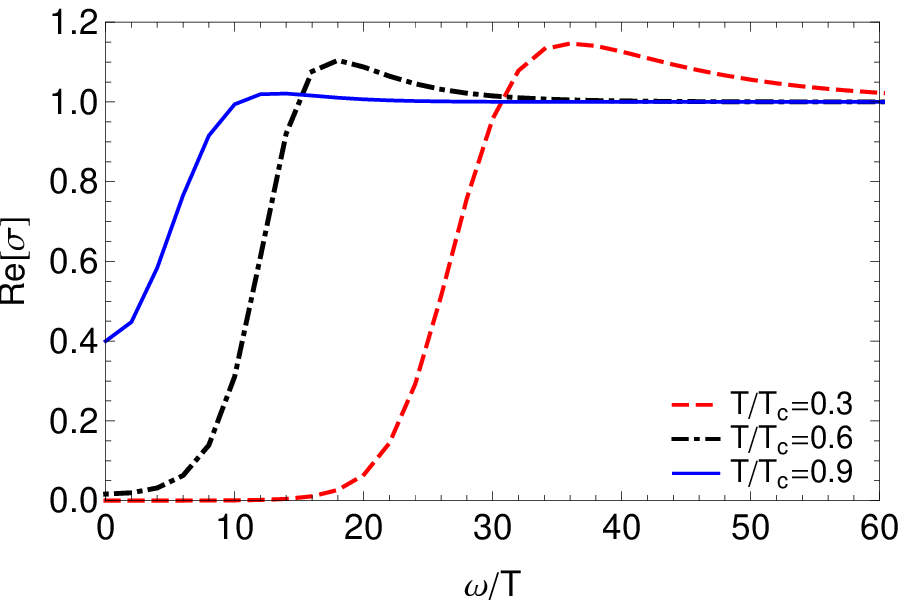}} \qquad %
\subfigure[~$ b=0.04$]{\includegraphics[width=0.4\textwidth]{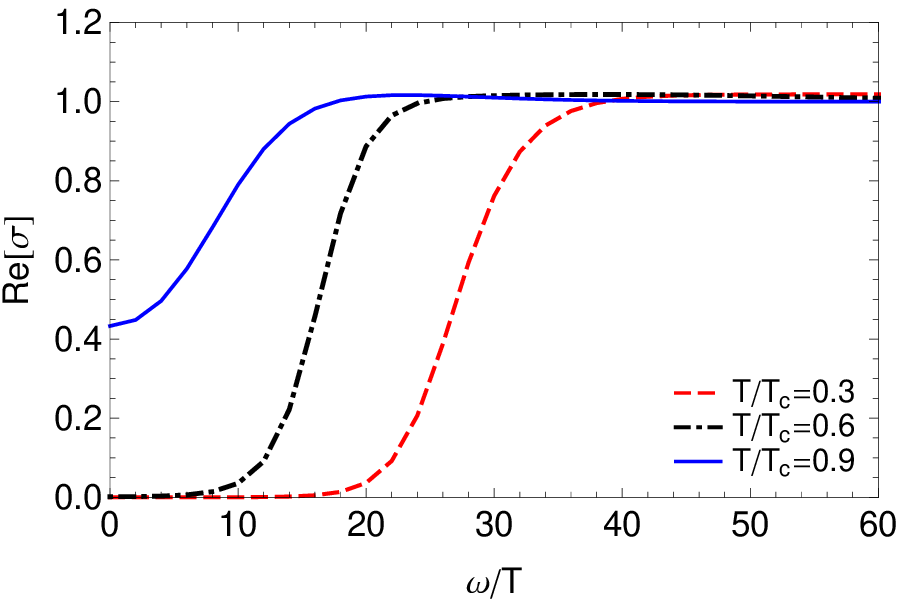}} \qquad %
\subfigure[~$b=0$]{\includegraphics[width=0.4\textwidth]{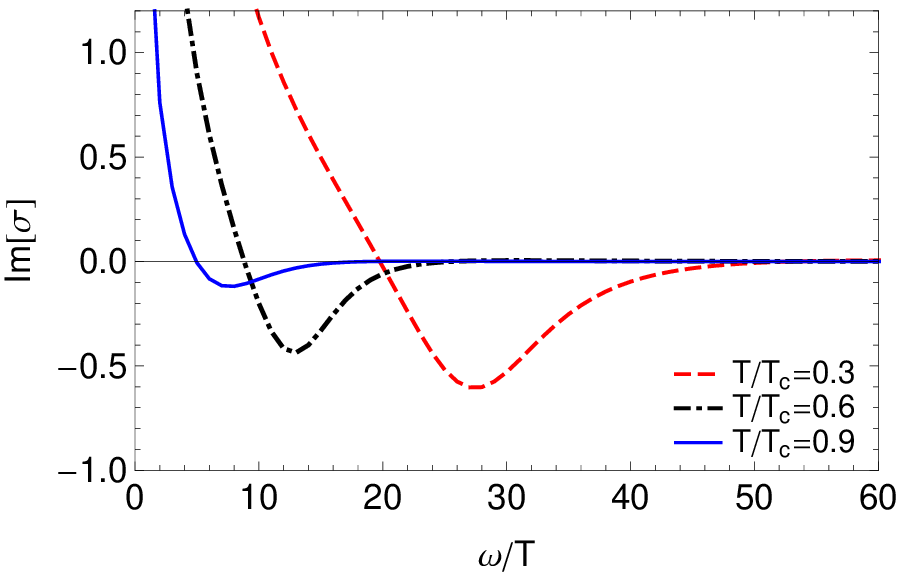}} \qquad %
\subfigure[~$ b=0.04$]{\includegraphics[width=0.4\textwidth]{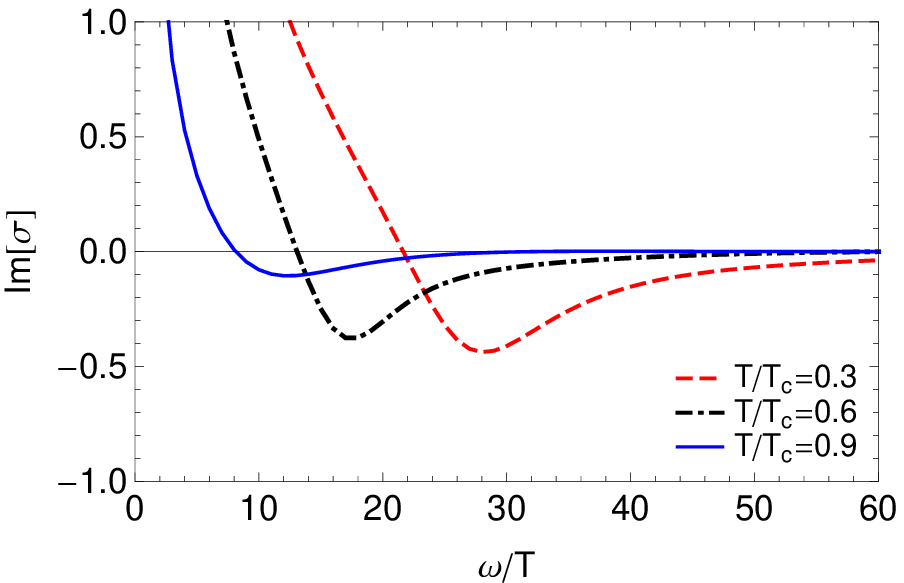}} \qquad %
\caption{The behavior of real and imaginary parts of conductivity
for $m^{2}=3/4$ in $d=4$.} \label{fig4}
\end{figure*}

\begin{figure*}[t]
\centering
\subfigure[~$b=0$]{\includegraphics[width=0.4\textwidth]{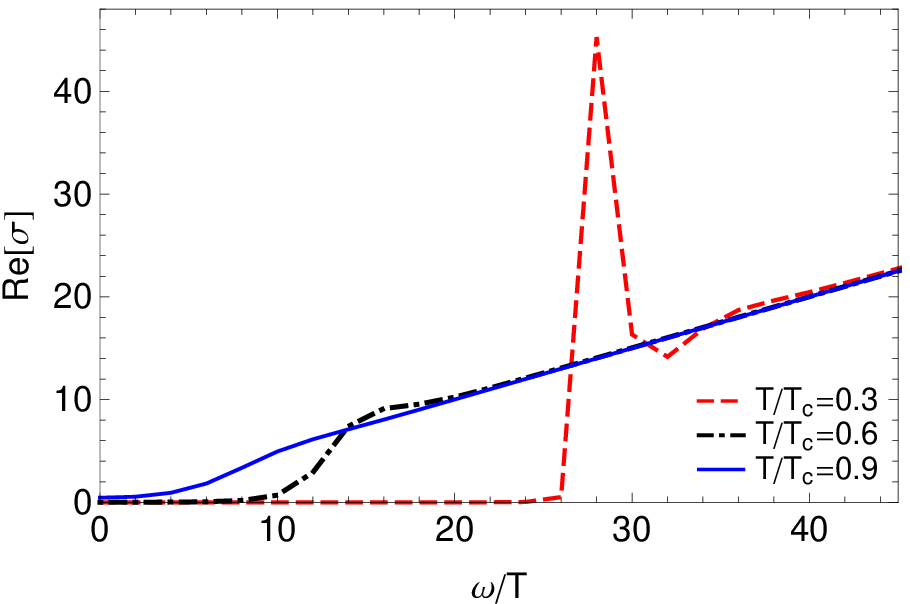}} \qquad %
\subfigure[~$ b=0.04$]{\includegraphics[width=0.4\textwidth]{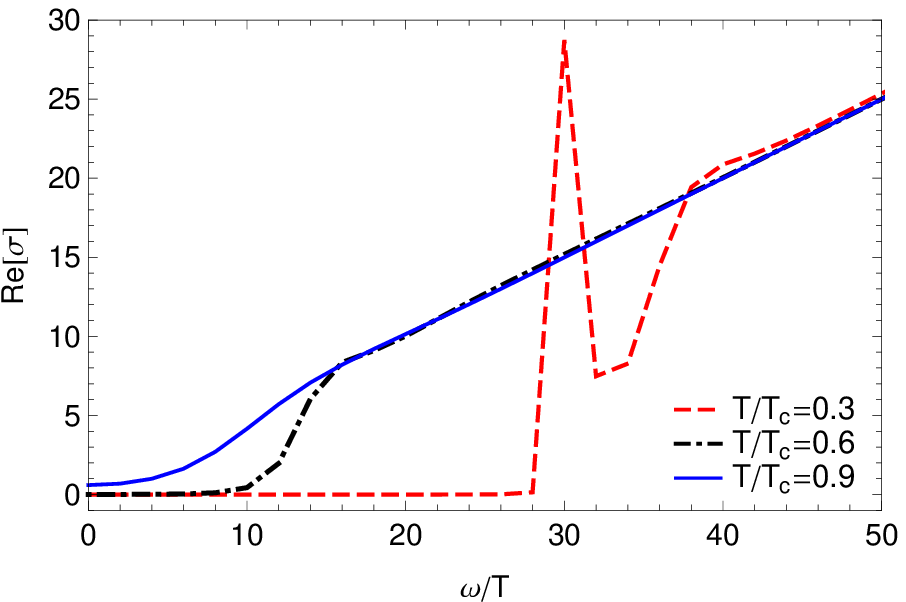}} \qquad %
\subfigure[~$b=0$]{\includegraphics[width=0.4\textwidth]{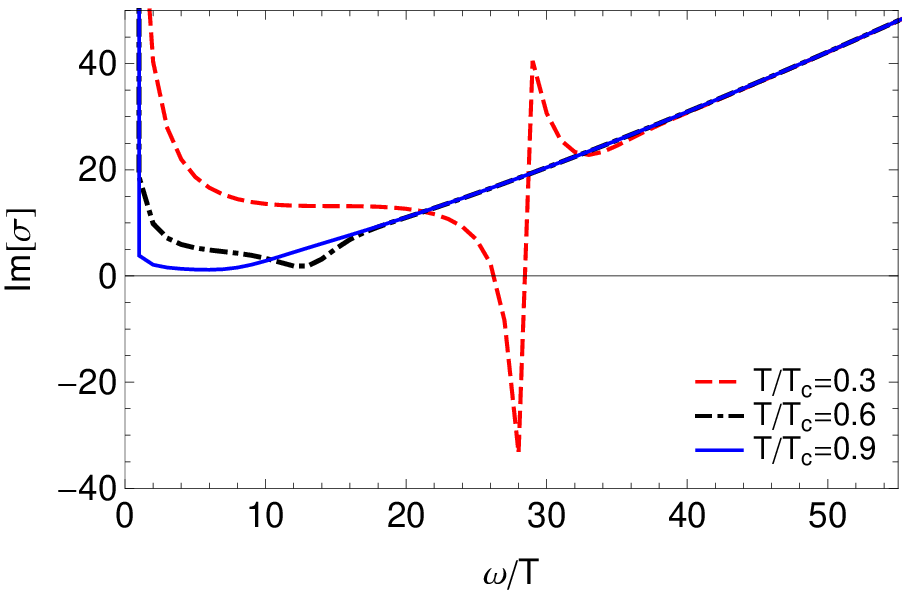}} \qquad %
\subfigure[~$ b=0.04$]{\includegraphics[width=0.4\textwidth]{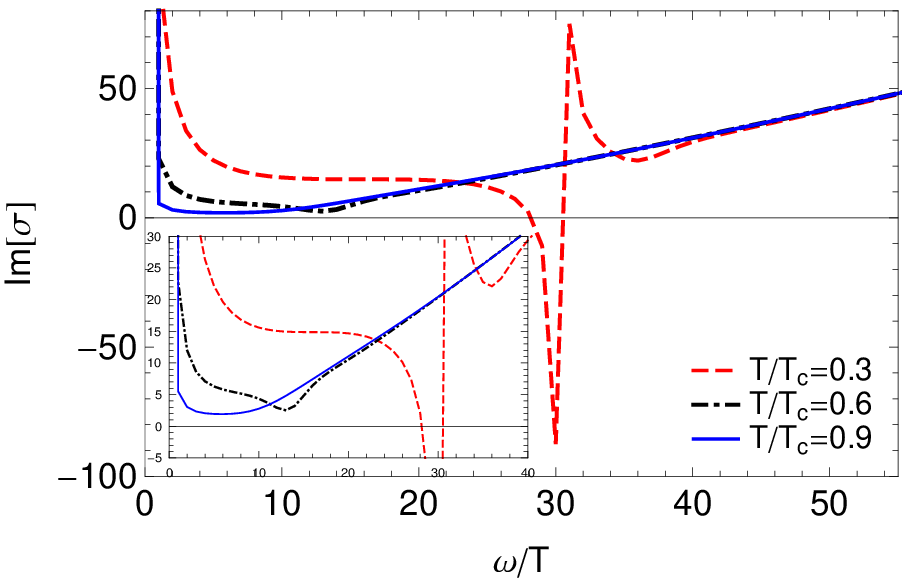}} \qquad %
\caption{The behavior of real and imaginary parts of conductivity
for $m^{2}=-3/4$ in $d=5$. } \label{fig8}
\end{figure*}

\begin{figure*}[t]
\centering
\subfigure[~$b=0$]{\includegraphics[width=0.4\textwidth]{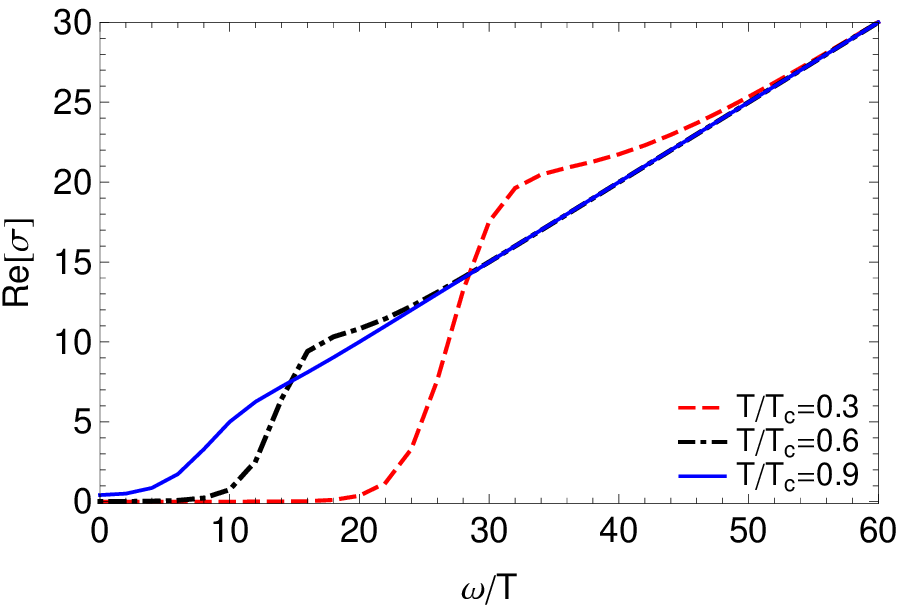}} \qquad %
\subfigure[~$ b=0.04$]{\includegraphics[width=0.4\textwidth]{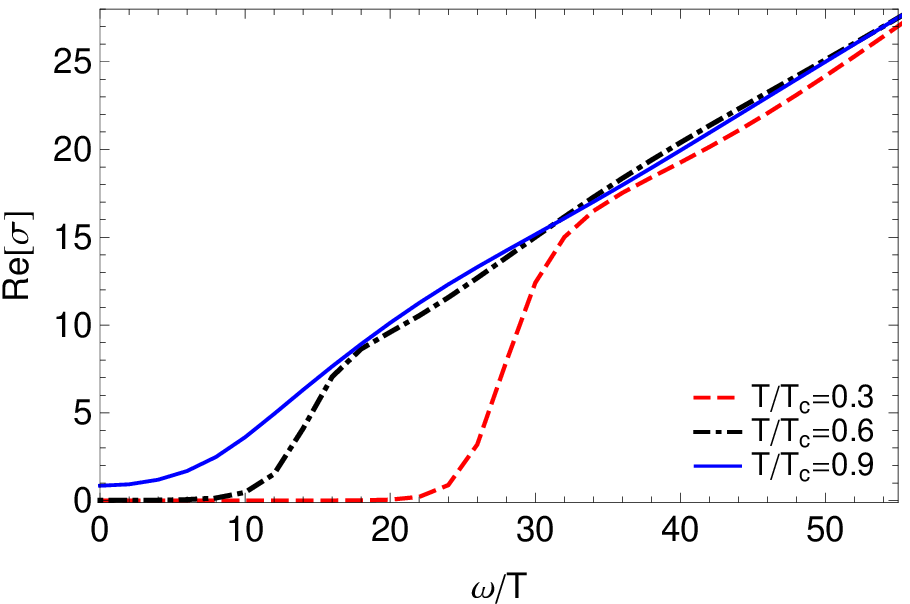}} \qquad %
\subfigure[~$b=0$]{\includegraphics[width=0.4\textwidth]{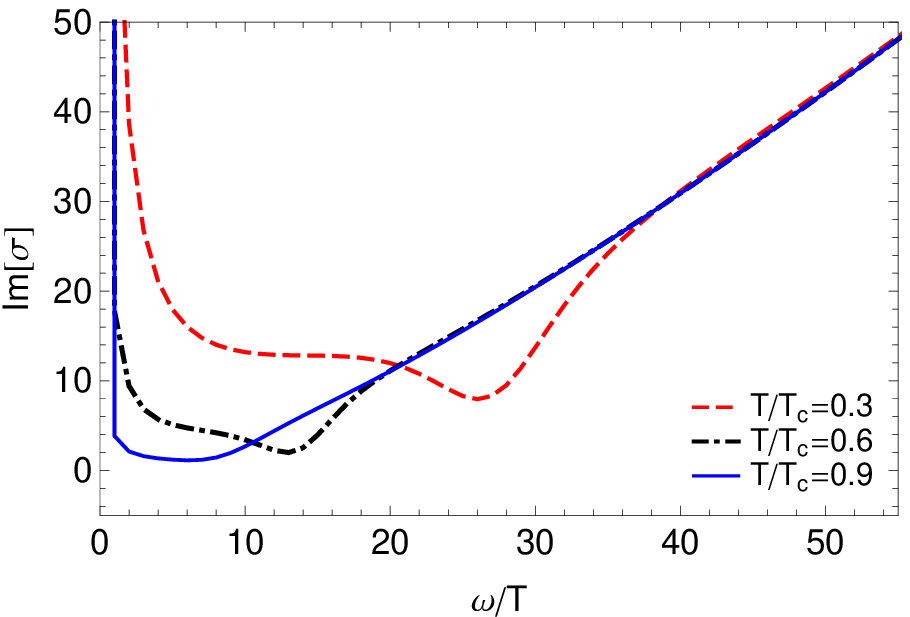}} \qquad %
\subfigure[~$ b=0.04$]{\includegraphics[width=0.4\textwidth]{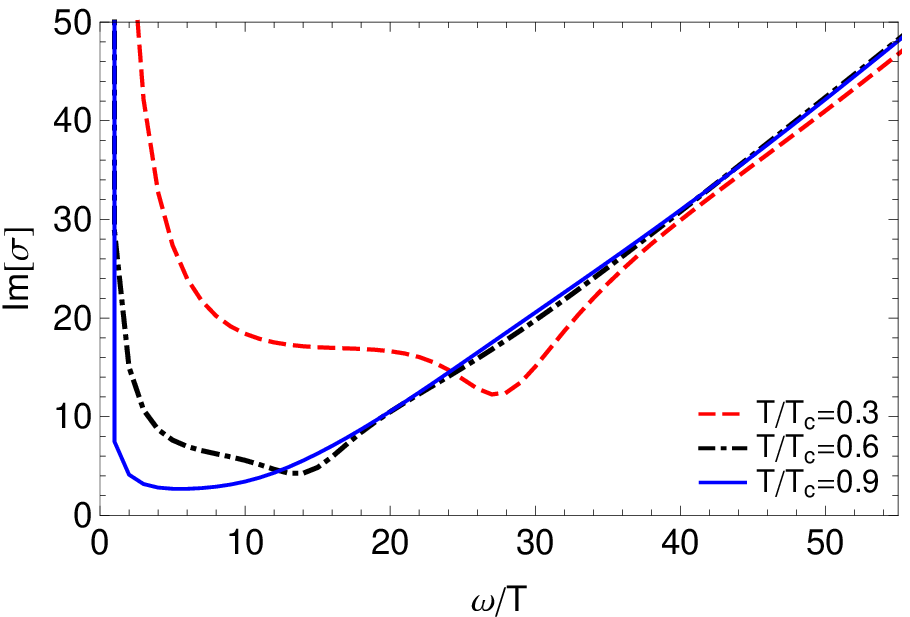}} \qquad %
\caption{The behavior of real and imaginary parts of conductivity
for $m^{2}=5/4$ in $d=5$. } \label{fig10}
\end{figure*}

\begin{figure*}[t]
\centering
\subfigure[~$b=0$]{\includegraphics[width=0.4\textwidth]{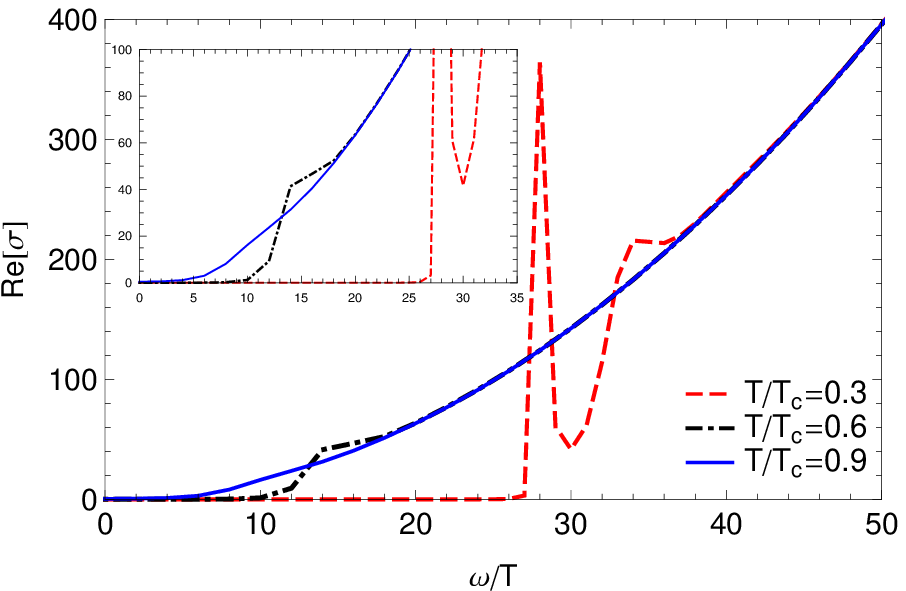}} \qquad %
\subfigure[~$ b=0.04$]{\includegraphics[width=0.4\textwidth]{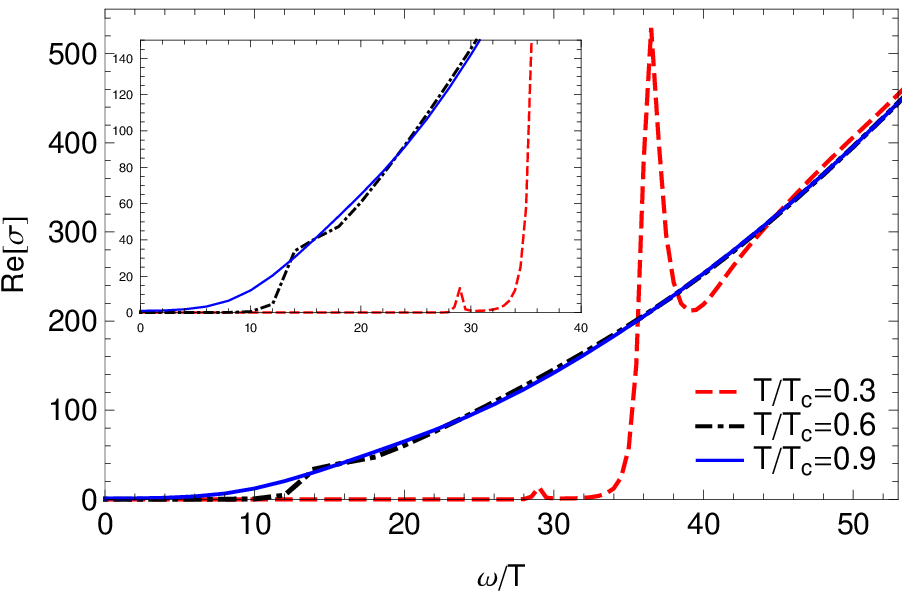}} \qquad %
\subfigure[~$b=0$]{\includegraphics[width=0.4\textwidth]{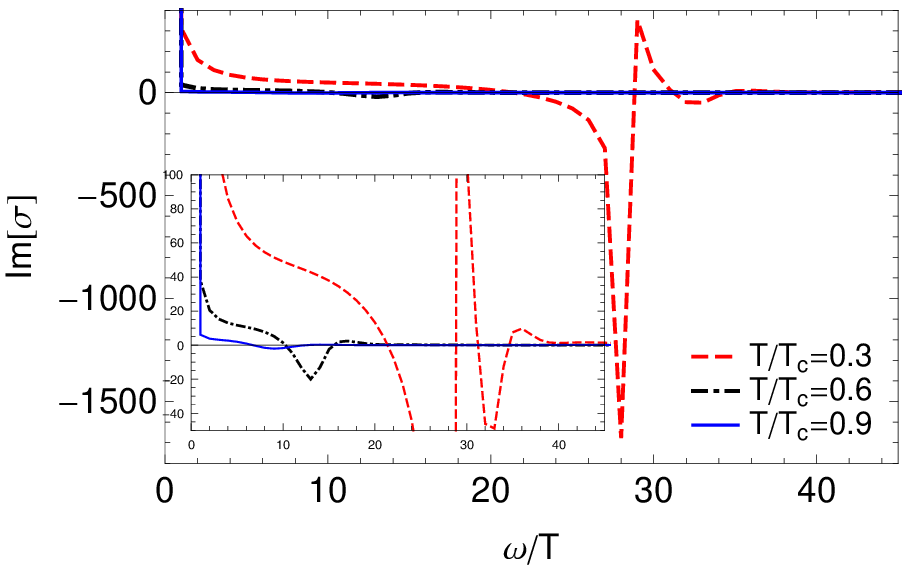}} \qquad %
\subfigure[~$ b=0.04$]{\includegraphics[width=0.4\textwidth]{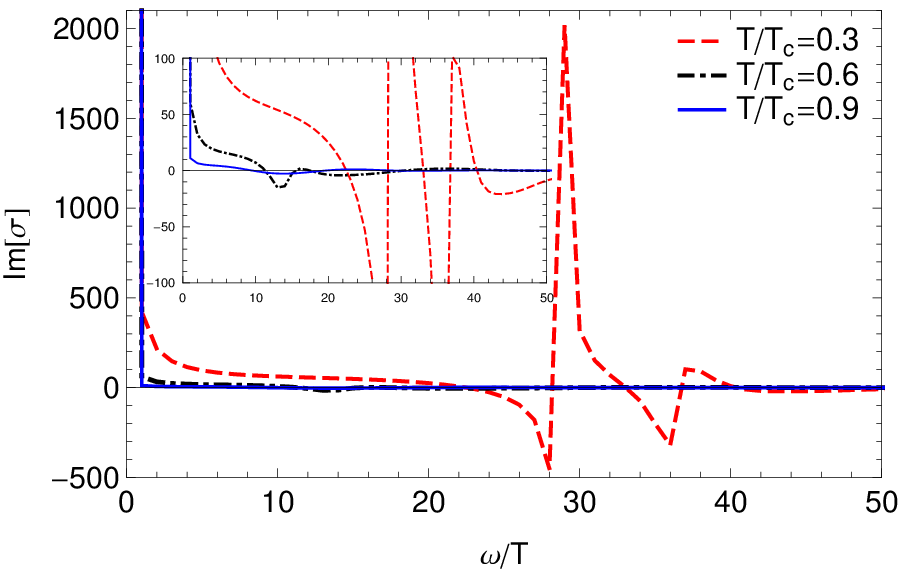}} \qquad %
\caption{The behavior of real and imaginary parts of conductivity
for $m^{2}=-5/4$ in $d=6$.} \label{fig13}
\end{figure*}

\begin{figure*}[t]
\centering
\subfigure[~$b=0$]{\includegraphics[width=0.4\textwidth]{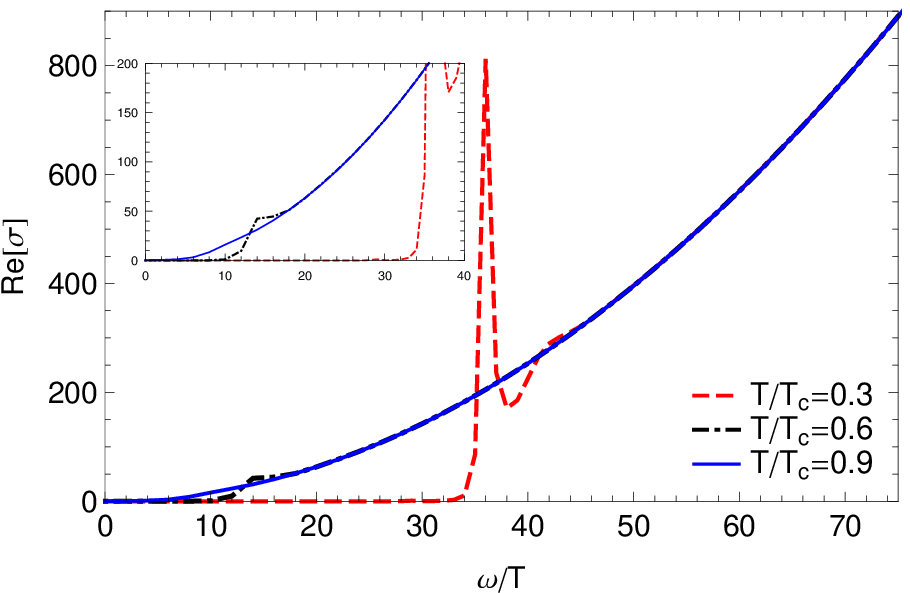}} \qquad %
\subfigure[~$ b=0.04$]{\includegraphics[width=0.4\textwidth]{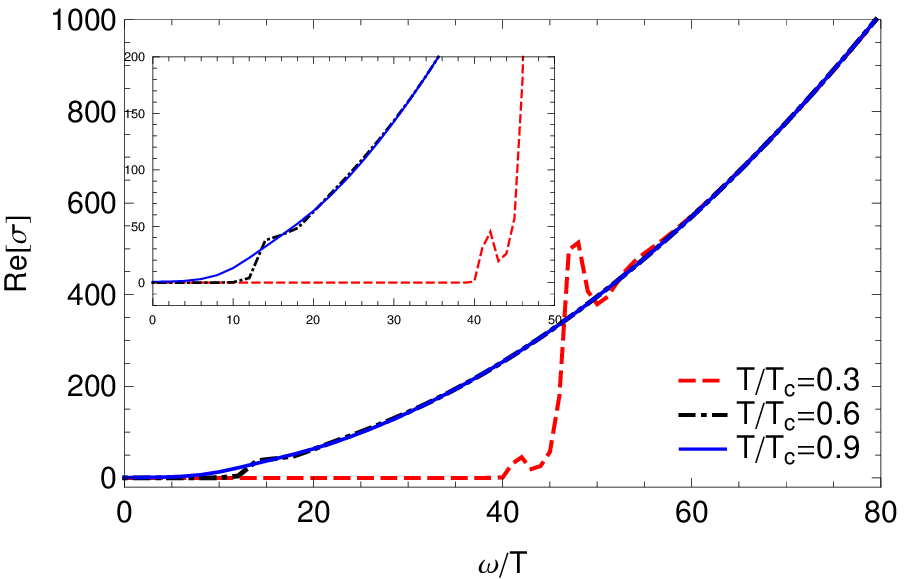}} \qquad %
\subfigure[~$b=0$]{\includegraphics[width=0.4\textwidth]{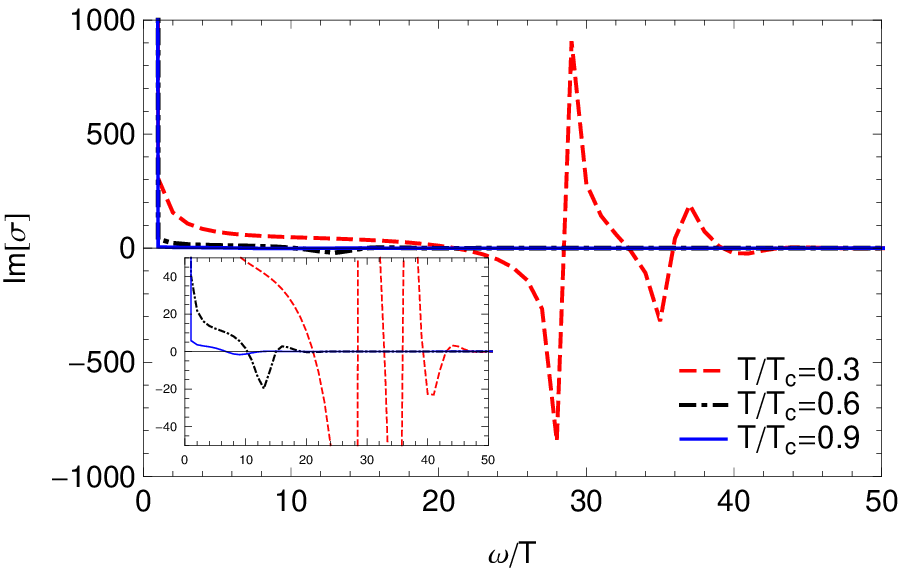}} \qquad %
\subfigure[~$ b=0.04$]{\includegraphics[width=0.4\textwidth]{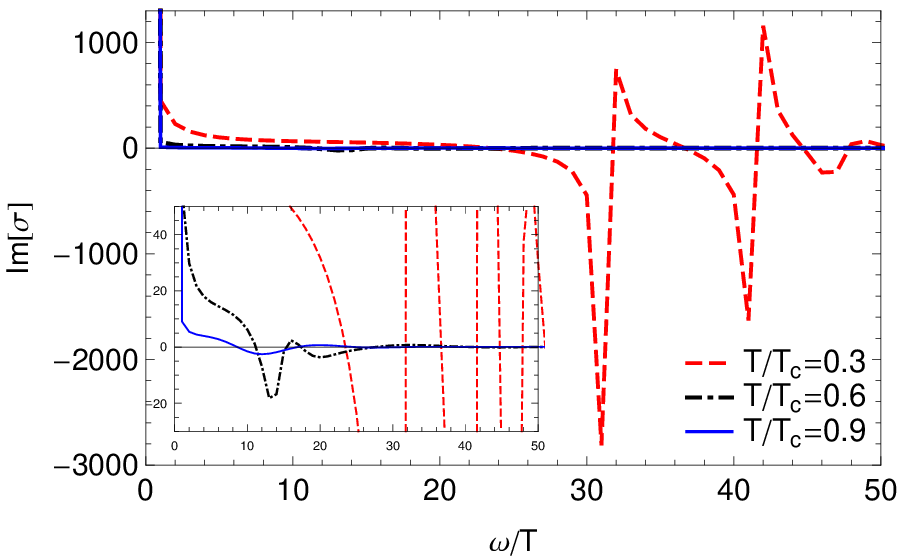}} \qquad %
\caption{The behavior of real and imaginary parts of conductivity
for $m^{2}=-2$ in $d=6$.} \label{fig15}
\end{figure*}

\begin{figure*}[t]
\centering
\subfigure[~$m^{2}=0,d=4$]{\includegraphics[width=0.3\textwidth]{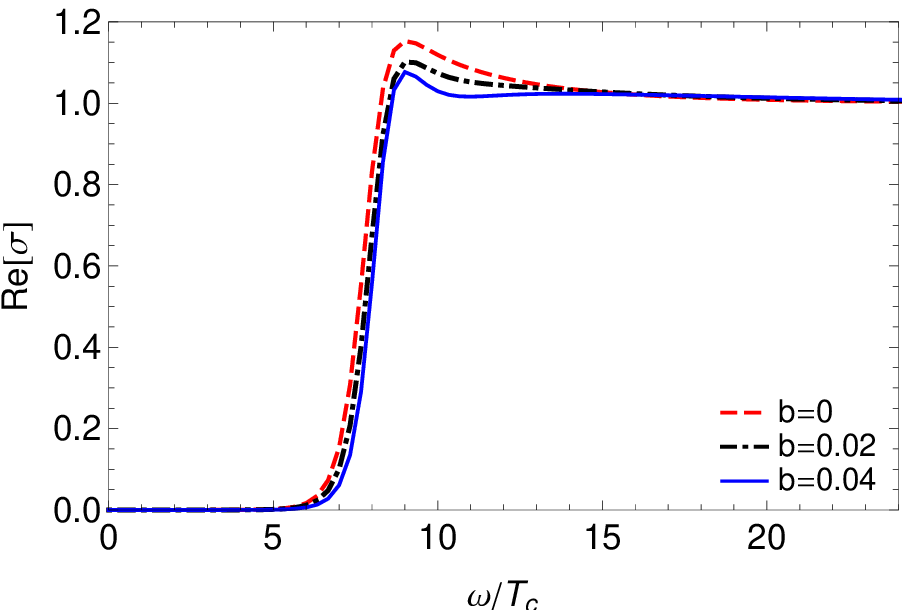}} \qquad %
\subfigure[~$m^{2}=3/4,d=4$]{\includegraphics[width=0.3\textwidth]{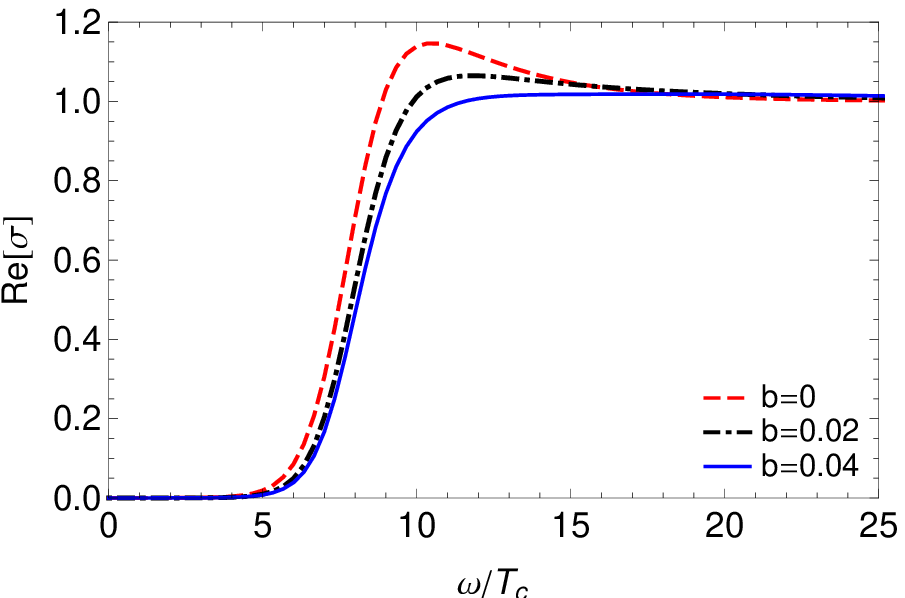}} \qquad %
\subfigure[~$m^{2}=-3/4,d=5$]{\includegraphics[width=0.3\textwidth]{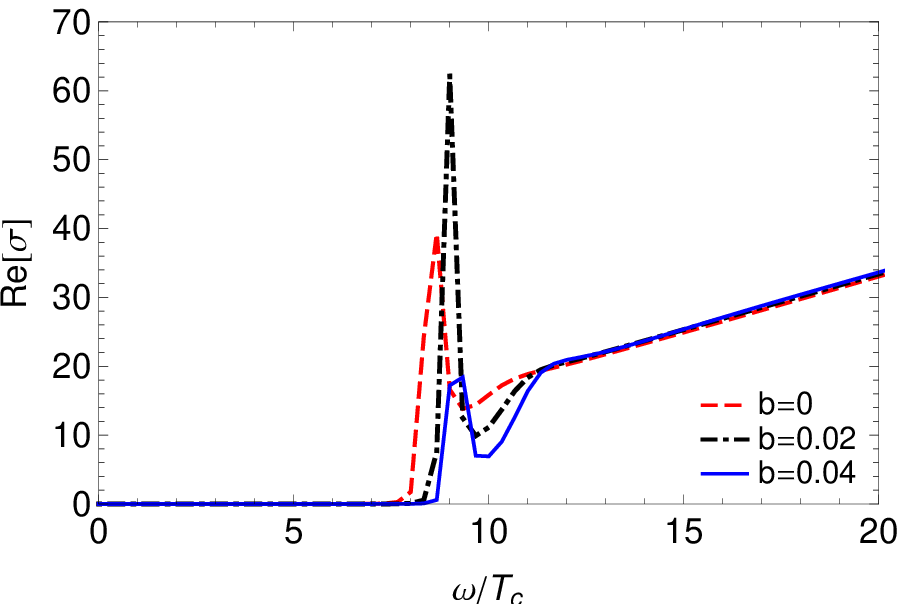}} \qquad %
\subfigure[~$m^{2}=5/4,d=5$]{\includegraphics[width=0.3\textwidth]{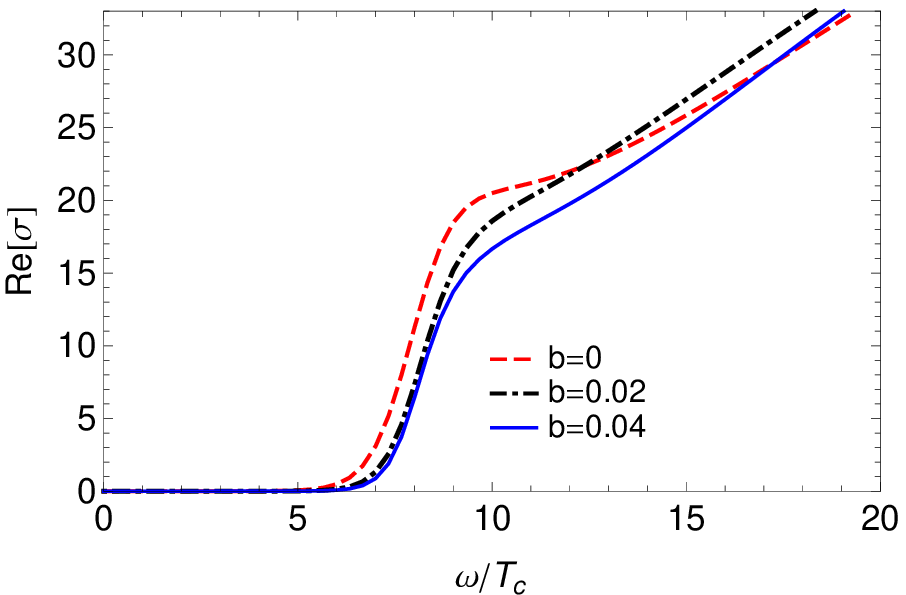}} \qquad %
\subfigure[~$m^{2}=-2,d=6$]{\includegraphics[width=0.3\textwidth]{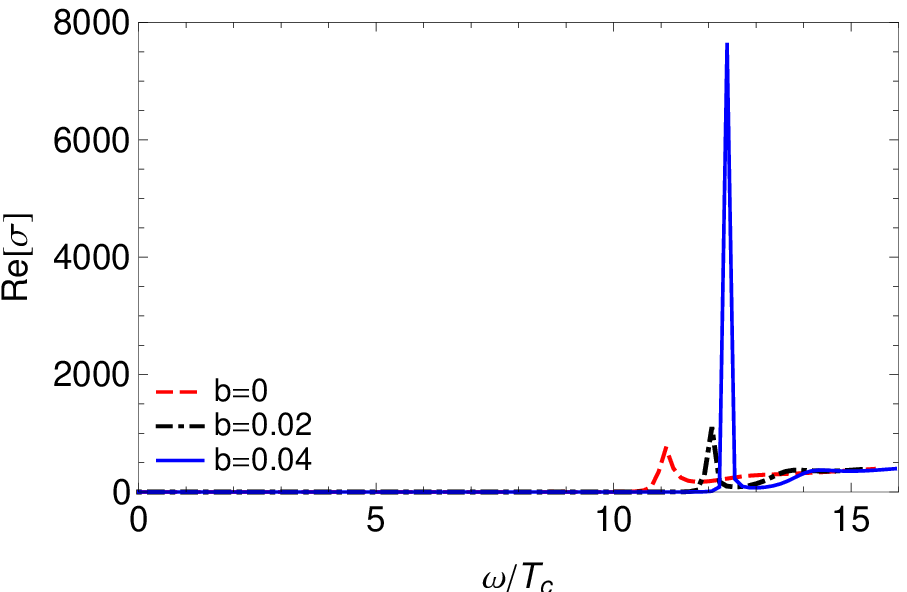}} \qquad %
\subfigure[~$m^{2}=-5/4,d=6$]{\includegraphics[width=0.3\textwidth]{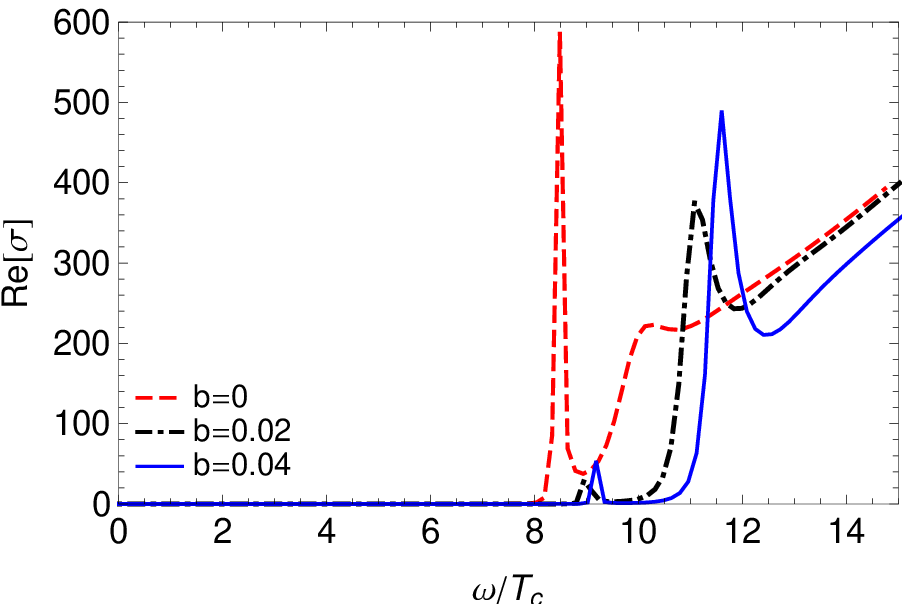}} \qquad %
\caption{The behavior of real parts of conductivity for
$T/T_{c}=0.3$.} \label{fig39a}
\end{figure*}

\begin{figure*}[t]
\centering
\subfigure[~$m^{2}=0,d=4$]{\includegraphics[width=0.3\textwidth]{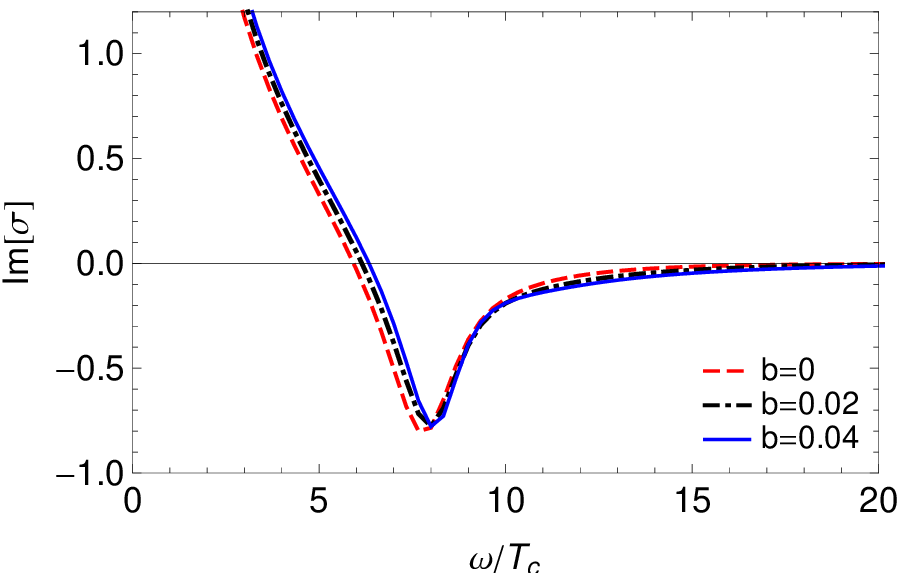}} \qquad %
\subfigure[~$m^{2}=3/4,d=4$]{\includegraphics[width=0.3\textwidth]{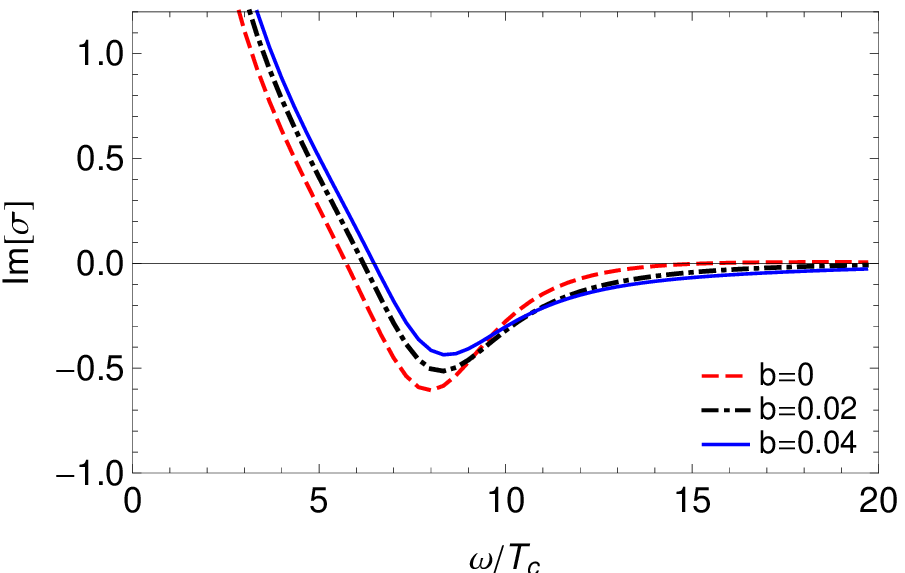}} \qquad %
\subfigure[~$m^{2}=-3/4,d=5$]{\includegraphics[width=0.3\textwidth]{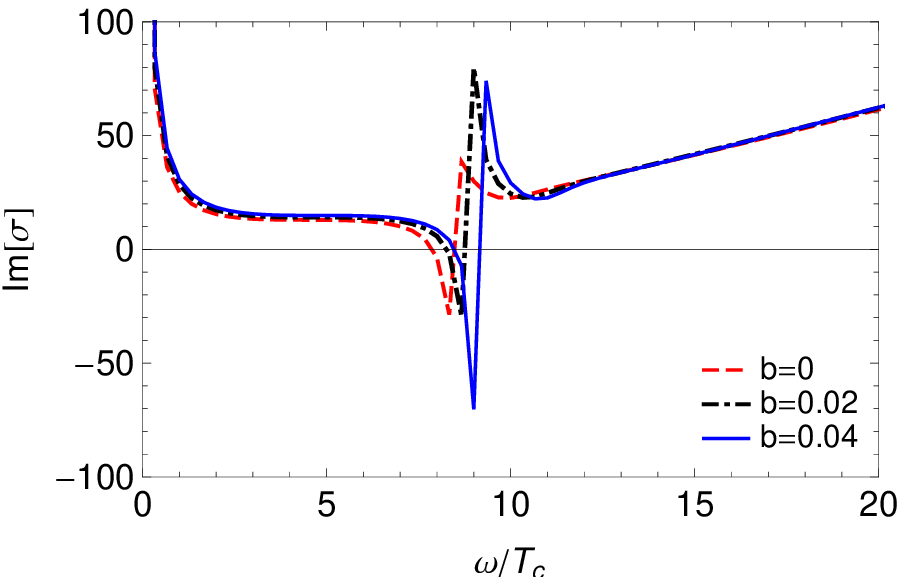}} \qquad %
\subfigure[~$m^{2}=5/4,d=5$]{\includegraphics[width=0.3\textwidth]{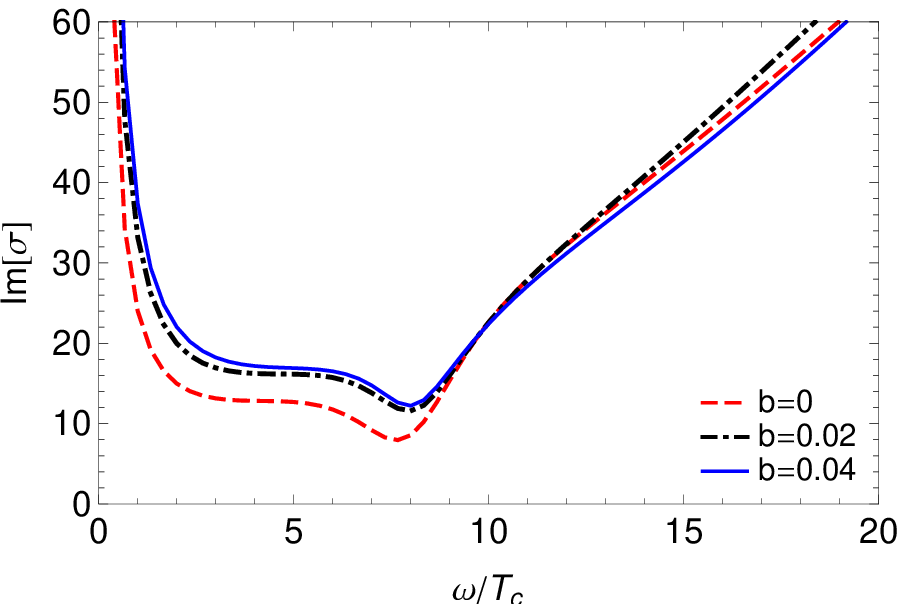}} \qquad %
\subfigure[~$m^{2}=-2,d=6$]{\includegraphics[width=0.3\textwidth]{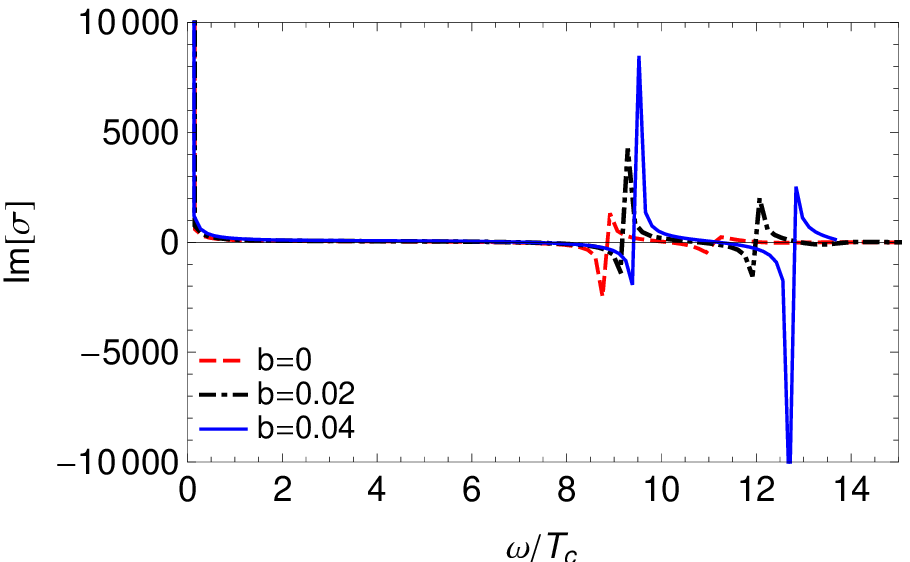}} \qquad %
\subfigure[~$m^{2}=-5/4,d=6$]{\includegraphics[width=0.3\textwidth]{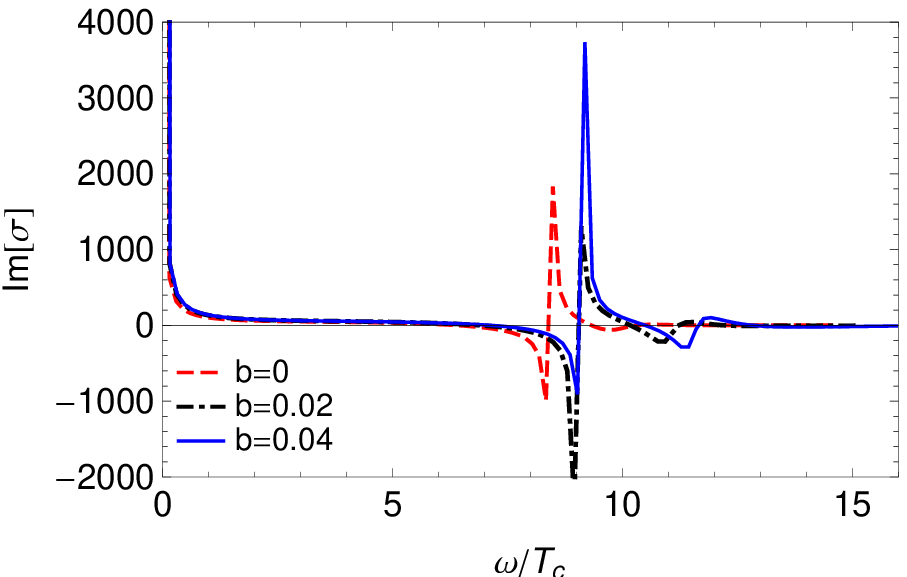}} \qquad %
\caption{The behavior of imaginary parts of conductivity for
$T/T_{c}=0.3$.} \label{fig40a}
\end{figure*}

In this section, we are going to calculate the electrical
conductivity as a function of frequency for this holographic
superconductor in the presence of nonlinear electrodynamics. To
follow our aim, we apply appropriate electromagnetic perturbation
as $\delta A_{y}=A_{y} e^{-i \omega t}$ on the black hole
background which based on the AdS/CFT duality corresponds to the
boundary electrical current. We choose this form of perturbation for simplicity
same as \cite{chaturverdip15,caipp,gaussp1}. The
consequence of turning on this perturbation in gauge field occurs
in $y$-component of equation (\ref{eqmax}) as
\begin{equation}\label{eqay}
A_y''(r)+\left[\frac{(d-4)}{r}+\frac{f'(r)}{f(r)}+\frac{2 b \phi
'(r) \phi ''(r)} {1+b \phi
'^2(r)}\right]A_y'(r)+\left[\frac{\omega ^2}{f^2(r)}-\frac{2 \rho
_{x}^2(r)}{r^2 f(r) \left(1+b \phi'^2(r)\right)}\right]A_y(r)=0,
\end{equation}
which has the asymptotic behavior as
\begin{equation}\label{eqasymay}
A_y''(r)+\frac{(d-2) }{r}A_y'(r)+\frac{\omega ^2}{r^4} A_y(r)=0,
\end{equation}
by considering $A^{(0)}$ and $A^{(1)}$ as constant parameters, we
have the following solution for $A_{y}$ asymptotically
\begin{equation} \label{aysol}
A_{y} = \left\{
\begin{array}{lr}
A^{(0)}+\frac{A^{(1)}}{r}+\cdots, & d=4\\
\bigskip\\
A^{(0)}+\frac{A^{(1)}}{r^2}+\frac{A^{(0)} \omega ^2 \log (\Lambda  r)}{2 r^2}+\cdots, & d=5\\
\bigskip\\
A^{(0)}+\frac{A^{(1)}}{r^3}+\frac{A^{(0)} \omega ^2}{2 r^2}+\cdots, & d=6\\
\end{array} \right.
\end{equation}%
where $\Lambda$ is an arbitrary constant. Equations
(\ref{eqasymay}) and (\ref{aysol}) are the same as corresponding
equations in $s$-wave case\cite{Doa}. According to the AdS/CFT
correspondence, the electrical current based on on-shell bulk
action $S_{o.s}$ and the Lagrangian of the matter field
$\mathcal{L}_{m}$ is defined by
\begin{equation} \label{eqj}
J=\frac{\text{$\delta $S}_{\text{bulk}}}{\text{$\delta $A}^{(0)}}=\frac{\text{$\delta $S}_{o.s}}
{\text{$\delta $A}^{(0)}}=\frac{\partial(\sqrt{-g}\mathcal{L}_{m})}{\partial A_y'}\vert r\rightarrow \infty,
\end{equation}
where
\begin{equation}\label{sos}
S_{o.s.}=\int_{r_+}^{\infty }\, dr \int \, {d}^{d-1}x \sqrt{-g}\mathcal{L}_{m},
\end{equation}
by inserting (\ref{aysol}) in (\ref{sos}), we arrive at
\begin{equation}\label{eqsos}
S_{o.s.}=- \frac{1}{2} \int \, {d}^{d-1}x  \left[A_{y}(r) f(r)
A_{y}'(r) \left(1+b \phi '^2(r)\right)\right] r^{d-4}.
\end{equation}
Pursuing the AdS/CFT framework, the electrical conductivity is
\begin{equation}\label{eqohm}
\sigma_{yy}=\frac{J_{y}}{E_{y}}, \  \  \  E_{y}=-\partial_{t} \delta A_{y}.
\end{equation}
Therefore, the electrical conductivity, based on holographic
approach by using Eqs. (\ref{eqj}), (\ref{sos}) and (\ref{eqohm})
and adding appropriate counterterms for $d=5$ and $6$ to remove
the divergency based on the re-normalization method
\cite{skenderis} is
\begin{equation} \label{conductivity}
\sigma_{yy} = \left\{
\begin{array}{lr}
\frac{A^{(1)}}{i \omega A^{(0)}}, & d=4\\
\bigskip\\
\frac{2 A^{(1)}}{i \omega A^{(0)}}+\frac{i \omega}{2}, & d=5\\
\bigskip\\
\frac{3 A^{(1)}}{i \omega A^{(0)}}, & d=6\\
\end{array} \right.
\end{equation}%
which is in complete agreement with the $\sigma_{xx}$ obtained in
\cite{Doa}. This shows that the calculation of $\sigma_{yy}$ in
holographic $p$-wave superconductor is the same as $\sigma_{xx}$
in holographic $s$-wave superconductor\cite{caipp}. In order to
investigate the trend of conductivity numerically, we impose the
ingoing wave boundary condition near the horizon for $A_{y}$ as
follows
\begin{equation}\label{ayexpand}
A_{y}(r)=f(r)^{-i \omega/ (4 \pi T)} \left[1+a(1-r)+b(1-r)+\cdots\right],
\end{equation}
where the Hawking temperature is expressed by $T$ and $a, b,
\cdots$ are coefficients which can be obtained by Taylor expansion
of Eq. (\ref{eqasymay}) around the horizon. Figs.
\ref{fig3}-\ref{fig15} illustrate the behavior of real and
imaginary parts of conductivity as a function of $\omega/T$ for
different values of mass and nonlinearity parameters in different
dimensions. In spite of the fact that there are obvious
differences in figures, they follow some universal behaviors.
First of all, the real part of conductivity is related to the
imaginary side based on the Kramers-Kronig relation which means
that the appearance of the delta function and pole in real and
imaginary parts of conductivity, respectively. Secondly, the
superconducting gap which appears below the critical temperature
becomes deeper and sharper by diminishing the temperature which
causes larger values of $\omega_{g}$. This fact approved the
results of previous section about condensation by going down the
temperature because we can interpret $\omega_{g}$ as the energy to
break the fermion pairs. So, the bigger values of $\omega_{g}$
leads to harder formation of fermion pairs  which hinders the
conductor/superconductor phase transition \cite{caipp}. At large
frequencies, the behavior of conductivity can be indicated to have
a power law behavior as $Re[\sigma]=\omega^{(d-4)}$ similar to
$s$-wave case \cite{Doa}. In addition, based on the BCS theory
$\omega_{g}\approx3.5T_{c}$ while in holographic setup the ratio
of gap frequency over critical temperature is found to be the
universal value around $8$ which hints to the fact that in BCS
theory the pairs couple to each other weakly with no interaction.
However, the holographic superconductors are strongly coupled.
This strong coupling is the reason that we use holographic model
to describe high temperature superconductors which exist in strong
coupling regime  \cite{caipp}. Due to describe details more, we
should say that the peak of real part of conductivity follows the
same trend as the minimum value of imaginary part by decreasing
the temperature. They shift toward the larger frequencies in all
dimensions. Our choice of the mass has a direct outcome on the
behavior of conductivity that becomes so obvious in some cases.
For example, in $d=5$ we face with sharp peak and deep minimum in
real and imaginary parts for $m^{2}=-3/4$ while for the case that
the vector field has a mass equals to $m^{2}=5/4$ we observe
smooth peak and minimum in two parts of conductivity. Overall, it
seems that  $T=0.3 T_{c}$ has a strong effect on conductivity in
$d=5$ and $6$. In order to study the effect of nonlinear
electrodynamics on the conductivity, we plot its behavior in fixed
temperature $T=0.3T_{c}$ in figs. \ref{fig39a}-\ref{fig40a}.
According to this graphs, the effects of nonlinearity parameter on
conductivity is straightly depend on the mass and dimension. For
example, in $d=4$ and for $d=5$ with $m^{2}=5/4$ increasing the
nonlinearity makes the peak and minimum parts more smooth and
flat. However, for other cases doesn't behave like that. In $d=6$,
increasing the nonlinearity causes smaller first gap and enlarging
the second one in real part of conductivity. In general, the gap
frequency is shifted in the presence of nonlinear electrodynamics.


\section{Holographic $p$-wave superconductor in Gauss-Bonnet gravity}
\subsection{Conductor/superconductor phase transition in holographic setup}\label{section3}
\begin{figure*}[t]
\centering
\subfigure[~$\alpha=0.08$]{\includegraphics[width=0.4\textwidth]{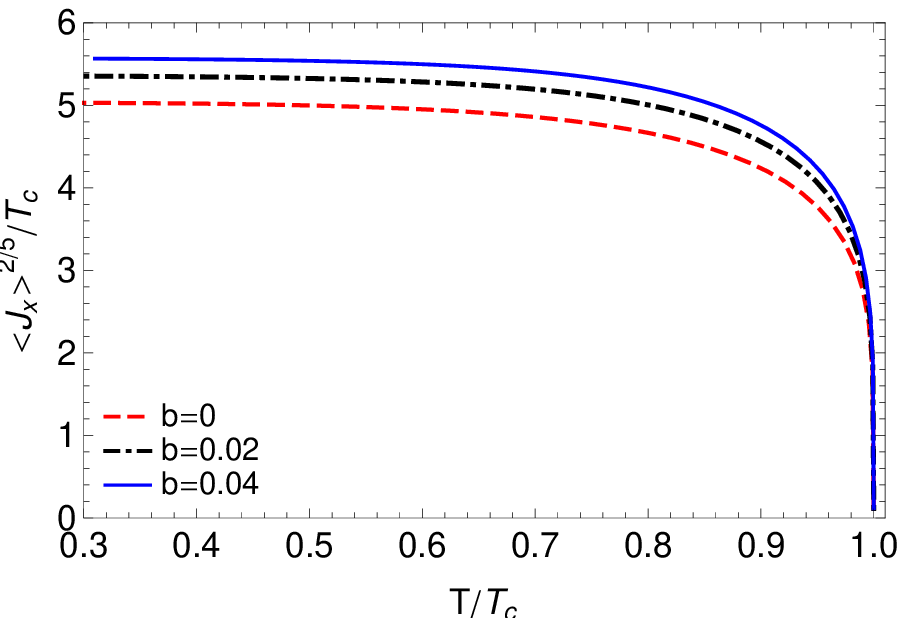}} \qquad %
\subfigure[~$\alpha=-0.08$]{\includegraphics[width=0.4\textwidth]{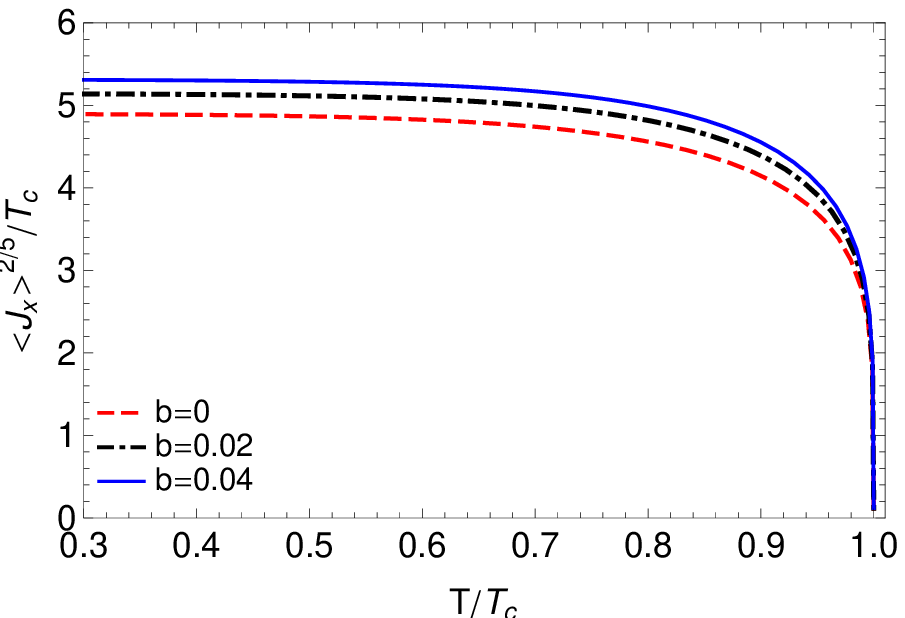}} \qquad %
\caption{The behavior of the condensation parameter as a function
of the temperature for different values of nonlinearity parameters
in $d=5$ with $\overline{m}^{2}=-3/4$.} \label{fig17}
\end{figure*}

\begin{figure*}[t]
\centering
\subfigure[~$b=0$]{\includegraphics[width=0.4\textwidth]{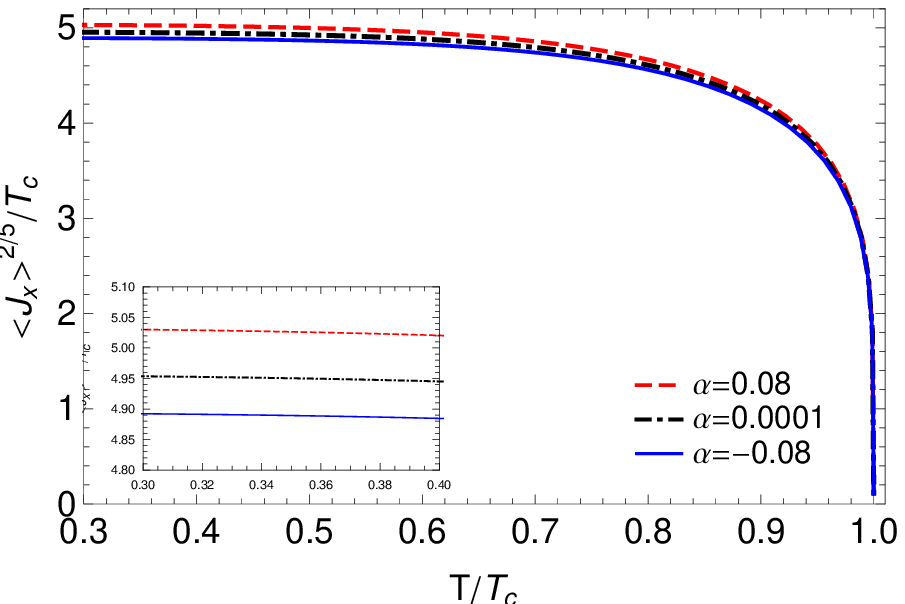}} \qquad %
\subfigure[~$ b=0.04$]{\includegraphics[width=0.4\textwidth]{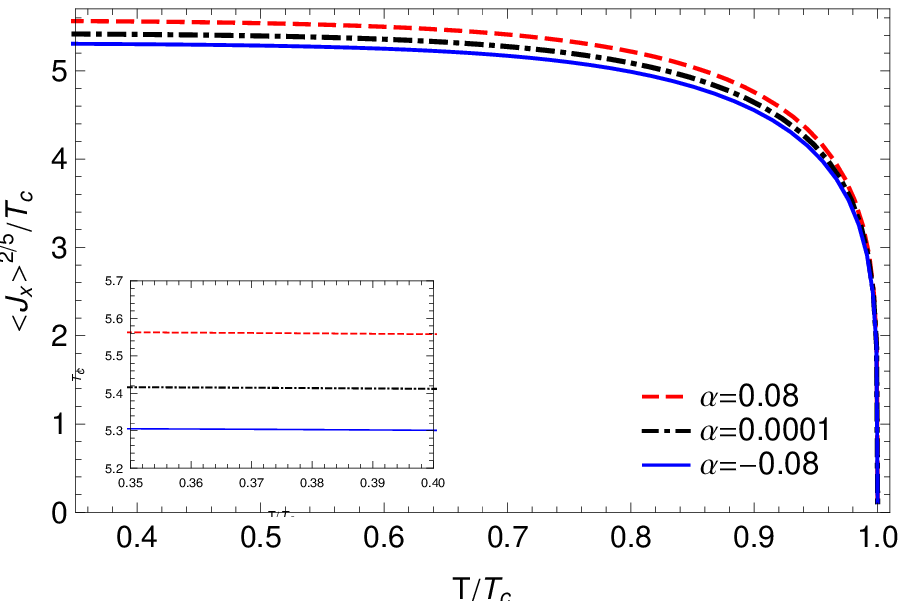}} \qquad %
\caption{The behavior of the condensation parameter as a function
of the temperature for different values of Gauss-Bonnet parameters
in $d=5$ with $\overline{m}^{2}=-3/4$.} \label{fig18}
\end{figure*}

\begin{figure*}[t]
\centering
\subfigure[~$\alpha=0.08$]{\includegraphics[width=0.4\textwidth]{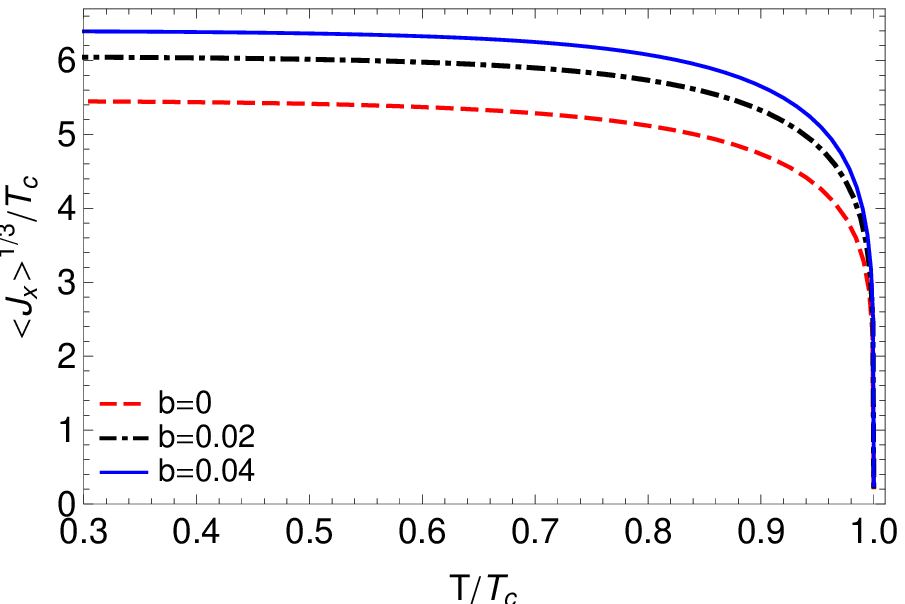}} \qquad %
\subfigure[~$\alpha=-0.08$]{\includegraphics[width=0.4\textwidth]{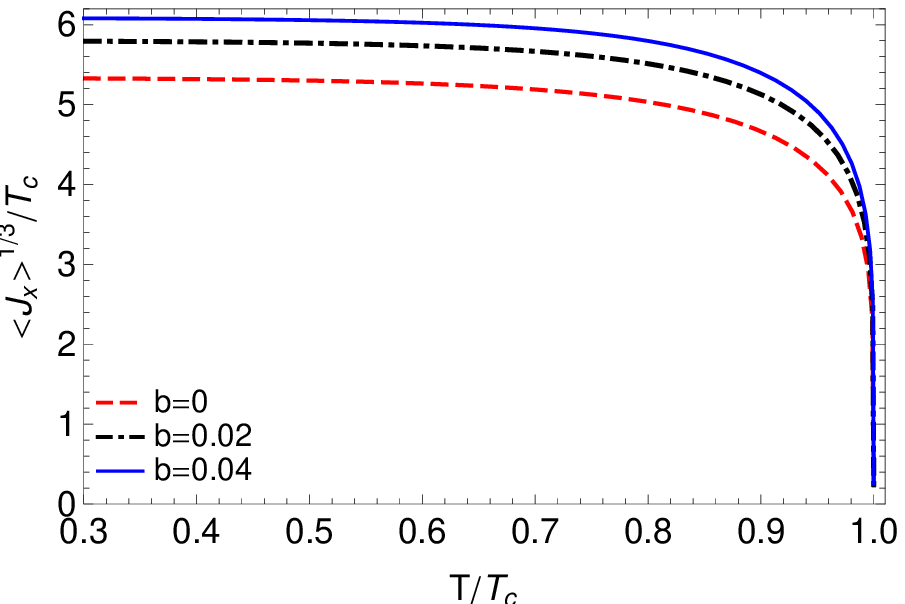}} \qquad %
\caption{The behavior of the condensation parameter as a function
of the temperature for different values of nonlinearity parameters
in $d=5$ with $\overline{m}^{2}=0$.} \label{fig19}
\end{figure*}

\begin{figure*}[t]
\centering
\subfigure[~$b=0$]{\includegraphics[width=0.4\textwidth]{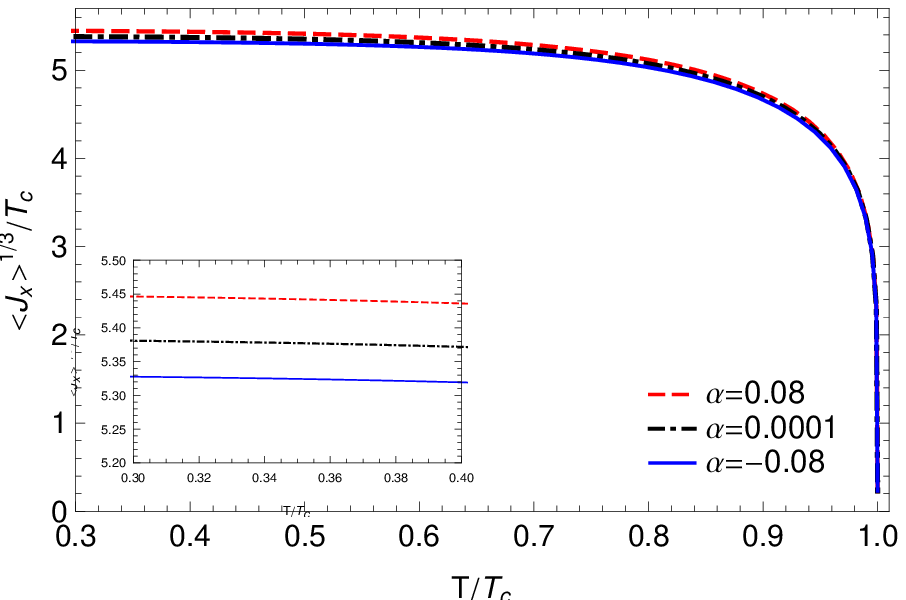}} \qquad %
\subfigure[~$ b=0.04$]{\includegraphics[width=0.4\textwidth]{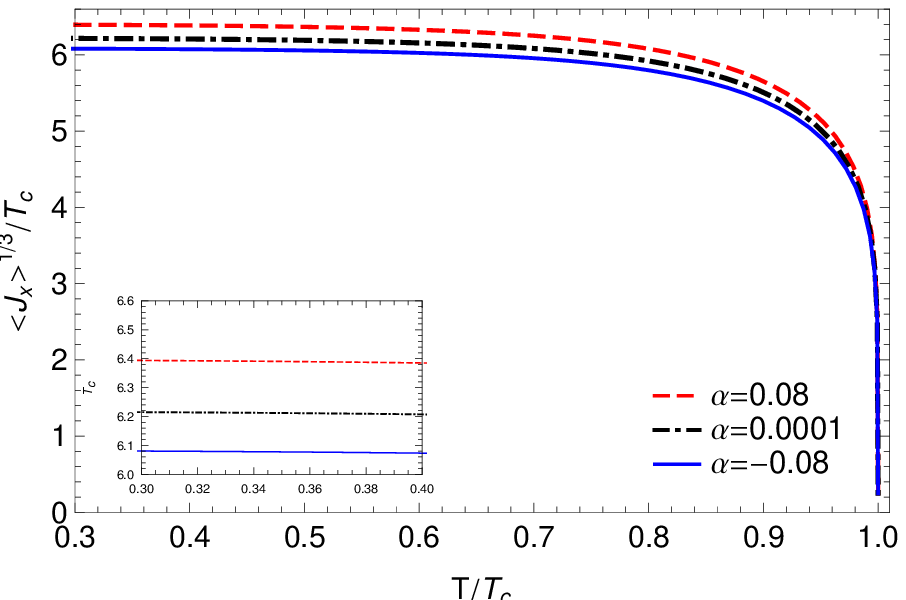}} \qquad %
\caption{The behavior of the condensation parameter as a function
of the temperature for different values of Gauss-Bonnet parameters
in $d=5$ with $\overline{m}^{2}=0$.} \label{fig20}
\end{figure*}

\begin{figure*}[t]
\centering
\subfigure[~$\alpha=0.08$]{\includegraphics[width=0.4\textwidth]{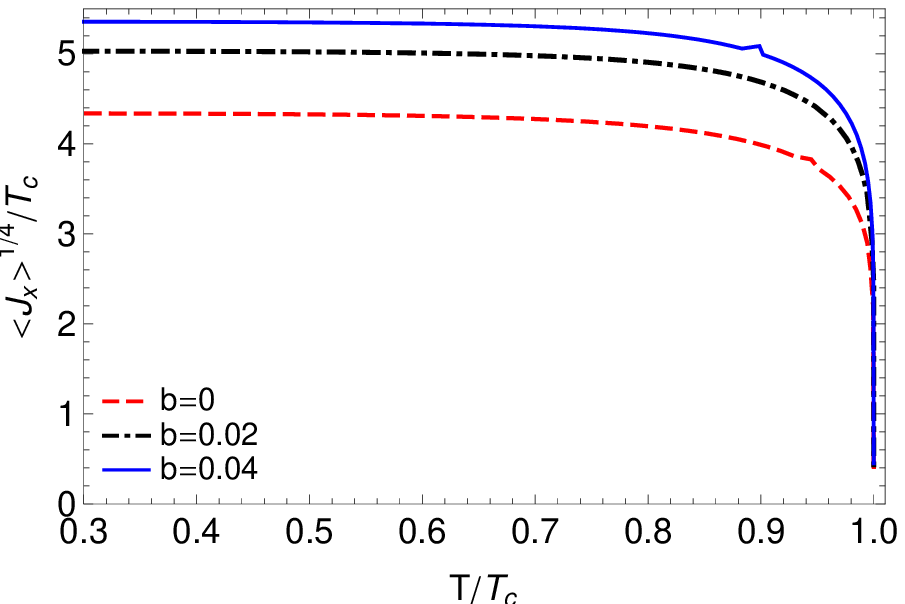}} \qquad %
\subfigure[~$\alpha=-0.08$]{\includegraphics[width=0.4\textwidth]{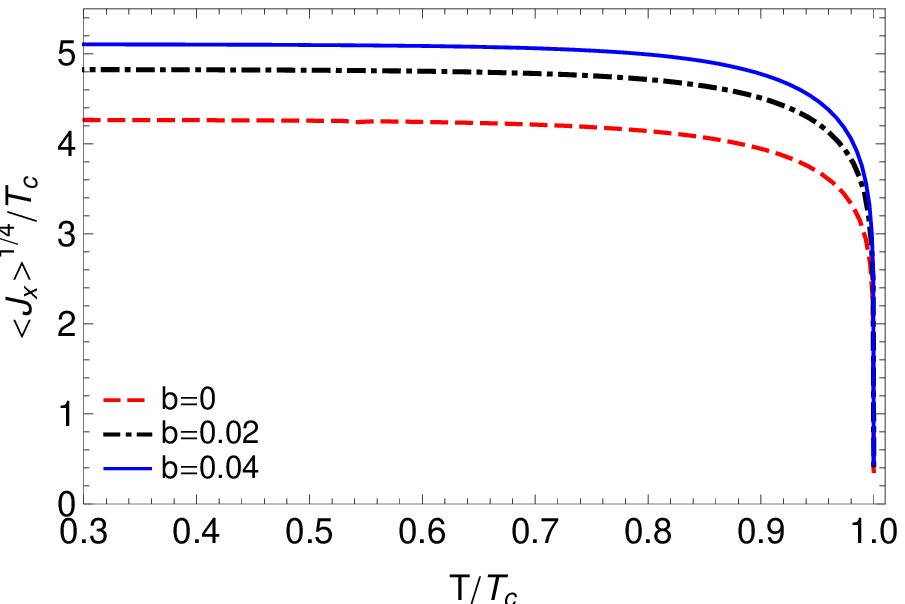}} \qquad %
\caption{The behavior of the condensation parameter as a function
of the temperature for different values of nonlinearity parameters
in $d=6$ with $\overline{m}^{2}=0$.} \label{fig019}
\end{figure*}

\begin{figure*}[t]
\centering
\subfigure[~$b=0$]{\includegraphics[width=0.4\textwidth]{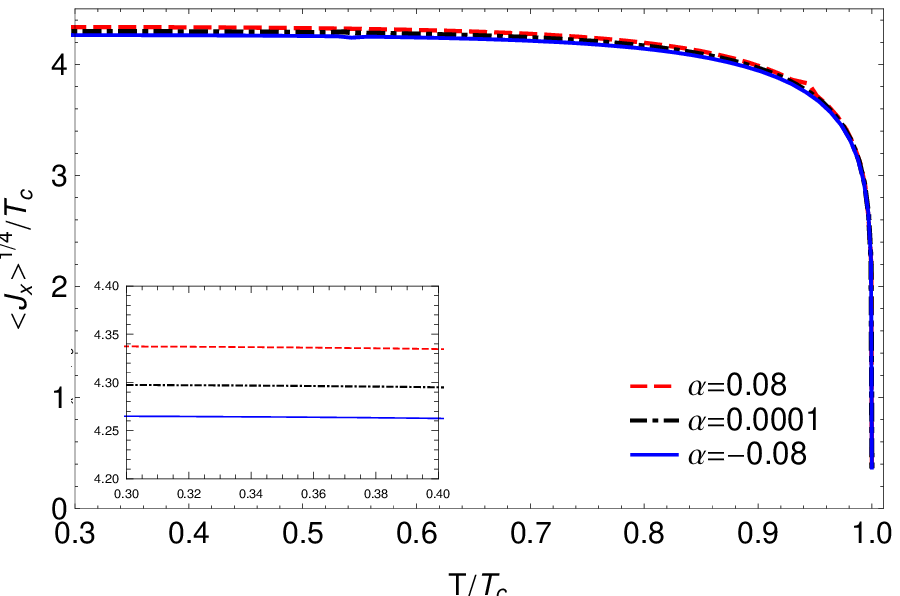}} \qquad %
\subfigure[~$ b=0.04$]{\includegraphics[width=0.4\textwidth]{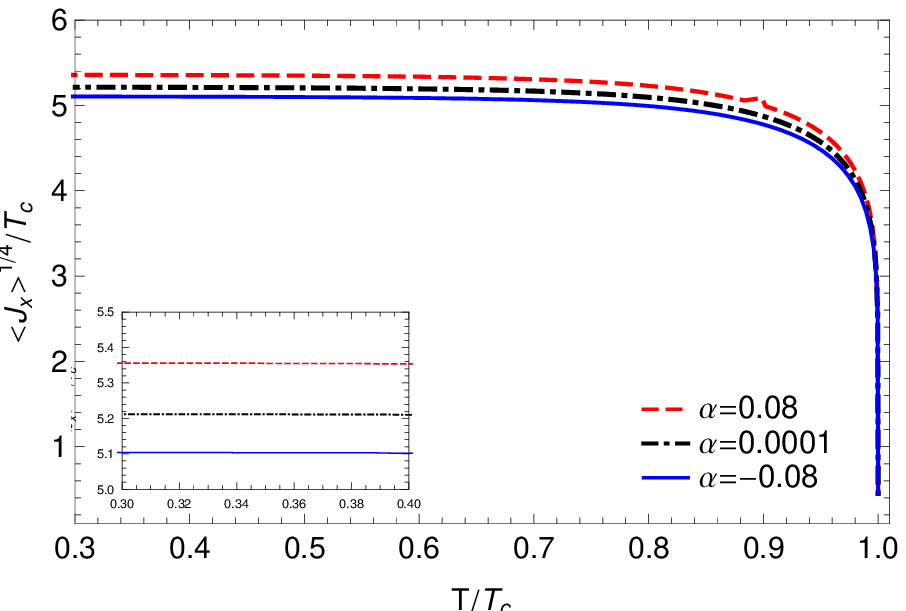}} \qquad %
\caption{The behavior of the condensation parameter as a function
of the temperature for different values of Gauss-Bonnet parameters
in $d=6$ with $\overline{m}^{2}=0$.} \label{fig020}
\end{figure*}

In this section, we want to study the condensation of the vector
field in the background of the AdS black holes with higher order
corrections both in the gravity and gauge field. We write down the
action in the following form
\begin{eqnarray}
&&S =\int d^{d}x\sqrt{-g} \left[\mathcal{L}_{G}+\mathcal{L}_{m}\right], \notag \\
&& \mathcal{L}_{G}= R-2 \Lambda+\frac{\alpha}{2}\left[R^{2}-4 R^{\mu\nu} R_{\mu \nu}
+R^{\mu \nu \rho \sigma} R_{\mu \nu \rho \sigma}\right] , \notag \\
&&\mathcal{L}_{m}=
\mathcal{L}_{\mathcal{NL}}-\frac{1}{2}\rho_{\mu\nu}^{\dagger}
\rho^{\mu\nu}-m^{2} \rho_{\mu}^{\dagger} \rho^{\mu} + i q \gamma
\rho_{\mu} \rho_{\nu}^{\dagger} F^{\mu\nu} ,\label{actgauss}
\end{eqnarray}%
where for same parameters as section \ref{section1}, we have
similar definition while the Gauss-Bonnet parameter, Ricci tensor
and Riemann curvature tensor are defined, respectively, by
$\alpha$, $R_{\mu \nu}$ and $R_{\mu \nu \rho \sigma}$. When
$\alpha\rightarrow0$, the above action reduces to Einstein one. We
take the line elements of the spacetime metric as
\begin{eqnarray} \label{metric2}
&&{ds}^{2}=-f(r){dt}^{2}+\frac{{dr}^{2}}{f(r)}+r^{2} \sum _{i=1}^{d-2}{dx_{i}}^{2}%
,\\
&&f(r)=\frac{r^2}{2 \alpha } \left[1-\sqrt{1-4 \alpha  \left(1-\frac{1}{r^{d-1}}\right)}\right],\label{eqfgauss} %
\end{eqnarray}%
where the function $f(r)$ has the asymptotic behavior as
\begin{equation}
f(r)=\frac{ r^2}{2 \alpha }\left[1-\sqrt{1-4 \alpha }\right].
\end{equation}
We can present the effective radius $L_{\text{eff}}$ for the AdS spacetime as \cite{caipp}
\begin{equation}
L_{\text{eff}}^2=\frac{2 \alpha }{1-\sqrt{1-4 \alpha }}.
\end{equation}
Based on the above equation, in order to have a well-defined
vacuum expectation value $\alpha\leq 1/4$ where the upper bound
$\alpha= 1/4$ is called Chern-Simon limit \cite{caipp}. Besides,
in the CFT side we consider the causality constraint on our choice
of Gauss-Bonnet parameter with $-7/36\leq \alpha \leq 9/100$ and
$-51/196\leq \alpha \leq32/256$ for AdS$5$/CFT$4$ and
AdS$6$/CFT$5$,
respectively\cite{caus1,caus2,caus3,caus4,caus5,caus6,caus7}. We
choose the vector and gauge fields same as equation (\ref{rhoA}).
By variation the equation (\ref{actgauss}) with respect to gauge
and vector fields, same equations as equations (\ref{eqmax}) and
(\ref{eqvector}) were obtained. Moreover, the equations
(\ref{eqphi}) and (\ref{eqrho}) with the asymptotic behavior as
(\ref{eqasym}) are recovered, finally. Equation (\ref{eqasym2}) in
Gauss-Bonnet gravity changes to
\begin{equation}\label{eqasym3}
\Delta _\pm=\frac{1}{2} \left[(d-3)\pm\sqrt{(d-3)^2+4 m^2 L_{\text{eff}}^{2}}\right],
\end{equation}
with the Breitenlohner-Freedman (BF) bound as
\begin{equation}
\overline{m}^{2}\geqslant - \frac{(d-3)^{2}}{4}, \  \   \    \ \overline{m}^{2}=m^{2} L_{\text{eff}}^{2}.
\end{equation}
In the Maxwell limit where $b\rightarrow0$, field equations turn
to corresponding equations in equation \cite{caipp}. In addition,
the Hawking temperature equals to equation (\ref{temp}). The
numerical results are listed in tables IV and V. In order to
compare our results with \cite{caipp} and \cite{gaussp1}, we
choose $\Delta_{+}=3/2$ and $2$ in $d=5$ dimension. There is a
perfect agreement between our results with corresponding cases in
\cite{caipp} and \cite{gaussp1}. However, because of the
difficulty of numerical solution, we can only do our study for one
value of mass in $d=6$. Tables IV and V give information about the
values of critical temperature $T_{c}$ based on $\rho^{1/(d-2)}$
for different effects of the mass and nonlinear parameters as well
as Gauss-Bonnet parameter. Increasing the Gauss-Bonnet coefficient
has the same effect as larger values of mass and nonlinear
parameters on the critical temperatures. We also face with
diminishing of critical temperature for stronger values of
$\alpha$ which makes the condensation harder to form by putting
off the appearance of hair in gravity side which corresponds to
superconducting phase in boundary field theory. Moreover, the
results in $d=5$ with $\overline{m}^{2}=-3/4$ and $\alpha=0.0001$
are alike the outcomes of corresponding case in Einstein gravity.
This is rooted in the fact that as said before in
$\alpha\rightarrow0$ limit, the Einstein case is regained.
Meanwhile, figs. \ref{fig17}-\ref{fig020} show the behavior of
condensation as a function of temperature in different choices of
mass, nonlinearity and Gauss-Bonnet parameters. Increasing each
one of theses three parameters makes the condensation value
raising. Furthermore, the effect of different values of
Gauss-Bonnet parameter becomes more and more obvious by growing
the nonlinear parameter $b$. Based on the obtained results and
same as \cite{caipp} the Gauss-Bonnet term doesn't change the
critical exponent of condensation which means that we face with
second order phase transition for all values of $\alpha$, the same
as section \ref{section1}.
\begin{table*}[t]
\label{tab4}
\begin{center}
\begin{tabular}{c|c|c|c|c|c|c|}
\cline{2-3}\cline{2-7}\cline{4-7}
& \multicolumn{2}{|c|}{$b=0$} & \multicolumn{2}{|c|}{$b=0.02$} &
\multicolumn{2}{|c|}{$b=0.04$} \\ \cline{2-3}\cline{2-7}\cline{4-7}
& $\overline{m}^{2}=-3/4$ & $\overline{m}^{2}=0$ & $\overline{m}^{2}=-3/4$ & $\overline{m}^{2}=0$ & $\overline{m}^{2}=-3/4$ & $\overline{m}^{2}=0$
\\ \hline
\multicolumn{1}{|c|}{$\alpha=0.08$} & $0.218$ $\rho^{1/3}$ & $ 0.195$  $\rho^{1/3}$ & $0.204$  $\rho^{1/3}$ & $%
0.174$  $\rho^{1/3}$ & $0.196$  $\rho^{1/3}$ & $0.163$  $\rho^{1/3}$ \\ \hline
\multicolumn{1}{|c|}{$\alpha=0.0001$} & $0.224$  $\rho^{1/3}$ & $0.200$  $\rho^{1/3}$ & $0.218$  $\rho^{1/3}$ & $%
0.181$  $\rho^{1/3}$ & $0.204$  $\rho^{1/3}$ & $0.171$  $\rho^{1/3}$ \\ \hline
\multicolumn{1}{|c|}{$\alpha=-0.08$} & $0.229$  $\rho^{1/3}$ & $0.205$  $\rho^{1/3}$ & $0.218$  $\rho^{1/3}$ & $%
0.187$  $\rho^{1/3}$ & $0.210$  $\rho^{1/3}$ & $0.177$  $\rho^{1/3}$ \\ \hline
\end{tabular}%
\caption{Numerical results for critical temperature $T_{c}$ in
$d=5$ for different values of mass, nonlinearity and Gauss-Bonnet
parameters.}
\end{center}
\end{table*}
\begin{table*}[t]
\label{tab5}
\begin{center}
\begin{tabular}{c|c|c|c|}
\cline{2-4}
& $b=0$ &$b=0.02$ & $b=0.04$ \\
\hline
\multicolumn{1}{|c|}{$\alpha=0.08$} & $0.267$ $\rho^{1/4}$& $0.228$ $\rho^{1/4}$ & $0.212$ $\rho^{1/4}$ \\
\hline
\multicolumn{1}{|c|}{$\alpha=0.0001$} & $0.273$ $\rho^{1/4}$ & $0.236$ $\rho^{1/4}$ & $0.221$ $\rho^{1/4}$ \\
\hline
\multicolumn{1}{|c|}{$\alpha=-0.08$} & $0.277$ $\rho^{1/4}$ & $0.243$ $\rho^{1/4}$ & $0.228$ $\rho^{1/4}$\\
\hline
\end{tabular}%
\caption{Numerical results for critical temperature $T_{c}$ with
$\overline{m}^{2}=0$ in $d=6$ for different values of nonlinearity
and Gauss-Bonnet parameters.}
\end{center}
\end{table*}


\subsection{Conductivity}\label{section4}
\begin{figure*}[t]
\centering
\subfigure[~$b=0$, $\alpha=0.08$]{\includegraphics[width=0.4\textwidth]{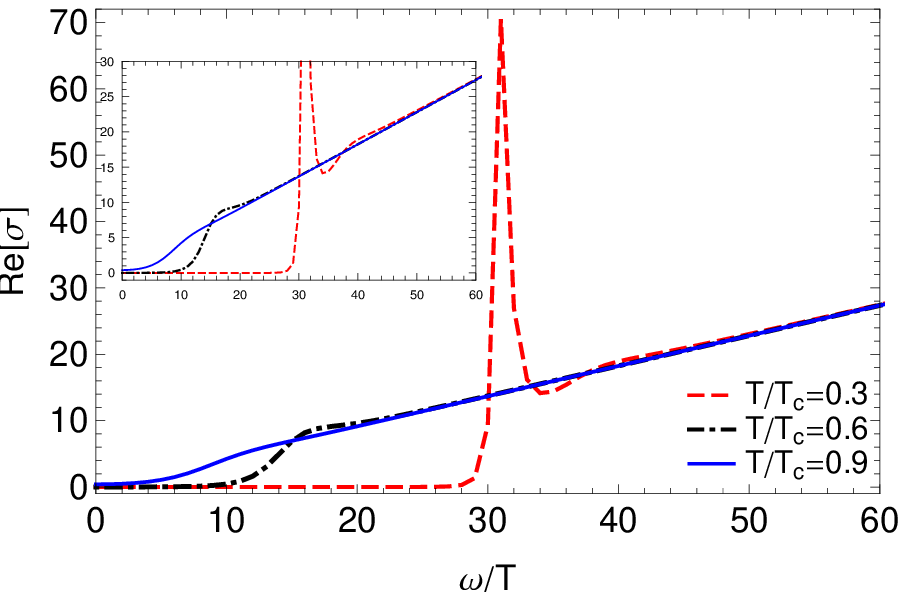}} \qquad %
\subfigure[~$ b=0.04$, $\alpha=0.08$]{\includegraphics[width=0.4\textwidth]{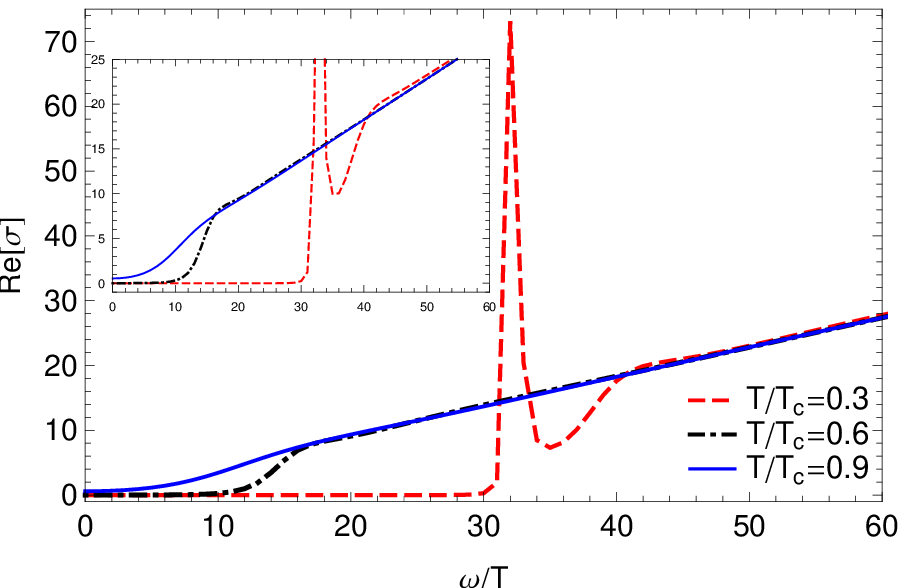}} \qquad %
\subfigure[~$b=0$, $\alpha=-0.08$]{\includegraphics[width=0.4\textwidth]{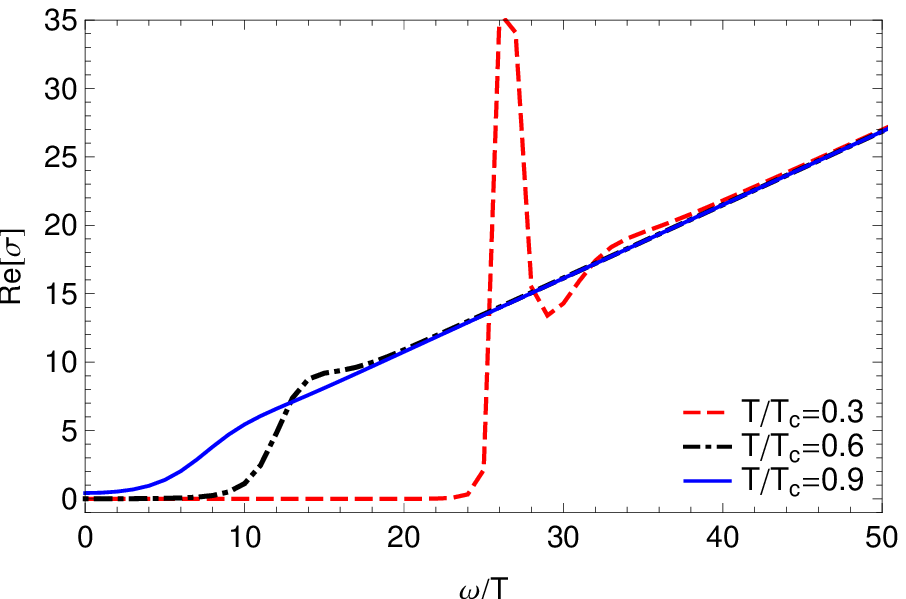}} \qquad %
\subfigure[~$ b=0.04$, $\alpha=-0.08$]{\includegraphics[width=0.4\textwidth]{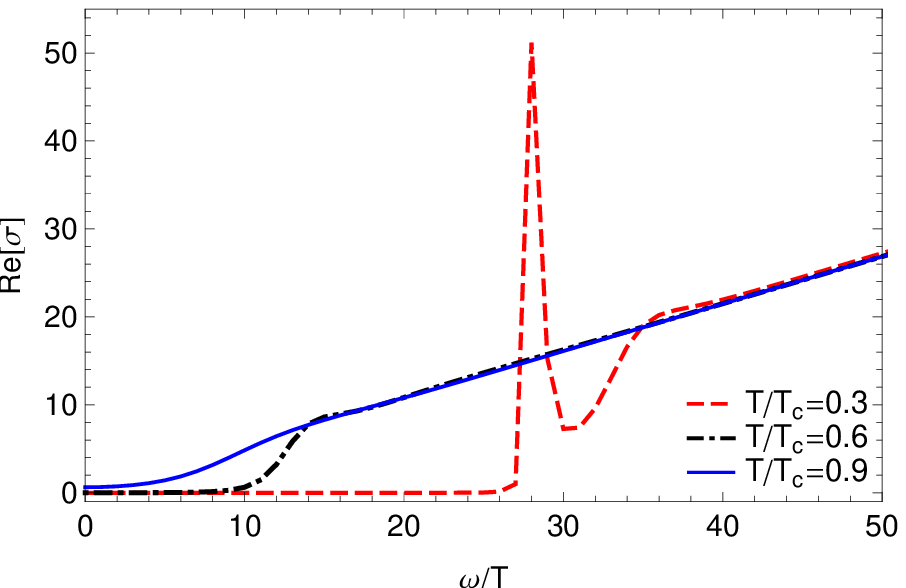}} \qquad %
\caption{The behavior of real parts of conductivity with
$\overline{m}^{2}=-3/4$ in $d=5$.} \label{fig21a}
\end{figure*}

\begin{figure*}[t]
\centering
\subfigure[~$b=0$, $\alpha=0.08$]{\includegraphics[width=0.4\textwidth]{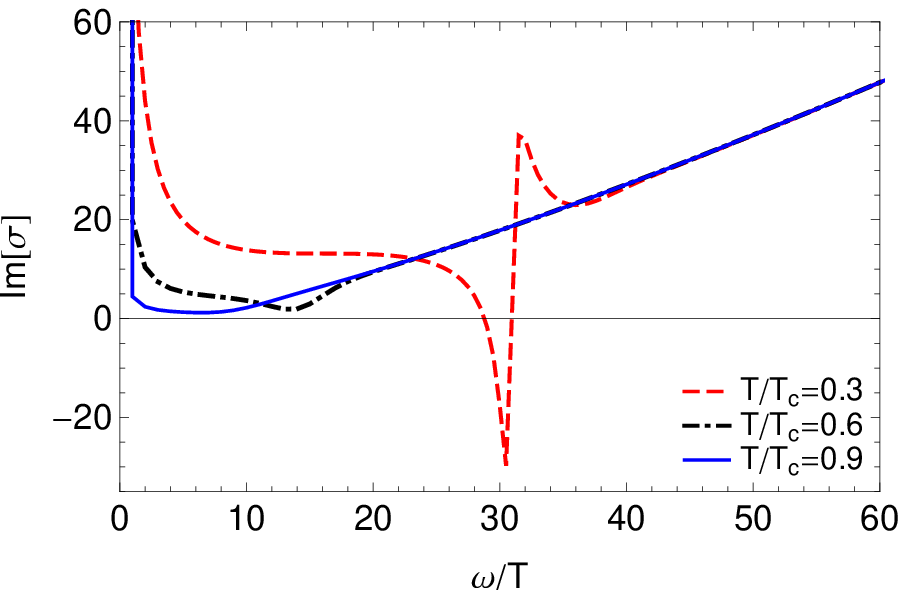}} \qquad %
\subfigure[~$ b=0.04$, $\alpha=0.08$]{\includegraphics[width=0.4\textwidth]{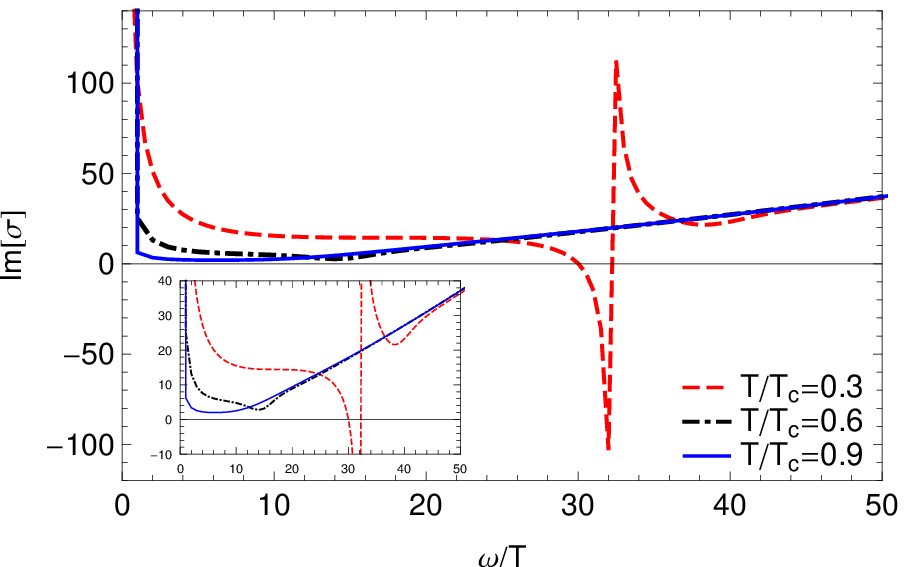}} \qquad %
\subfigure[~$b=0$, $\alpha=-0.08$]{\includegraphics[width=0.4\textwidth]{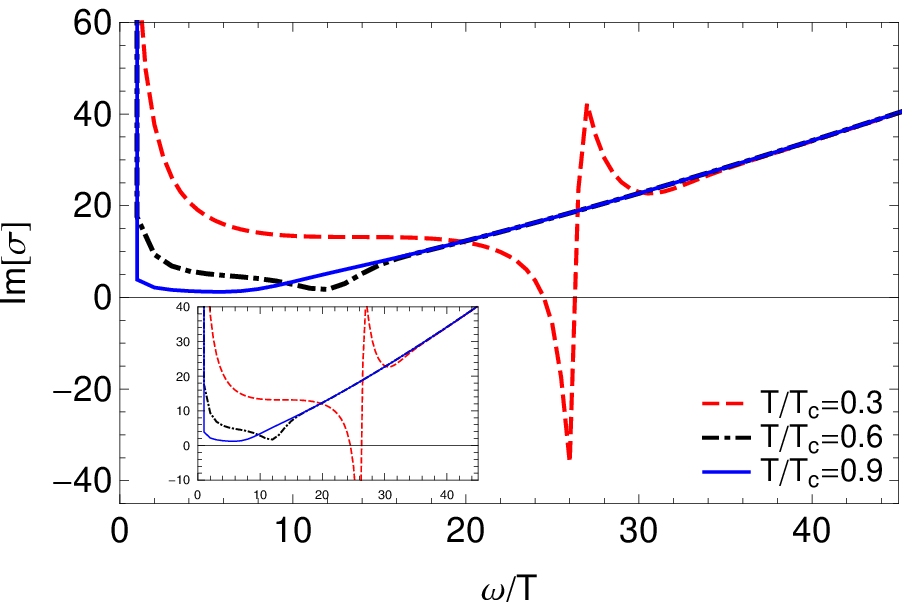}} \qquad %
\subfigure[~$ b=0.04$, $\alpha=-0.08$]{\includegraphics[width=0.4\textwidth]{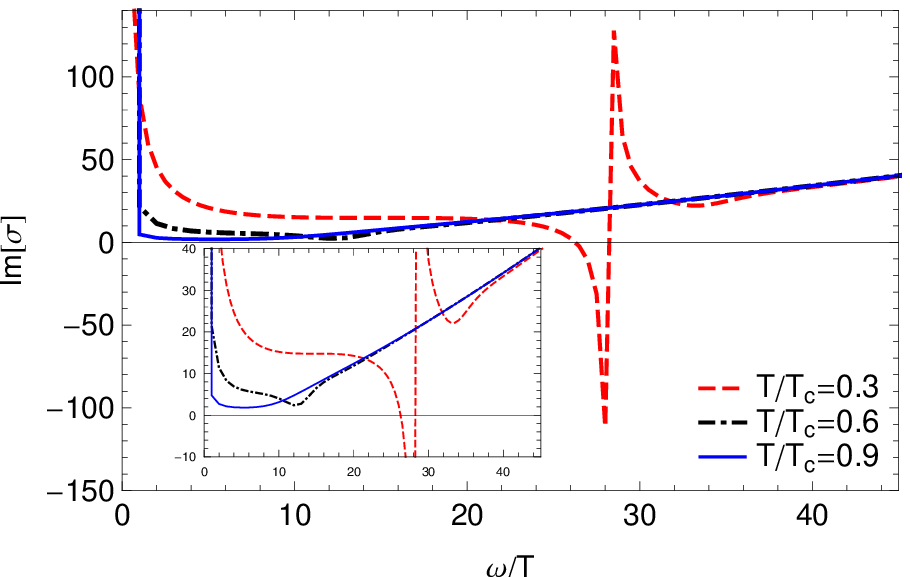}} \qquad %
\caption{The behavior of imaginary parts of conductivity with
$\overline{m}^{2}=-3/4$ in $d=5$.} \label{fig24}
\end{figure*}

\begin{figure*}[t]
\centering
\subfigure[~$b=0$, $\alpha=0.08$]{\includegraphics[width=0.4\textwidth]{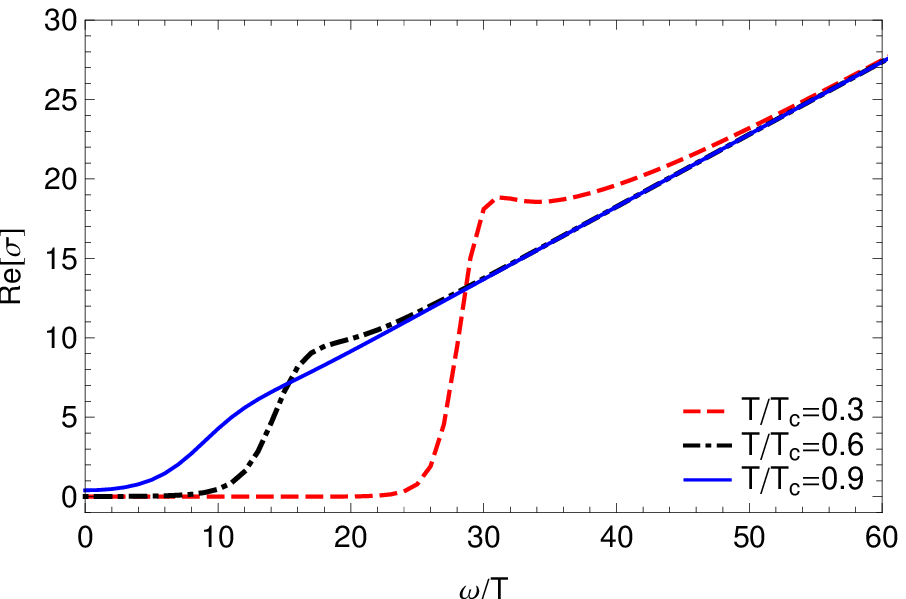}} \qquad %
\subfigure[~$ b=0.04$, $\alpha=0.08$]{\includegraphics[width=0.4\textwidth]{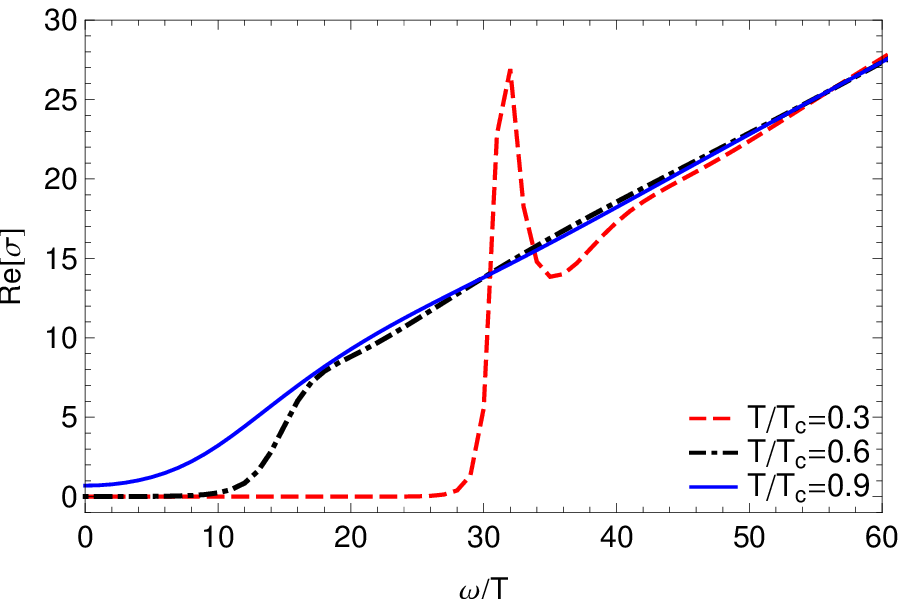}} \qquad %
\subfigure[~$b=0$, $\alpha=-0.08$]{\includegraphics[width=0.4\textwidth]{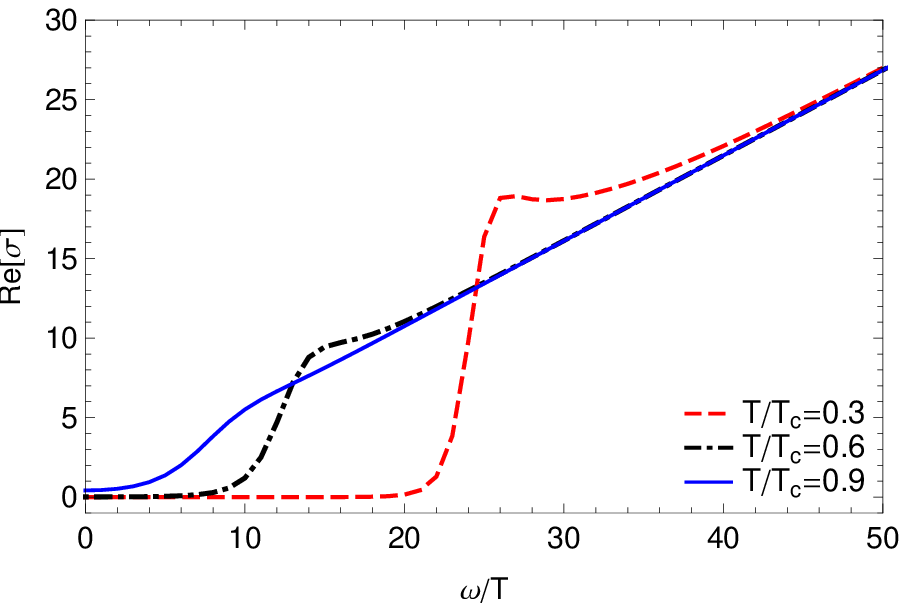}} \qquad %
\subfigure[~$ b=0.04$, $\alpha=-0.08$]{\includegraphics[width=0.4\textwidth]{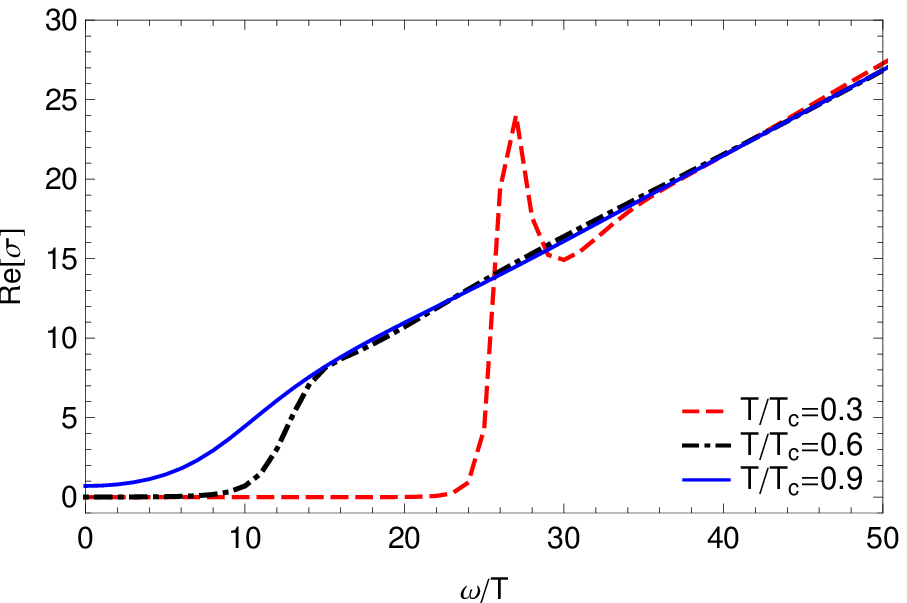}} \qquad %
\caption{The behavior of real parts of conductivity with
$\overline{m}^{2}=0$ in $d=5$.} \label{fig27}
\end{figure*}

\begin{figure*}[t]
\centering
\subfigure[~$b=0$, $\alpha=0.08$]{\includegraphics[width=0.4\textwidth]{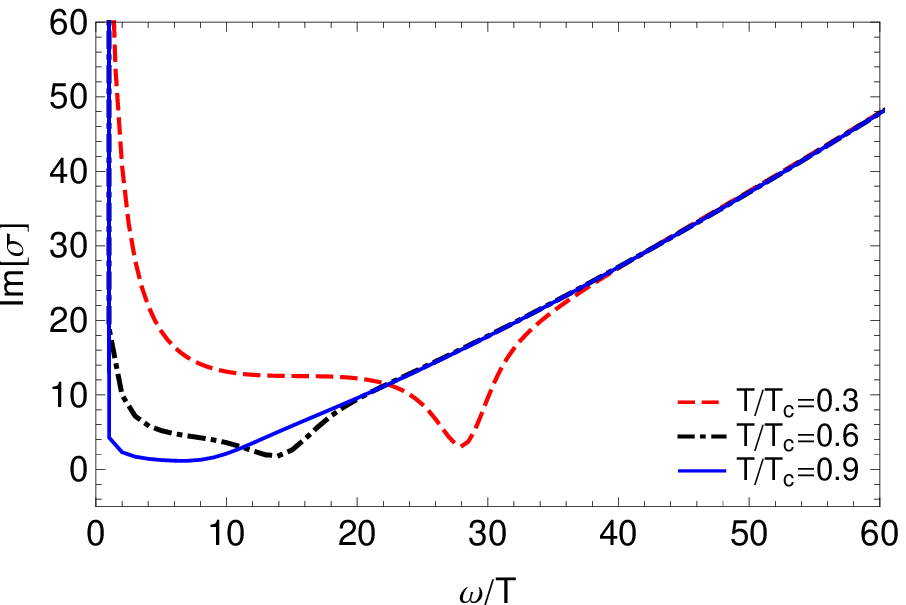}} \qquad %
\subfigure[~$ b=0.04$, $\alpha=0.08$]{\includegraphics[width=0.4\textwidth]{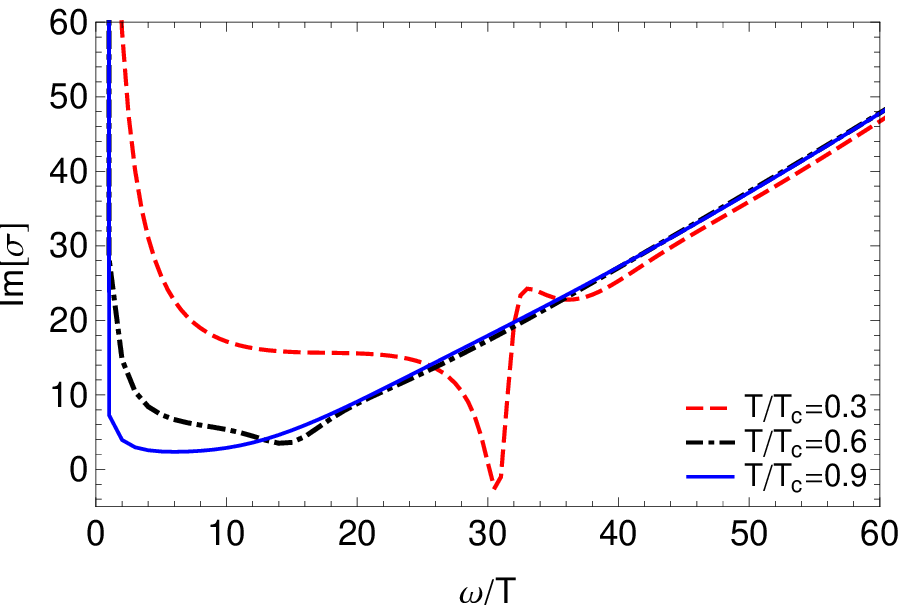}} \qquad %
\subfigure[~$b=0$, $\alpha=-0.08$]{\includegraphics[width=0.4\textwidth]{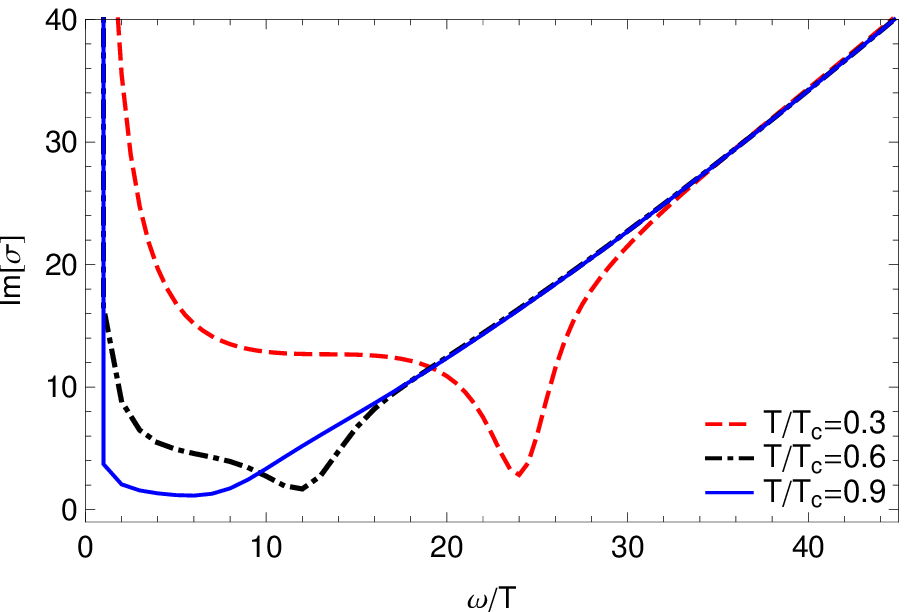}} \qquad %
\subfigure[~$ b=0.04$, $\alpha=-0.08$]{\includegraphics[width=0.4\textwidth]{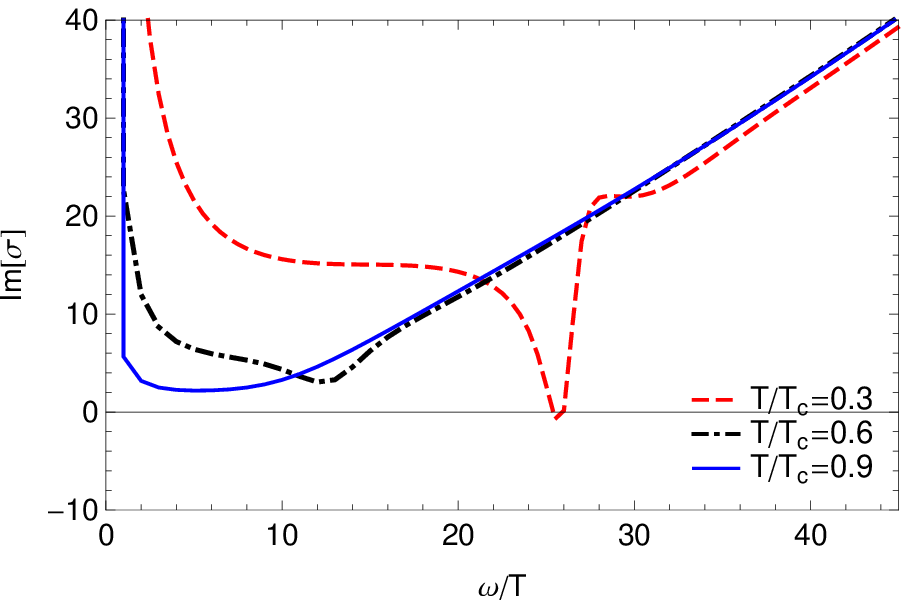}} \qquad %
\caption{The behavior of imaginary parts of conductivity with $\overline{m}^{2}=0$ in $d=5$}
\label{fig30}
\end{figure*}

\begin{figure*}[t]
\centering
\subfigure[~$b=0$, $\alpha=0.08$]{\includegraphics[width=0.4\textwidth]{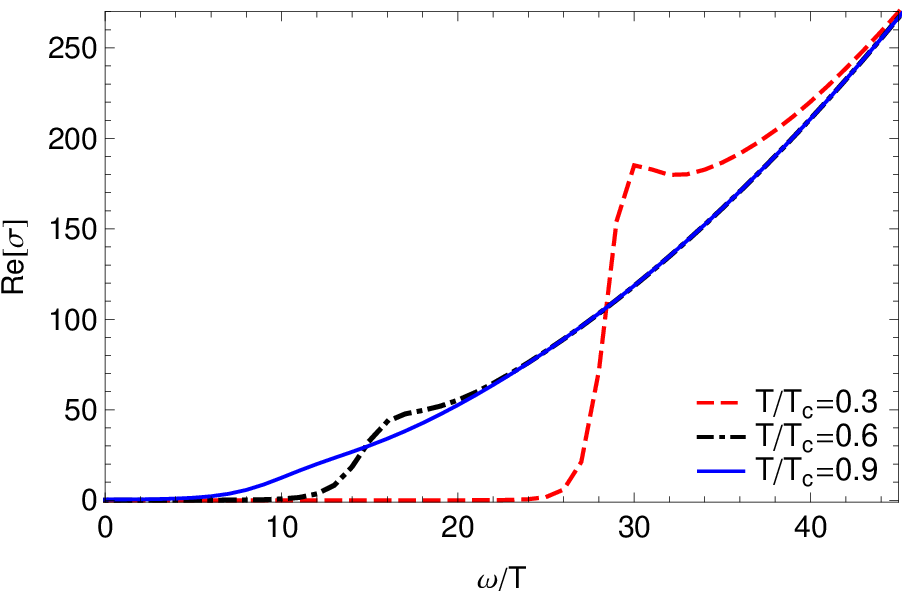}} \qquad %
\subfigure[~$ b=0.04$, $\alpha=0.08$]{\includegraphics[width=0.4\textwidth]{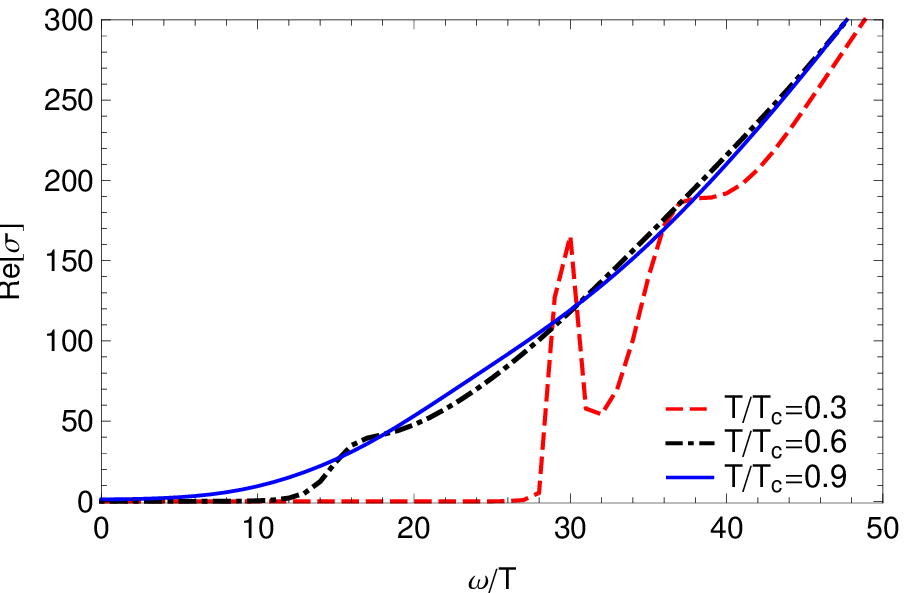}} \qquad %
\subfigure[~$b=0$, $\alpha=-0.08$]{\includegraphics[width=0.4\textwidth]{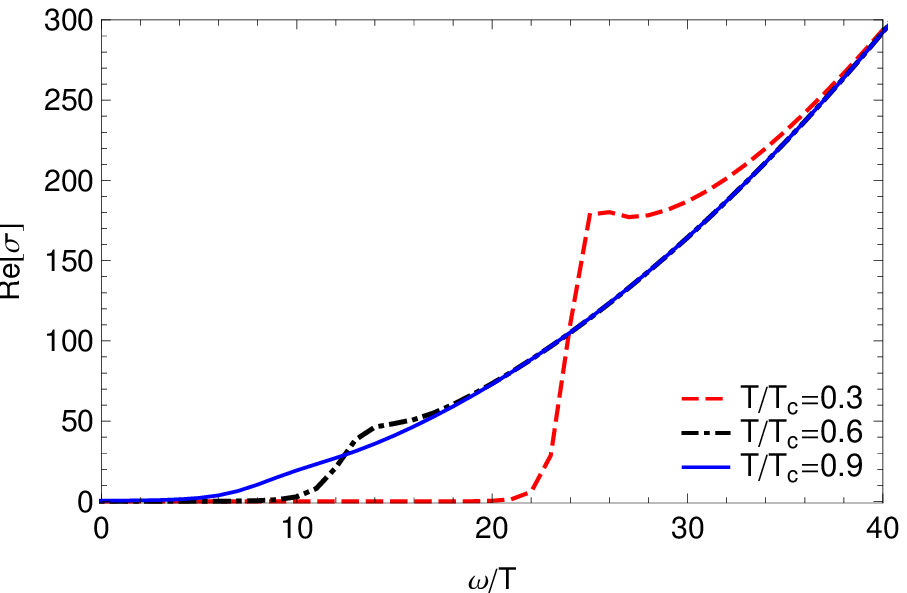}} \qquad %
\subfigure[~$ b=0.04$, $\alpha=-0.08$]{\includegraphics[width=0.4\textwidth]{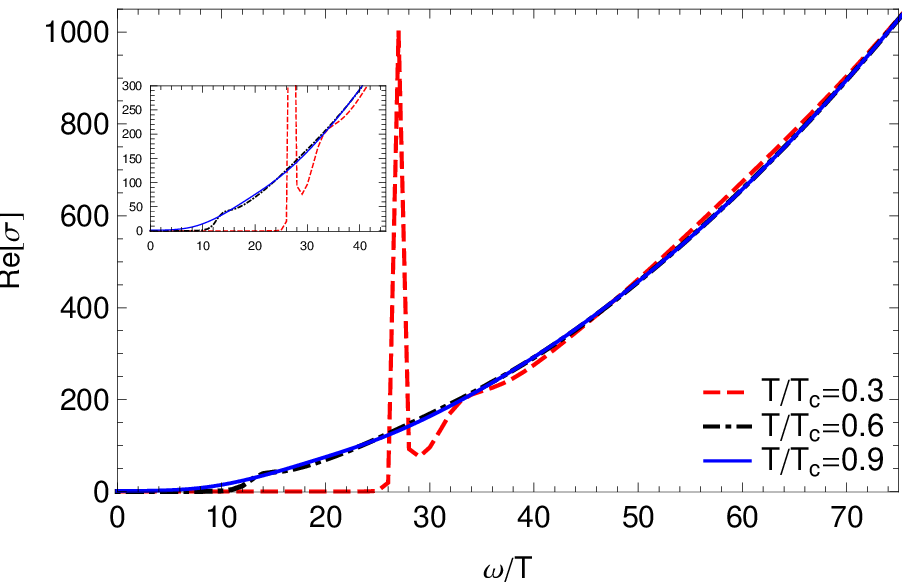}} \qquad %
\caption{The behavior of real parts of conductivity with $\overline{m}^{2}=0$ in $d=6$}
\label{6d}
\end{figure*}

\begin{figure*}[t]
\centering
\subfigure[~$b=0$, $\alpha=0.08$]{\includegraphics[width=0.4\textwidth]{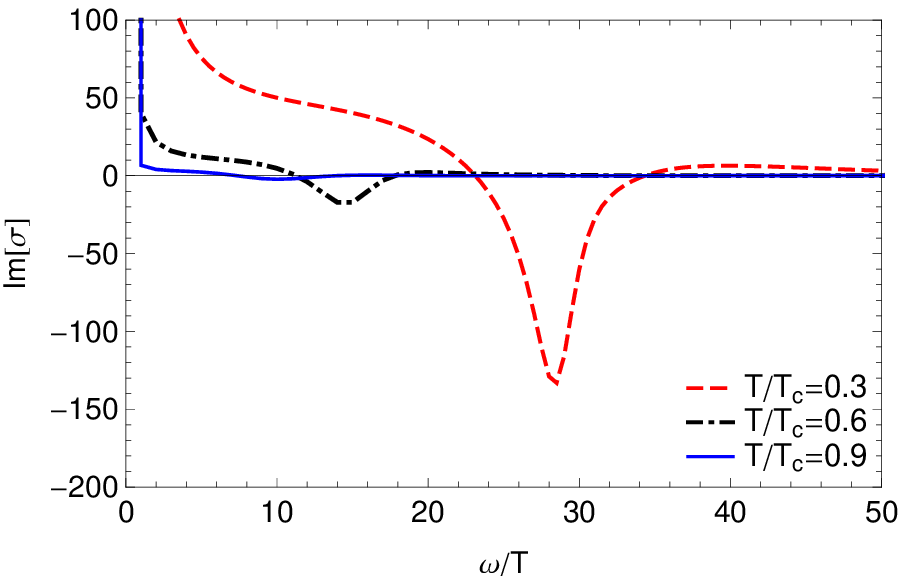}} \qquad %
\subfigure[~$ b=0.04$, $\alpha=0.08$]{\includegraphics[width=0.4\textwidth]{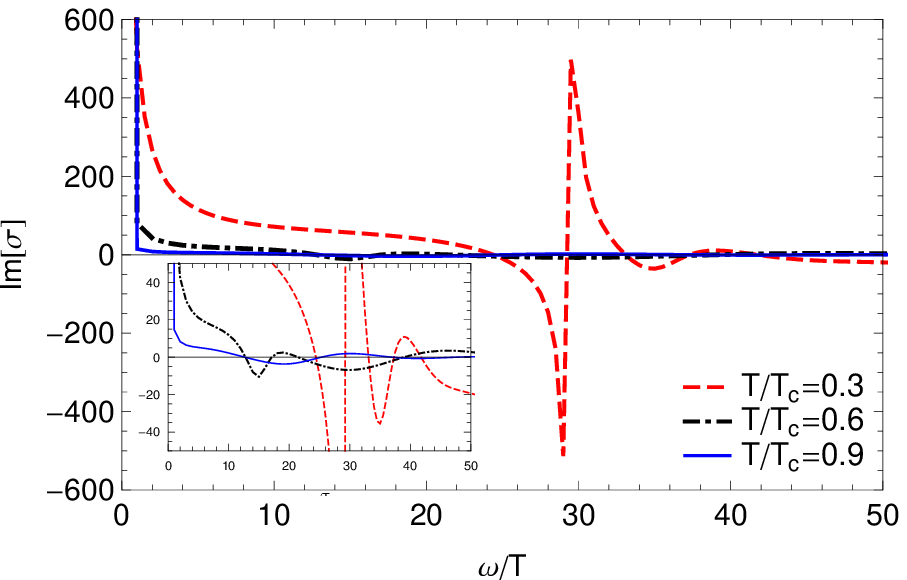}} \qquad %
\subfigure[~$b=0$, $\alpha=-0.08$]{\includegraphics[width=0.4\textwidth]{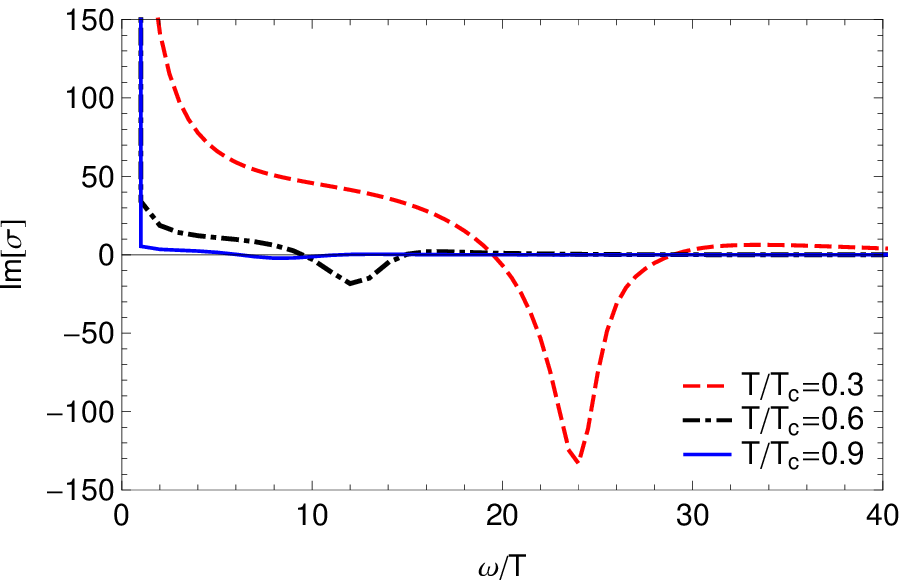}} \qquad %
\subfigure[~$ b=0.04$, $\alpha=-0.08$]{\includegraphics[width=0.4\textwidth]{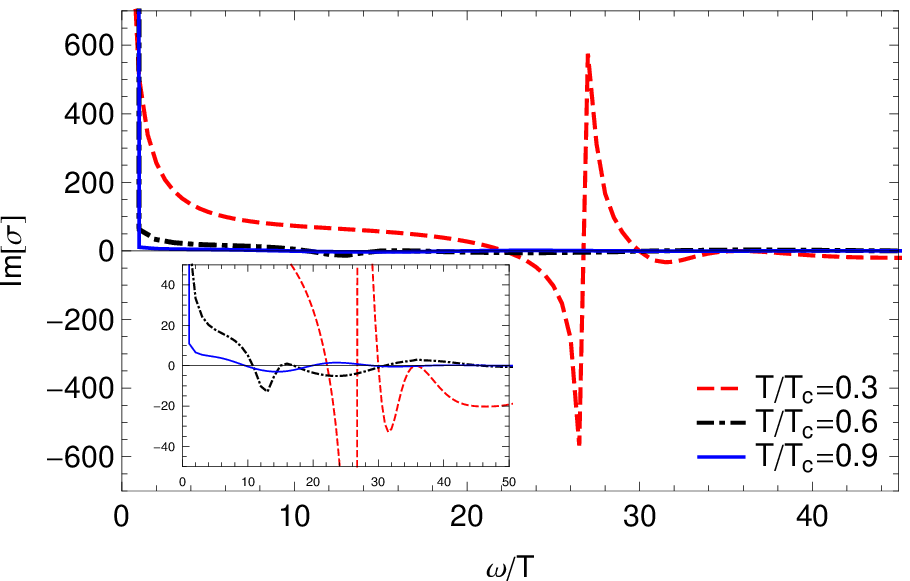}} \qquad %
\caption{The behavior of imaginary parts of conductivity with
$\overline{m}^{2}=0$ in $d=6$.} \label{6dd}
\end{figure*}

\begin{figure*}[t]
\centering
\subfigure[~$\alpha=0.08$]{\includegraphics[width=0.4\textwidth]{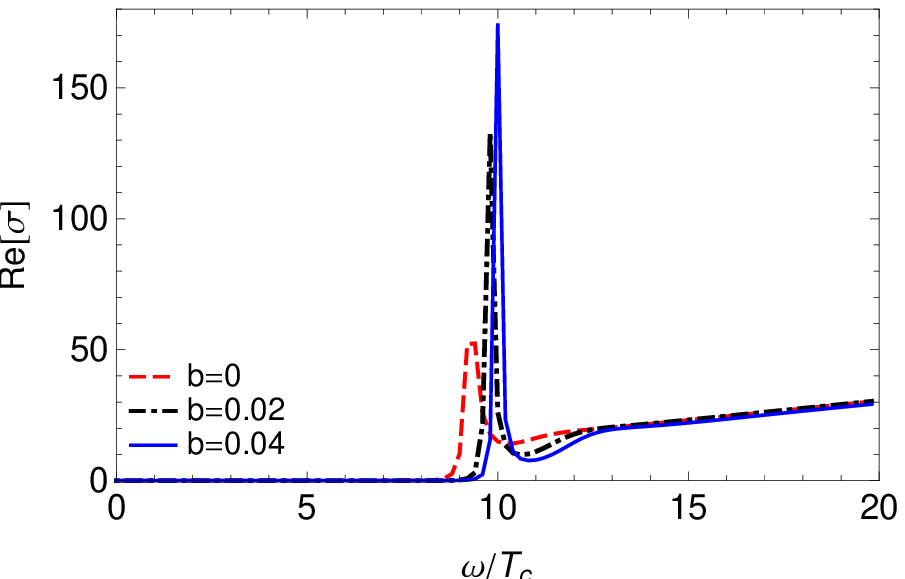}} \qquad %
\subfigure[~$\alpha=-0.08$]{\includegraphics[width=0.4\textwidth]{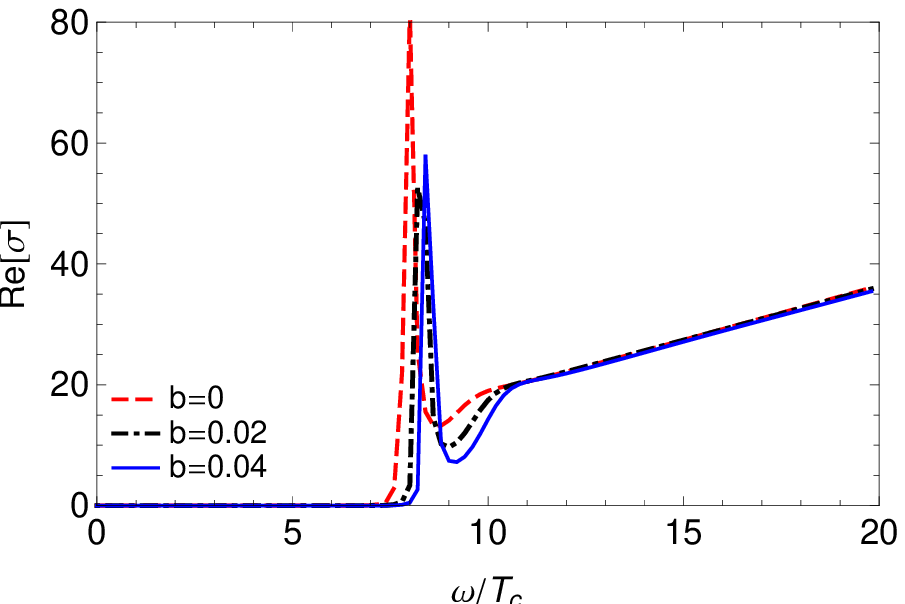}} \qquad %
\subfigure[~$\alpha=0.08$]{\includegraphics[width=0.4\textwidth]{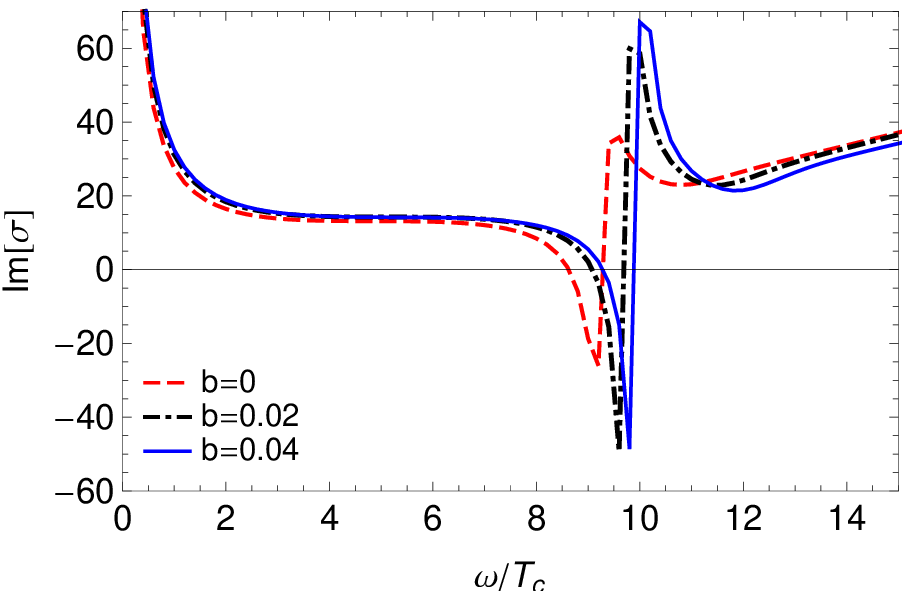}} \qquad %
\subfigure[~$\alpha=-0.08$]{\includegraphics[width=0.4\textwidth]{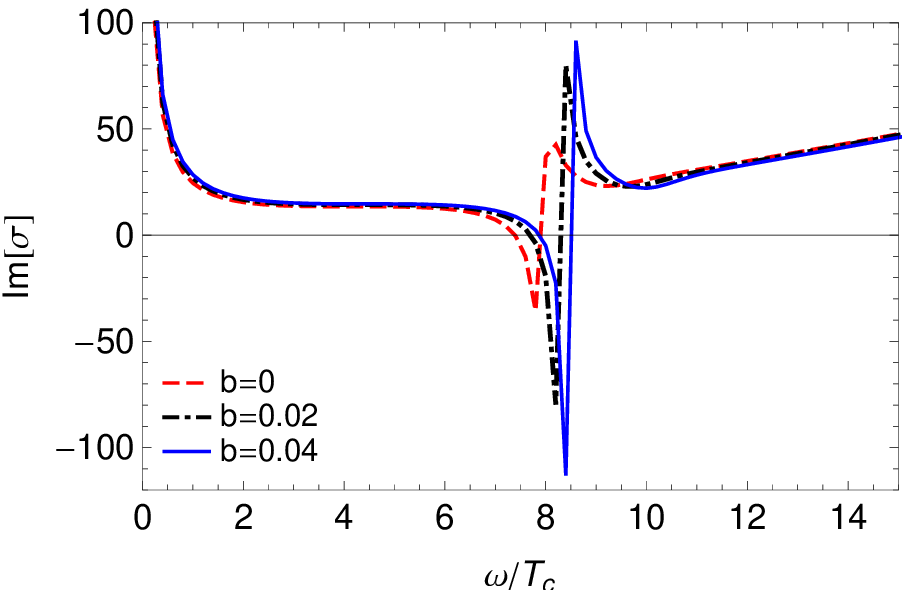}} \qquad %
\caption{The behavior of real and imaginary parts of conductivity
with $\overline{m}^{2}=-3/4$ and $T/T_{c}=0.3$ in $d=5$.}
\label{fig36a}
\end{figure*}

\begin{figure*}[t]
\centering
\subfigure[~$b=0$]{\includegraphics[width=0.4\textwidth]{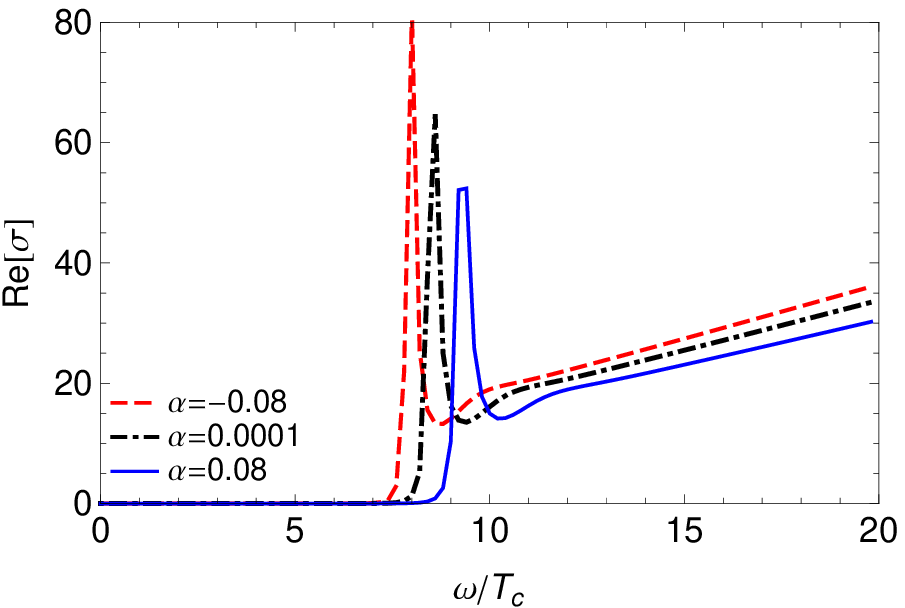}} \qquad %
\subfigure[~$ b=0.04$]{\includegraphics[width=0.4\textwidth]{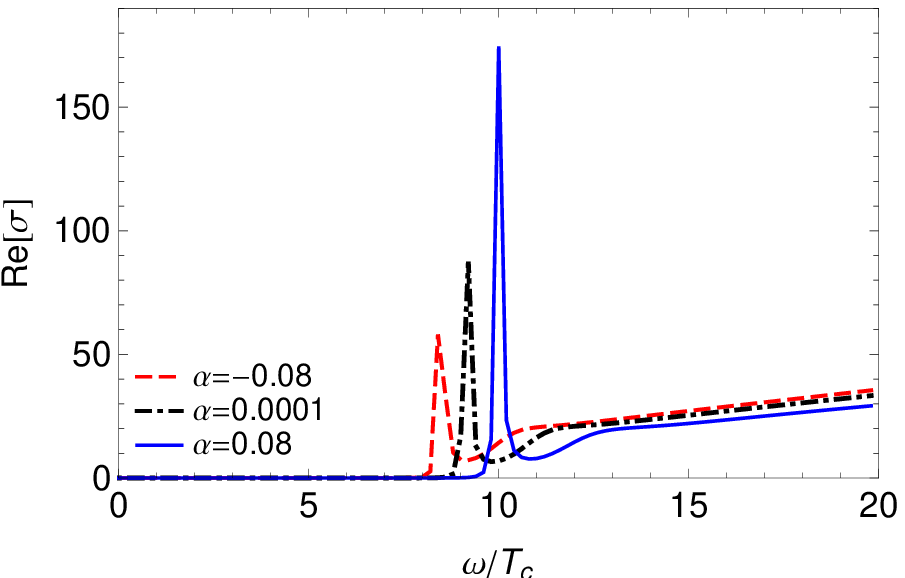}} \qquad %
\subfigure[~$b=0$]{\includegraphics[width=0.4\textwidth]{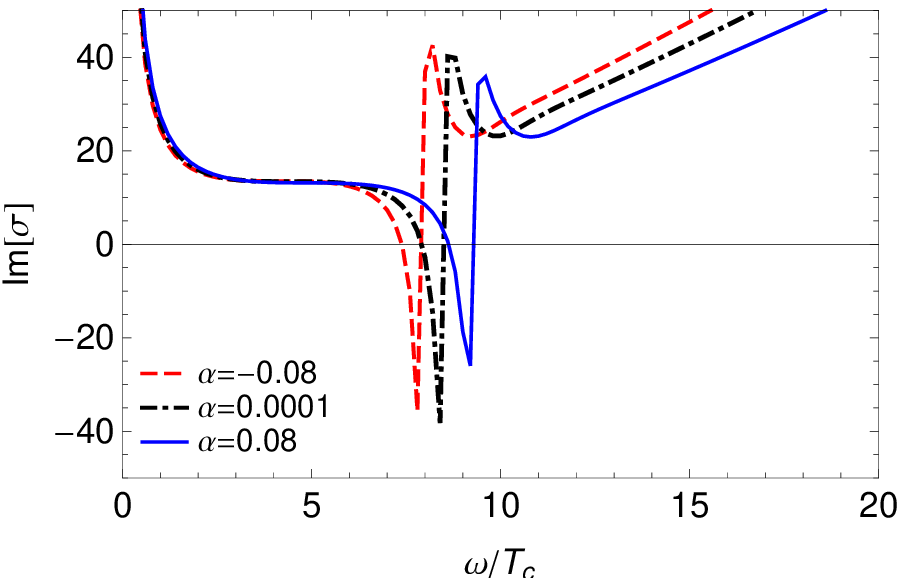}} \qquad %
\subfigure[~$ b=0.04$]{\includegraphics[width=0.4\textwidth]{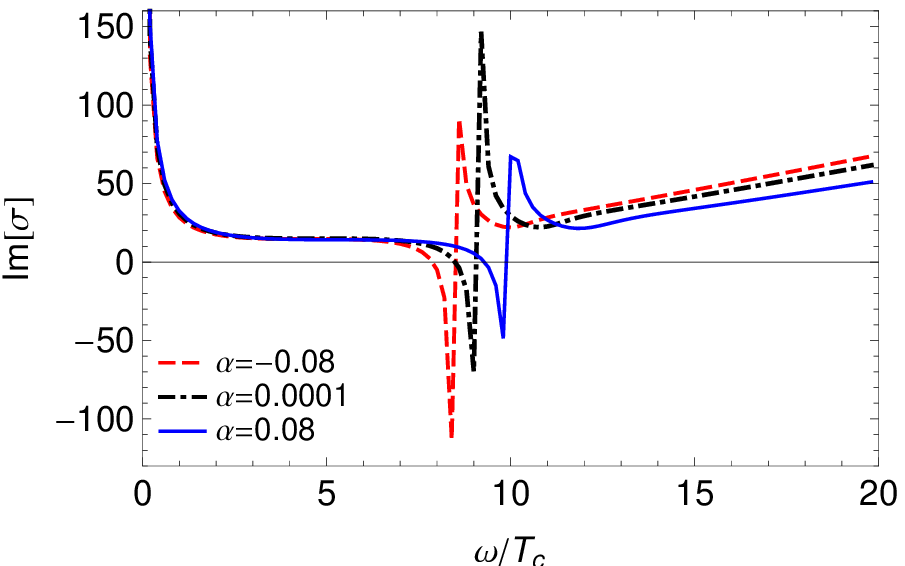}} \qquad %
\caption{The behavior of real and imaginary parts of conductivity
for $\overline{m}^{2}=-3/4$ and $T/T_{c}=0.3$ in $d=5$.}
\label{fig37}
\end{figure*}

\begin{figure*}[t]
\centering
\subfigure[~$\alpha=0.08$]{\includegraphics[width=0.4\textwidth]{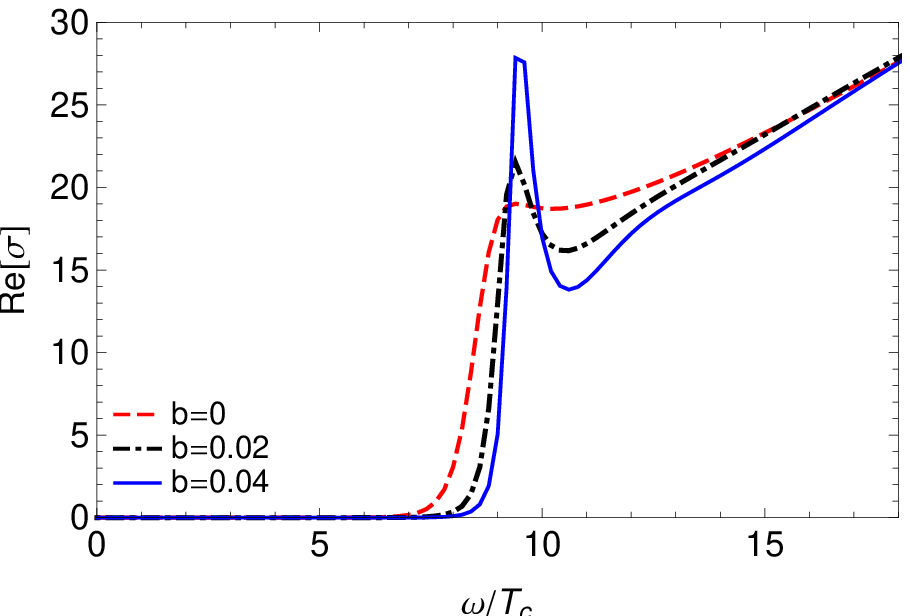}} \qquad %
\subfigure[~$\alpha=-0.08$]{\includegraphics[width=0.4\textwidth]{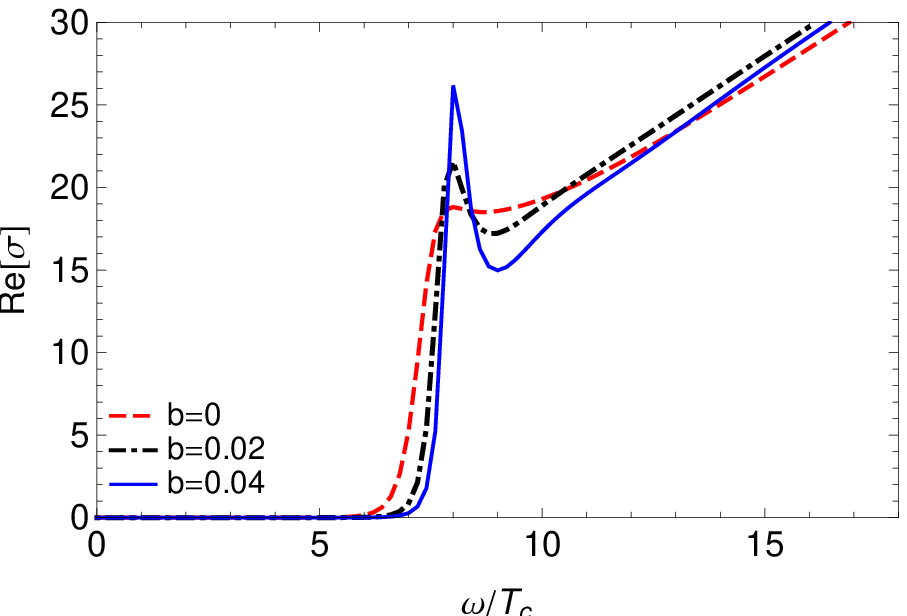}} \qquad %
\subfigure[~$\alpha=0.08$]{\includegraphics[width=0.4\textwidth]{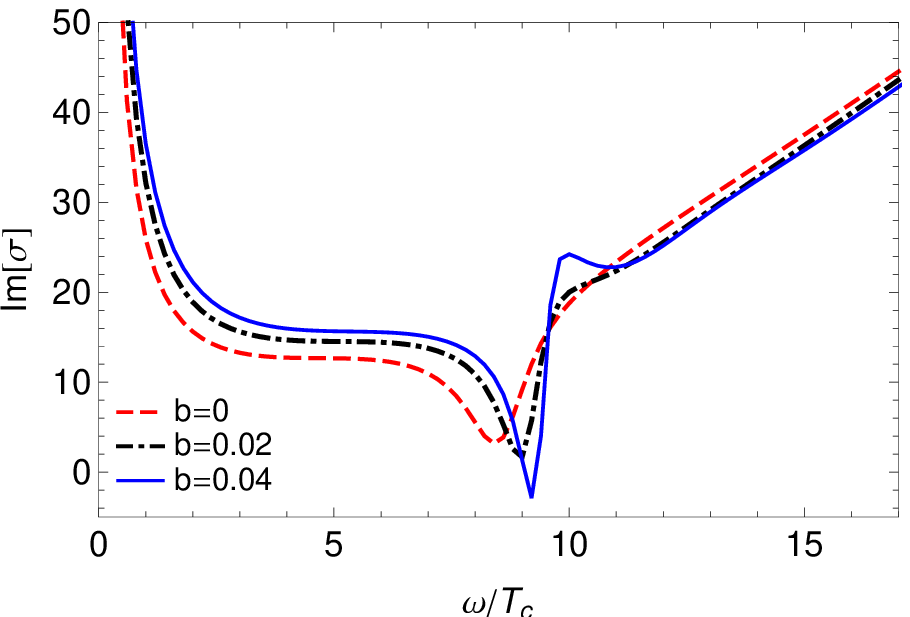}} \qquad %
\subfigure[~$\alpha=-0.08$]{\includegraphics[width=0.4\textwidth]{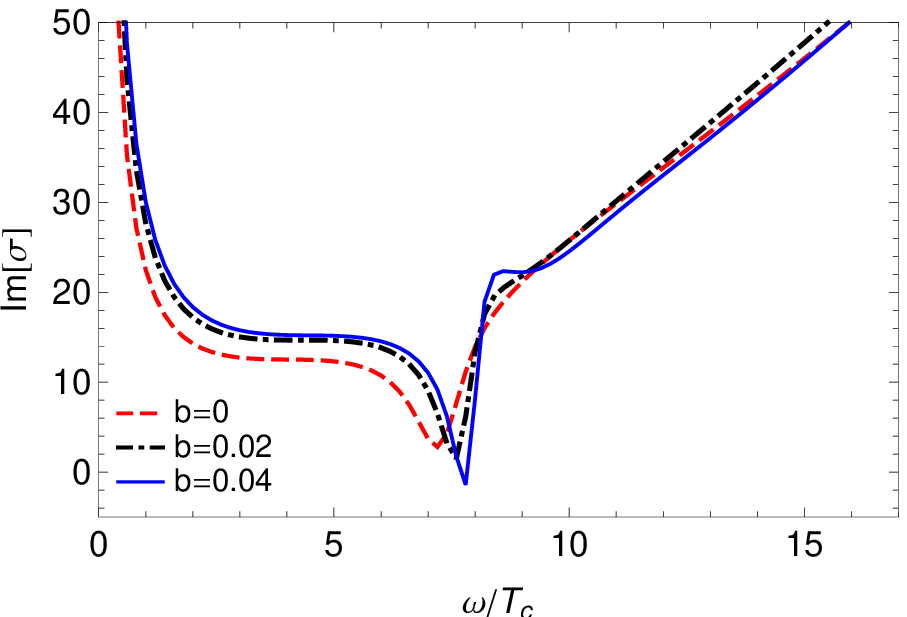}} \qquad %
\caption{The behavior of real and imaginary parts of conductivity
for $\overline{m}^{2}=0$ and $T/T_{c}=0.3$ in $d=5$.}
\label{fig38}
\end{figure*}

\begin{figure*}[t]
\centering
\subfigure[~$b=0$]{\includegraphics[width=0.4\textwidth]{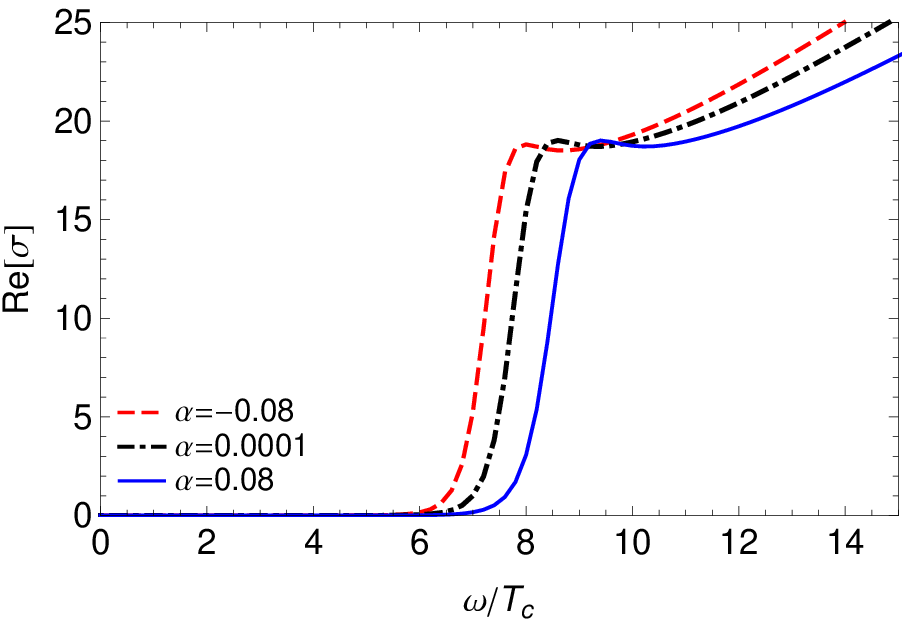}} \qquad %
\subfigure[~$ b=0.04$]{\includegraphics[width=0.4\textwidth]{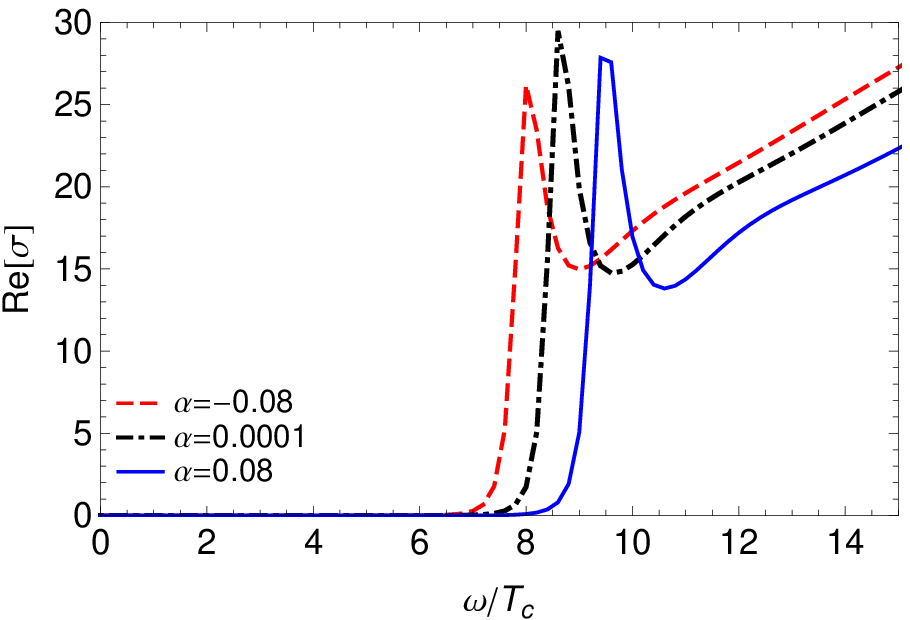}} \qquad %
\subfigure[~$b=0$]{\includegraphics[width=0.4\textwidth]{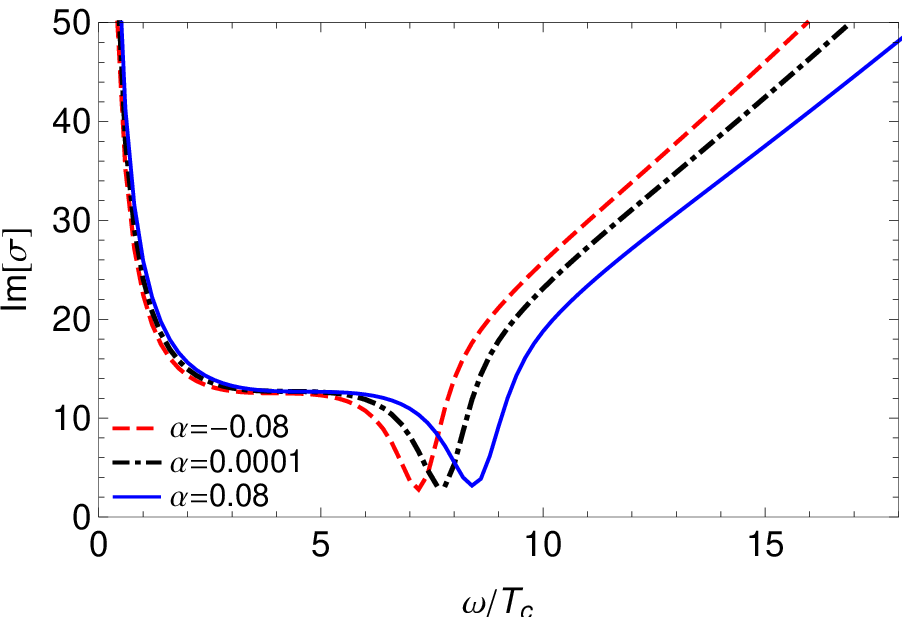}} \qquad %
\subfigure[~$ b=0.04$]{\includegraphics[width=0.4\textwidth]{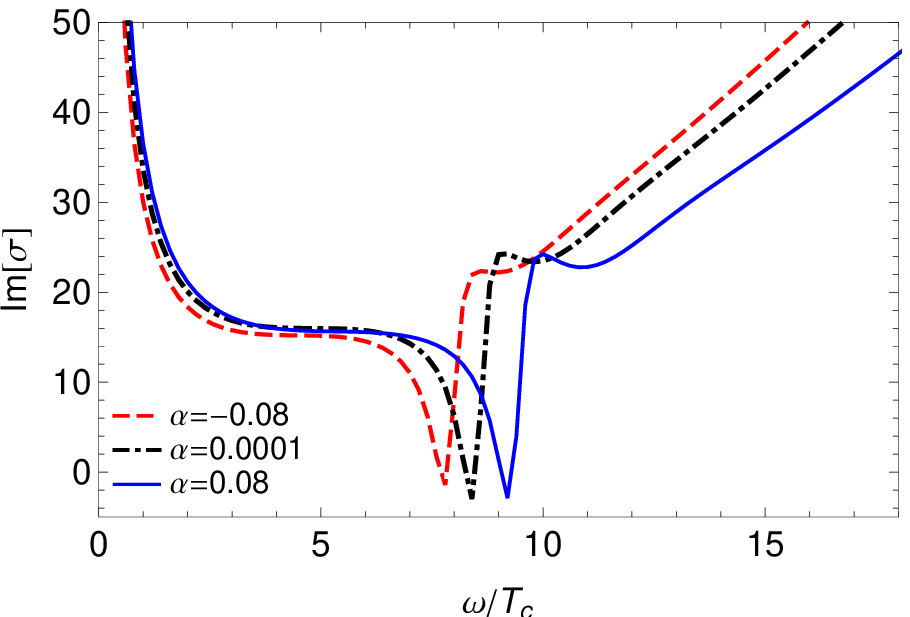}} \qquad %
\caption{The behavior of real and imaginary parts of conductivity
for $\overline{m}^{2}=0$ and $T/T_{c}=0.3$ in $d=5$.}
\label{fig39}
\end{figure*}

In order to study the electrical conductivity for holographic
$p$-wave superconductor in the Gauss-Bonnet gravity, we follow the
same approach as section \ref{section2}. By choosing the same
perturbation as $\delta A_{y}=A_{y} e^{-i \omega t}$, we obtain
the linearized equation as (\ref{eqay}) with the asymptotic
behavior ($r\rightarrow\infty$) as
\begin{equation}\label{eqasymaygauss}
A_y''(r)+\frac{(d-2) }{r}A_y'(r)+\frac{\omega ^2 L_{\text{eff}}^4}{r^4} A_y(r)=0,
\end{equation}
which has the solution as follows
\begin{equation} \label{aysolgauss}
A_{y} = \left\{
\begin{array}{lr}
A^{(0)}+\frac{A^{(1)}}{r^2}+\frac{A^{(0)} \omega ^2 L_{\text{eff}}^4 \log (\Lambda  r)}{2 r^2}+\cdots, & d=5\\
\bigskip\\
A^{(0)}+\frac{A^{(1)}}{r^3}+\frac{A^{(0)} \omega ^2 L_{\text{eff}}^4}{2 r^2}+\cdots, & d=6\\
\end{array} \right.
\end{equation}%
which is similar to equation (\ref{aysol}) except the
$L_{\rm{eff}}^4$ term which shows the influence of Gauss-Bonnet
gravity. By pursuing the identical procedure as section
\ref{section2}, equations (\ref{eqj}), (\ref{sos}), (\ref{eqsos})
and (\ref{eqohm}) are regained. So, the electrical conductivity
based on holographic approach in Gauss-Bonnet gravity can be
expressed by
\begin{equation} \label{conductivitygauss}
\sigma_{yy} = \left\{
\begin{array}{lr}
\frac{2 A^{(1)}}{i \omega L_{\text{eff}}^2 A^{(0)}}+\frac{i \omega L_{\text{eff}}^2}{2}, & d=5\\
\bigskip\\
\frac{3 A^{(1)}}{i \omega L_{\text{eff}}^2 A^{(0)}}, & d=6\\
\end{array} \right.
\end{equation}%
Again like section \ref{section2}, definition of $\sigma_{yy}$ is
same as $\sigma_{xx}$ in Ref. \cite{Afsoon} which confirms the
fact that calculation of $\sigma_{yy}$ in holographic $p$-wave
superconductors is similar to $\sigma_{xx}$ in holographic
$s$-wave superconductors. Next, by expanding $A_{y}$ as equation
(\ref{ayexpand}) we can plot the behavior of real and imaginary
parts of conductivity as a function of frequency for different
values of the nonlinearity, mass and Gauss-Bonnet parameters in
$d=5$ and $6$ in figs. \ref{fig21a}-\ref{6dd}. Although the
differences in figures, they show the universality trend same as
conductivity behavior in the Einstein gravity which explained in
section \ref{section2}. In $\omega\rightarrow0$ limit, the delta
function behavior of real part of conductivity is related to
imaginary part which has a pole in this region through
Kramers-Kronig relation. The infinite DC conductivity is a tail of
superconducting phase. On the other hand, at large frequency
regime the behavior of real part of conductivity can be expressed
by $Re[\sigma]=\omega^{(d-4)}$. Furthermore, for holographic
$p$-wave superconductor in Gauss-Bonnet gravity
$\omega_{g}\approx8T_{c}$ which is much more than the BSC value
($3.5$). This difference is originated from the fact that the
strong interaction governs in holographic superconductors similar
to \citep{caipp}. However, we observe the deviation from $8$ by
increasing the nonlinearity as well as Gauss-Bonnet parameters.
Increasing the nonlinearity and Gauss-Bonnet parameters or
decreasing the temperature while the other two components are
fixed, shifts the maximum and minimum values of real and imaginary
parts toward larger frequencies. Based on \cite{caipp} for the
fixed values of temperature, real and imaginary parts of
conductivity diminish by enlarging the Gauss-Bonnet effect. It's
true in some cases in the presence of Maxwell electrodynamics. For
example, the behavior of real parts of conductivity in $d=5$ with
$\overline{m}^{2}=-3/4$ confirms this idea while in the same
dimension with $\overline{m}^{2}=0$, it doesn't follow the same
trend. However, this idea doesn't govern more in nonlinear regime.
In general, we can say that the gap frequency depends on mass,
nonlinearity and Gauss-Bonnet parameters in each dimension. The
dependence of conductivity to nonlinearity and Gauss-Bonnet
parameters are shown in figs. \ref{fig36a}-\ref{fig39}.



\section{summary and conclusion}\label{section5}
In this work, by applying AdS/CFT correspondence in higher
dimensional spacetime, we have analyzed the behavior of
holographic $p$-wave superconductor by considering the higher
order corrections in gravity as well as gauge field side. First of
all, we studied the condensation of vector field in the presence
of nonlinear correction to the electrodynamics as given in Eq.
(\ref{eqnon}). This equation can be considered as the leading
order expansion terms of the well-known Born-Infeld, Logarithmic
and Exponential nonlinear electrodynamics. After finding equations
of motion in Einstein gravity, we solved them numerically by
applying suitable conditions and investigated the effect of mass,
dimension and nonlinear parameters. We found out the relation
between critical temperature $T_{c}$ and $\rho^{1/(d-2)}$ in all
cases. Then, we plotted the behavior of condensation as a function
of temperature. Based on the obtained results, we found out that
increasing the value of the mass as well as nonlinearity decreases
the critical temperature. This makes the condensation harder to
form. Also by taking look at graphs, we understand that the
condensation value enlarges for stronger effect of the mass and
nonlinearity which means that vector hair faces with difficulty to
occur. It was argued in \cite{cai13p,chaturverdip15} that the
holographic $p$-wave superconductors undergo first order phase
transition instead of usual second type in some situations but we
didn't observe such a behavior.

Next, in section \ref{section2} we have numerically investigated
the behavior of electrical conductivity as a function of
frequency. For this purpose, we applied an electromagnetic
perturbation as $\delta A_{y}=A_{y} e^{-i \omega t}$ in gravity
side which corresponds to electrical current in CFT part. We
presented the electrical conductivity formula in $d=4, 5$ and $6$
and graphs of real and imaginary parts. The conductivity differs
based on our choice of mass, dimension and nonlinearity. However,
some global trends were observed. Firstly, the real and imaginary
parts follow the Kramers-Kronig relation by having a delta
function and divergence behavior in the low frequency regime.
Infinite DC conductivity is a feature of superconducting phase.
Secondly, at large enough frequencies, the behavior of the real
part can be interpreted by $Re[\sigma]=\omega^{(d-4)}$. Thirdly,
the ratio of $\omega_{g}/T_{c}$ in all cases is much larger that
the BCS value ($3.5$) because holographic superconductors are
strongly coupled. In holographic setup in many cases we found
$\omega_{g}\simeq 8 T_{c}$ while a deviation from this value
occurred by increasing the dimension and nonlinearity effect. The
presence of nonlinear electrodynamics shifts the gap frequency
toward larger values. In section \ref{section3}, with the same
procedure as section \ref{section1}, we found the ratio of
$T_{c}/\rho^{1/(d-2)}$ numerically and achieved the trend of
condensation versus temperature for different values of mass,
nonlinearity and Gauss-Bonnet parameters in different dimensions
for holographic $p$-wave superconductor in Gauss-Bonnet gravity.
Increasing the effect of mass and nonlinear parameter follows the
same behavior as Einstein case. Furthermore, going up the
Gauss-Bonnet parameter $\alpha$ hinders the superconducting phase
by diminishing the critical temperature. Besides, the Gauss-Bonnet
term $\alpha$ doesn't change the order of phase transition. In
section \ref{section4}, the electrical conductivity of
$(d+1)$-dimensional holographic $p$-wave superconductor in
Gauss-Bonnet gravity with higher order corrections in gravity and
gauge fields was studied. Same as section \ref{section2}, the
conductivity formula and behavior of real and imaginary parts
impressed by mass, nonlinearity and Gauss-Bonnet parameters in
different dimensions were obtained. The universal behaviors same
as Einstein case were achieved. In addition, increasing the effect
of nonlinear and Gauss-Bonnet parameters or decreasing the
temperature shifts the gap energy toward larger frequencies. In
general, the gap frequency $\omega_{g}$ is depended on mass,
nonlinearity and Gauss-Bonnet terms. It would be interesting to
investigate the effect of backreaction in this case.

\begin{acknowledgments}
We thank Shiraz University Research Council. The work of AS has
been supported financially by Research Institute for Astronomy and
Astrophysics of Maragha (RIAAM), Iran.
\end{acknowledgments}


\end{document}